\PassOptionsToPackage{x11names}{xcolor}
\documentclass[twocolumn]{aastex631}
\usepackage{mathtools}
\usepackage{amsmath}
\usepackage{graphicx}
\usepackage{appendix}
\usepackage{amssymb}
\usepackage{float}
\usepackage{multirow}
\usepackage{booktabs}
\usepackage{colortbl}
\usepackage{tablefootnote}
\usepackage{makecell}
\graphicspath{{}}
\hypersetup{colorlinks=true, citecolor=blue, linkcolor=cyan, urlcolor=magenta}
\usepackage[T1]{fontenc}
\usepackage{bm}
\usepackage{afterpage}

\begin{document}

\title{Investigating quenching in Recently Quenched Elliptical galaxies with HI studies}

\author{Deepak~K.~Deo}
\affiliation{Department of Physics \& Astronomy, University of Missouri-Kansas City, Kansas City, MO 64110, USA}
\email{dd9wn@umsystem.edu}

\author{Daniel~H.~McIntosh}
\affiliation{Department of Physics \& Astronomy, University of Missouri-Kansas City, Kansas City, MO 64110, USA}

\author{Sravani~Vaddi}
\affiliation{Arecibo Observatory, NAIC, HC3 Box 53995, Arecibo, Puerto Rico, PR 00612, USA}

\author{Kameswara~B.~Mantha}
\affiliation{School of Physics \& Astronomy, University of Minnesota, Twin Cities, 116 Church St SE, Minneapolis, MN, 55455}

\author{Ruta~Kale}
\affiliation{National Centre for Radio Astrophysics, Tata Institute of Fundamental Research, S. P. Pune University Campus, Ganeshkhind, Pune 411007, India}

\author{Alfonso~G.~Franco}
\affiliation{Rockhurst University, Department of Physics, 1100 Rockhurst Road, Kansas City, MO 64110, USA}

\author{Paul~Rulis}
\affiliation{Department of Physics \& Astronomy, University of Missouri-Kansas City, Kansas City, MO 64110, USA}

\begin{abstract}

Recently Quenched Elliptical galaxies (RQEs) represent a critical phase in the transition from star-forming to quiescent galaxies. However, the mechanisms driving their quenching remain elusive. We conduct a multi-wavelength analysis of 155 RQEs, along with their precursors (preRQEs) and descendants (postRQEs), focusing on their neutral hydrogen (HI) content and star formation properties. Contrary to conventional quenching models emphasizing gas depletion, RQEs retain substantial HI reservoirs ($f_{\text{gas}} \geq 17\%$), suggesting that quenching is not primarily driven by gas exhaustion.
We identify a critical halo mass threshold at $\log M_{\text{halo}} = 12.1 M_{\odot}$, delineating different evolutionary pathways for RQEs. This threshold aligns with the transition from cold-mode to hot-mode gas accretion in theoretical models. RQEs in lower-mass halos ($\log M_{\text{halo}} < 12.1 M_{\odot}$) likely experience rapid quenching, possibly initiated by major mergers, followed by brief AGN activity and sustained LINER emission. We propose two evolutionary pathways: (a) rapid quenching via major mergers followed by AGN/LINER activity and passive evolution, and (b) rapid quenching followed by rejuvenation through minor mergers before evolving into more massive, long-term quenched ellipticals.
These results challenge the conventional understanding of galaxy quenching, especially in low-density environments where RQEs typically reside. Our findings suggest that while RQEs may follow a rapid quenching pathway, their evolution is influenced by interactions between gas accretion modes, feedback mechanisms, and environmental factors. Future observations with advanced radio interferometers like SKA will be crucial for elucidating the quenching mechanisms in RQEs and their role in galaxy evolution.

\end{abstract}


\section{Introduction} \label{sec:intro}

Galaxies are complex systems that evolve over cosmic time, undergoing various physical processes including quenching—the cessation of star formation often due to the depletion, heating, or stripping of cold neutral hydrogen (HI) gas. This transformation converts galaxies from blue, gas-rich, star-forming types to red, gas-poor, quiescent types, creating a prominent bimodality in the optical color-magnitude and stellar mass-star formation rate planes \citep{2004ApJ...600..681B, 2007ApJS..173..267S, 2007ApJ...670..156D}.
While blue galaxies are typically Late Type Galaxies (LTGs), mainly spirals, red galaxies are Early Type Galaxies (ETGs) typically consisting of Elliptical (E) or S0 populations. Approximately 6\% of ETGs in the low-redshift universe ($0.02 < z < 0.05$) are blue ETGs, representing a transitional phase between star-forming spirals and quiescent ellipticals \citep{2004ApJ...601L.127F, 2009MNRAS.396..818S, 2010A&A...515A...3H, 2014MNRAS.442..533M, 2018MNRAS.475..788M}. 
Among these, recently quenched ellipticals (RQEs)—a subset of high-mass blue ETGs ($M_{\star}\geq 2\times 10^{10}M_{\odot}$) identified at low redshifts ($0.02 \lesssim z \lesssim 0.08$) by \cite{2014MNRAS.442..533M} using data from the Sloan Digital Sky Survey Data Release 4 (SDSS DR4; \citet{2006ApJS..162...38A})—are particularly intriguing.
These RQEs, typically residing at the centers of small groups (median halo mass $M_{\text{halo}} \simeq 10^{12.2} M_{\odot}$), present a unique opportunity to study quenching in environments usually conducive to star formation \citep{2009A&A...498..407G}. 
Given the correlation between cold HI gas—the raw fuel for star formation—and quenching, this study investigates the HI reservoirs and star formation characteristics of RQEs using a multi-wavelength approach. By comparing RQEs with their potential evolutionary counterparts—both star-forming precursors and long-quenched descendants—we aim to uncover insights into the mechanisms driving their unexpected quenching and trace their evolutionary path.

Recently quenched ellipticals (RQEs), primarily located at the centers of low-mass dark matter (DM) halos—with over 90\% residing in halos of $M_{\text{halo}} \leq 10^{12.7} M_{\odot}$—constitute a particularly intriguing type of quenched galaxy. Serendipitously discovered by \cite{2014MNRAS.442..533M} (hereafter M14) during their search for galaxies that have recently undergone major mergers, RQEs in the local universe ($z\leq 0.08$) seem to represent a population that has experienced rapid quenching on timescales of a few hundred million years, consistent with the post-major merger models of \cite{1988ApJ...325...74S} and \cite{2008ApJS..175..356H}. The likelihood of a major merger as the quenching mechanism is supported by the very young light-weighted stellar age of RQEs (Age$_{lw} \leq 3$ Gyr), causing them to appear predominantly blue in both color-mass and color-color ($urz$) diagrams. This indicates a significant young stellar population from recent, substantial star formation, followed by rapid quenching. Furthermore, their elliptical morphology suggests a major-merger origin \citep{1972ApJ...178..623T, 1977egsp.conf..401T, 2003ApJ...597..893N, 1988ApJ...331..699B, 1992ApJ...393..484B, 1992ApJ...400..460H, 1993ApJ...409..548H}. M14 proposes that they are plausible 'first-generation' ellipticals formed through major mergers, which also accounts for their very young light-weighted stellar age. Intriguingly, RQEs are found at the centers of small DM halos (median halo mass $M_{\text{halo}} \simeq 10^{12.2} M_{\odot}$), where ongoing star formation would typically be expected due to active cold gas accretion. This indicates that the process responsible for their quenching was disruptive enough to interfere with cold gas accretion.

Galaxies with ongoing star formation typically exhibit a young light-weighted stellar age ($\lesssim 3$ Gyr), often due to the presence of bright young stars \cite{2005MNRAS.362...41G, 2010MNRAS.404.1775T}. When galaxies display a similar young light-weighted stellar age and appear blue in optical colors yet lack active star formation, it suggests a recent cessation of star formation \cite{2009MNRAS.396..818S}. The major merger model proposed by \cite{2008ApJS..175..356H} predicts the possibility of observing such young blue elliptical galaxies that are quenched as long as they possess alive $A$ stars before transitioning into evolved, red-and-dead, typical ellipticals \cite{2015MNRAS.451..433H, 2023A&A...671A.166G}. Post-starburst galaxies ($E+A$), characterized by blue optical colors without ongoing star formation, are also often the result of major mergers \cite{1996ApJ...466..104Z}.

The elliptical morphology of RQEs might also stem from major mergers of spiral galaxies, leading to the formation of `first-generation' ellipticals, as suggested by M14. Major mergers can disrupt the disk structures of spiral galaxies, resulting in ellipticals \cite{1972ApJ...178..623T, 2003LNP...626..327B}. However, not all RQEs are necessarily first-generation ellipticals. Young blue elliptical galaxies could also result from one or more minor mergers \citep{2009MNRAS.394.1713K, 2017A&A...598A..45G, 2023MNRAS.520.5870G}. A study using $N-$body simulations showed that repeated minor mergers, with mass ratios ranging from 4:1 to 50:1, can transform galaxies into elliptical-like structures, challenging the belief that major mergers are the sole pathway to forming elliptical galaxies \cite{2007A&A...476.1179B}.

Central galaxies in low-density environments or small dark matter halos (\(\log M_{\text{halo}} \lesssim 13.5 M_{\odot}\)) are typically expected to actively form new stars by accreting star-forming cold neutral Hydrogen (HI) gas \cite{2009A&A...498..407G}. Yet, this is not the case for RQEs, which are unusually quenched. The exact cause of their quenching, whether due to a major merger or another mechanism, remains unclear. Quenching can occur primarily due to the exhaustion, heating, or stripping of cold HI gas. This paper addresses the HI properties of RQEs, aiming to unveil clues to their puzzling quiescence. We also compare the RQEs to their plausible precursors and descendants to understand the evolutionary trend in relation to their HI content.

Current quenching mechanisms in the literature fail to fully explain the unusual quenching observed in RQEs residing at the center of small groups (hereafter central RQEs). For instance, strong outflows from AGN feedback can initially quench galaxies in central group environments with halo masses less than $10^{13} M_{\odot}$, but this mechanism struggles to sustain quenching due to the expected ongoing accretion of gas \citep{2006MNRAS.365...11C, 2012MNRAS.425.2027K}. Gravitational quenching, maintained by gravitational heating, is effective primarily in halos larger than $10^{12.85} M_{\odot}$ \citep{2008MNRAS.383..119D, 2013MNRAS.429.3353N}, while quenching due to supernova feedback is relevant in halos smaller than $10^{11} M_{\odot}$, outside the range in which central RQEs exist. Halo quenching could be a plausible explanation for quenching in centrals of small groups if additional feedback from AGN can keep the gas from cooling down and forming stars \cite{2012MNRAS.427.1816G}. However, only 6\% of central RQEs host AGN, specifically Seyferts, which are known to have short active periods, typically less than or equal to 100 million years \citep{2004cbhg.symp..169M, 2006ApJS..166....1H}. This brief active period is insufficient compared to the RQE phase—assumed to be on the order of an A-type star's lifetime—suggesting that Seyfert activity is likely not connected to the quenching process in central RQEs.

To constrain the quenching mechanism at play in central RQEs, it is crucial to examine the properties of the star-forming cold gas within these galaxies. Studying the total mass, morphology, and kinematics of the cold gas can offer significant insights into the quenching process. For example, a disturbed gas morphology may point to recent mergers as the cause of quenching. Additionally, analyzing gas distribution and kinematics—such as inflows or outflows—can reveal whether quenching is occurring inside-out or outside-in \citep{Lin_2019}. Additionally, the presence and properties of hot halo gas and its potential impact on the infall of cooler gas can suggest a halo quenching scenario \citep{2012MNRAS.427.1816G}.

Ideally, one would study these properties for molecular hydrogen (H$_{2}$) gas, as it is directly responsible for star formation. However, direct observation of H$_{2}$ is severely limited due to its lack of a permanent dipole moment, making its dipole rotational transitions highly forbidden. As an alternative, the rotational transition lines of carbon monoxide (CO) molecules are often used as a proxy for studying H$_{2}$ \citep{Bolatto_2013}. In the absence of CO data, neutral atomic hydrogen (HI) can be studied as it serves as the raw fuel for star formation. HI converts to H$_{2}$ in regions where the column density and metallicity are high enough to shield H$_{2}$ from photodestruction by interstellar ultraviolet photons, thereby initiating the star formation process \citep{Krumholz_2012}. Thus, neutral HI can be regarded as the reservoir for future star formation \citep{2015MNRAS.450..618K}, and its content can be investigated to identify clues that can constrain plausible mechanisms responsible for quenching.

For instance, a few HI studies on quenched ETGs residing in low-density environments have revealed the presence of large reservoirs of HI gas implying mechanisms that do not remove star-forming cold gas away from the galaxy could be responsible for quenching [e.g., \citep{2010MNRAS.409..500O, 2012MNRAS.422.1835S}].
Excess HI with insufficient star formation could also be linked to a minor merger scenario as shown by a HI study on passive galaxies via JVLA (Jansky Very Large Array) and uGMRT (Upgraded Giant Meterwave Radio Telescope) by \citep{2016MNRAS.462..382G, 2018MNRAS.476..896G}. They found that the HI gas in these galaxies is in the form of a regularly rotating disc and that the source of this gas could be a minor merger in the past.
Another empirical HI study on quenched galaxies by \cite{2009MNRAS.400.1225C} revealed that they could have large amounts of HI gas if they rejuvenate their gas reservoir, which can happen via a gas-rich major or minor merger or through accretion of gas from the surrounding intergalactic medium (IGM).

Simulation studies on mass-selected samples, including both ellipticals and spirals, regardless of their environment (central or satellite), have predicted that high-mass galaxies of similar stellar mass range as RQEs are on average 75\% deficient in HI gas \citep{2014Natur.509..177V}. 
However, simulations that take into account the role of the environment have shown that centrals of similar mass ranges as RQEs ($10^{10} - 10^{11} M_{\odot}$) tend to contain their HI reservoirs, with only a small fraction of them being devoid of HI [\cite{2013MNRAS.434.2645D}]. 
These contrasting findings further motivate our investigation into the HI content of RQEs, aiming to determine whether they are HI deficient or rich and how this relates to their star formation quenching.
 
The paper is organized as follows: Sec. \ref{sec:sample} discusses the sample of RQEs and comparable subpopulation we are considering for this study and how they were selected. Sec. \ref{sec:HIdata} presents information on what kind of HI data is available for our sample and how they were retrieved. Sec.\ref{sec:HIanalyses} performs HI analysis and presents results related to various analyses related to gas content and star formation activity. Sec. \ref{sec:dis} explores the implications of these results and suggests avenues for future research. The paper concludes with a summary of our findings in Sec. \ref{sec:con}.
Throughout the paper, we adopt a standard flat cosmology with $H_{0} = 70$ km/s/Mpc, $\Omega_{\Lambda} = 0.7$, and $\Omega_{m}=0.3$. All magnitudes are reported in the AB system, as defined by \cite{1983ApJ...266..713O}.

\section{The sample}\label{sec:sample}
M14 utilized the galaxy samples and data from the SDSS (Sloan Digital Sky Survey) fourth data release (DR4, \cite{2006ApJS..162...38A}) to identify a potentially evolutionarily interesting subpopulation of elliptical galaxies in the present-day universe. Comprising 2\% of blueETGs with a central bulge ($R_{90}/R_{50}$)\footnote{ratio of radii enclosing 90\% and 50\% of the $r$ Petrosian flux for galaxies in SDSS \citep{2001AJ....122.1238S, 2001AJ....122.1861S, 2003AJ....125.1682N}.} $\geq 2.6$, within the specified stellar mass range ($2\times 10^{10} - 9.7\times 10^{11} M_{\odot}$) and redshift range ($0.02\lesssim z \lesssim 0.08$), RQEs exhibit substantial evidence, such as no or very weak H$\alpha$ emission (H$\alpha$ equivalent width, EW $\leq 2$ Å) and a young light-weighted stellar age ($\leq 3$ Gyr), indicating recent quenching of their star formation.
As such, M14 argued, these galaxies can explain 24\% of the buildup of comparable red-and-dead galaxies in the last 3 billion years.
M14 found that 90\% of these RQEs were the central (most massive) galaxy in small dark-matter halos (residing in $M_{\text{halo}}\leq1.5\times10^{13.5}M_{\odot}$, 90\% of which in $M_{\text{halo}}\leq5\times10^{12}M_{\odot}$), and argued that such RQEs are good candidates for recent major-merger remnants or so-called ‘first-generation' elliptical galaxies (see also \citet{2015MNRAS.451..433H}). 
To improve our understanding of galaxy quenching at the centers of small halos that should remain actively star-forming, we focus our cold gas study on the 155 central RQEs from M14.
Additionally, we select 239 star-forming ellipticals and 301 long-quenched ellipticals as comparable subpopulations from the SDSS that are meticulously matched to RQEs across various parameters such as stellar mass range, halo mass range, redshift range, morphology, central concentration, and environment (centrals), with the exception of star formation activity to compare the results of our primary objects (central RQEs) with them. Comparable subpopulation are selected in way that they represent a sample of plausible actively star-forming precursors to RQEs (preRQE) and long-quenched descendant to RQEs (postRQE). In the following subsections, we discuss in detail how RQEs and comparable subpopulations were selected and how they compare to each other in various properties. This meticulously constructed sample allows us to trace the evolutionary path of RQEs, focusing on their HI properties, and to better understand the mechanisms driving their quenching in environments that would typically support active star formation.

\subsection{RQE selection}\label{sec:rqesel}
RQEs were serendipitously discovered by M14 while searching for plausible recent post-major mergers, aiming to constrain the role that major mergers play in the migration of galaxies from blue-starforming phase to red-and-dead phase. M14 first identified blue ETGs in the low-z universe, from which they identified ellipticals, followed by quenched ellipticals, and eventually centrals. Below, we explain how M14 briefly defined blue ETGs from a large sample, used visual classification to identify ellipticals and post-mergers, utilized emission line data to identify quenched ellipticals, and then used a group catalog from \cite{2007ApJ...671..153Y} (hereafter Y07) to separate them into satellites and centrals.

M14 identified 172 RQEs from a complete, mass-selected ($M_{\star} \geq 2\times 10^{10} M_{\odot}$), and redshift-limited ($0.02 \lesssim z \lesssim 0.08$) sample of 63,454 galaxies (hereafter \textit{sample1}) using the New York University Value-Added Galaxy Catalogue (NYU-VAGC; \cite{2005AJ....129.2562B}) based on SDSS Data Release 4 (\cite{2006ApJS..162...38A}). They calculated stellar masses utilizing rest-frame $(g-r)$ colors, absolute $r$-band magnitudes, and stellar mass-to-light ratios from \cite{2003ApJS..149..289B}. By applying an empirical color cut $^{0.1}(g-r) \leq 0.81 + 0.1[log_{10}(M_{gal,\star}/M_{\odot}h^{-2}) - 10.0]$ to select blue-cloud galaxies, and a central concentration cut ($C_r = R_{90}/R_{50} \ge 2.6$; \cite{2001AJ....122.1861S}) to isolate spheroid-dominated systems, they identified 8403 unusually blue early-type galaxies (ETGs).

M14 then employed visual classification to identify ellipticals (E; 1368), peculiar ellipticals (pE; 124), and spheroidal post-mergers (SPM; 110) among the 8403 blue ETGs, resulting in 1602 plausible merger remnants (see M14 for details). They used the BPT diagram (\cite{1981PASP...93....5B}) to classify blue ellipticals (Es and pEs) into star-forming and quiescent galaxies, following methods by \cite{2010ApJ...719..415P} (hereafter PG10) to identify spectroscopically quiescent ellipticals, including those with no detectable H${\alpha}$ line (flux S/N < 3) or weak H${\alpha}$ emission (EW $\leq 2\AA$). Their $(u-r)$ versus $(r-z)$ colors were analyzed to confirm the non-star-forming nature and identify Seyfert/LINER systems lacking star formation (\cite{Holden_2012}). Speculating that spectroscopically quiescent blue ellipticals, with or without AGN, had experienced recent star formation that was subsequently quenched, M14 used light-weighted stellar ages (Age$_{lw}$) from \cite{2005MNRAS.362...41G} to identify 172 recently quenched ellipticals (Age$_{lw} \leq 3$ Gyr) in their mass-limited and volume-limited sample.

Finally, M14 crossmatched the 172 RQEs with the group catalog developed by Y07 to separate them into satellites and centrals. Y07 identifies the centrals of galaxy groups using a halo-based group finder, which includes the Friends-of-Friends (FOF) algorithm for initial group identification and a ranking method based on luminosity and halo mass to determine the central galaxy. The central galaxy is typically the most massive or luminous galaxy in the group, not necessarily located at the geometric center of the dark matter halo. Based on the crossmatch, M14 found that 155 of the 172 RQEs were centrals in the Y07 group catalog, and the rest were identified as satellites (galaxies orbiting around the centrals). These 155 RQE centrals were found to reside in low-mass (median $M_{\text{halo}}=1.5 \times 10^{12} M_{\odot}$) dark-matter halos and are the primary focus of this paper.

\subsection{Comparable Subpopulations}
To better understand the HI content and potential quenching mechanisms in RQEs, we selected two comparison groups from SDSS: star-forming ellipticals, termed plausible RQE precursors (preRQEs), and long-quenched ellipticals, termed plausible RQE descendants (postRQEs). These comparison groups were matched to RQEs in terms of their stellar mass range, halo mass range, environment (i.e. centrals), redshift range, and morphology (Ellipticals with $C_r \geq 2.6$), but differed in star formation activity to ensure a plausible evolutionary link from star-forming to recently quenched to long-quenched phases. Comparing RQEs with preRQEs helps investigate the conditions leading to quenching, examining whether the quenching was due to heating, removal, or consumption of HI gas. In contrast, comparing RQEs with postRQEs provides insights into the duration and stability of the quenching process and the mechanisms that could sustain quenching for a long time. This approach aims to constrain the potential quenching mechanisms in RQEs, thereby providing valuable insights into their evolutionary stages and quenching processes.

\begin{figure*}
    \centering
    \includegraphics[width=1\textwidth]{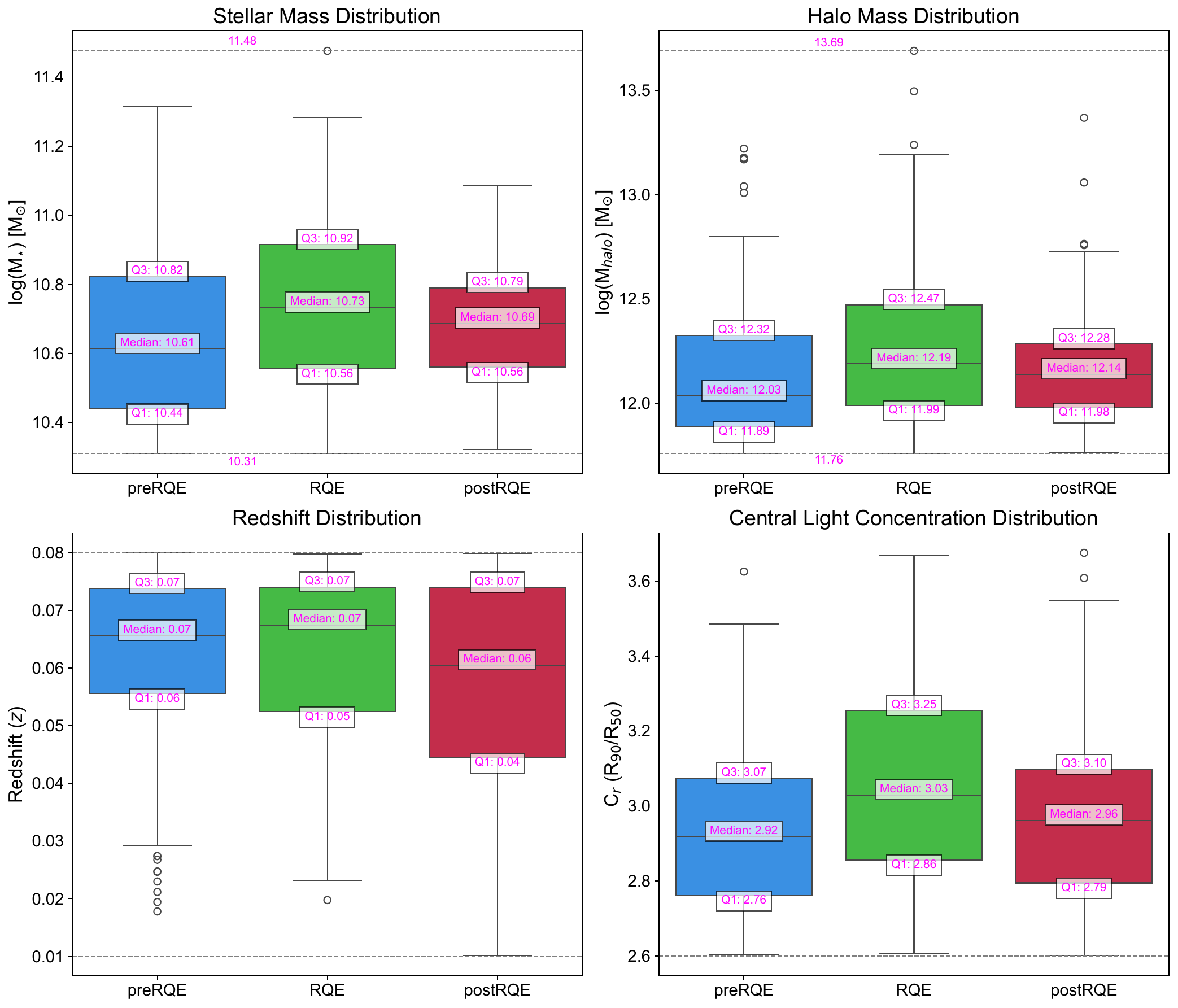}
    \caption{Distribution of stellar mass, halo mass, redshift, and central light concentration (reflecting the prominence of the central bulge) for RQEs and comparable subpopulations. The figure presents boxplots comparing the properties of RQEs with preRQEs and postRQEs. The boxes represent the IQR (interquartile range), illustrating the distribution of the central $50\%$ of the population. Whiskers extend to 1.5 times the IQR, and outliers are shown as points beyond the whiskers. Medians are annotated within the boxes, and Q1 and Q3 values are indicated, highlighting the distribution and any differences or similarities among the subpopulations. Dashed lines indicate selection cuts (related to RQEs) to which preRQEs and postRQEs were matched.}
    \label{fig:sample_comp}
\end{figure*}

\begin{figure*}
    \centering
    \includegraphics[scale=0.8]{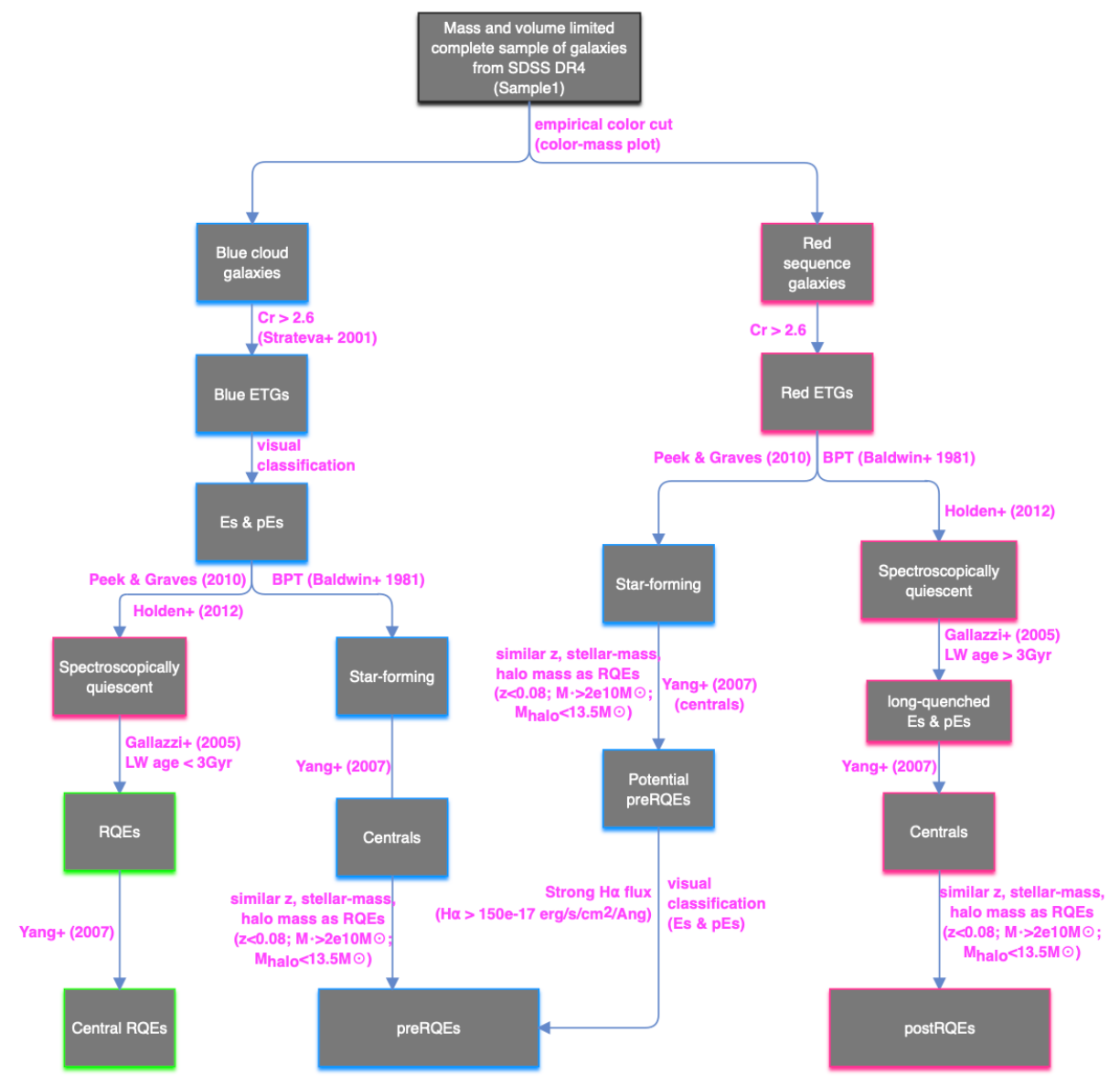}
    \caption{Flowchart illustrating the selection process for RQEs and control samples (preRQEs and postRQEs).}
    \label{fig:flowchart}
\end{figure*}

Details of the selection of preRQE and postRQE are provided in the next two subsections, and a simplified visual representation of the selection process for all three populations is shown in Figure \ref{fig:flowchart}. Figure \ref{fig:sample_comp} presents boxplots comparing the distributions of preRQEs and postRQEs with RQEs across various parameters. The boxes represent the interquartile range (IQR), illustrating the distribution of the central 50\% of the population. Medians are annotated within the boxes, and $Q1$ ($25\%-$tile) and $Q3$ ($75\%-$tile) values are indicated, highlighting the distribution and any differences or similarities among the subpopulations. Although the selection cuts for the subpopulations were matched to RQEs across various properties, the distributions do not need to match exactly. However, the distributions are generally similar. Given the overlap in IQRs and the closeness of median values, it is reasonable to say that preRQEs and postRQEs are quite similar to RQEs in the distribution of these four properties. Specifically, in terms of median stellar mass, postRQEs are more similar to RQEs, as is the case with halo mass and central light concentration. In contrast, the redshift distribution of preRQEs is more similar to that of RQEs. The differences observed are relatively small, generally less than 4\%, and are not significant enough to suggest a substantial deviation between the groups. Therefore, preRQEs and postRQEs can be considered generally similar to RQEs in all four properties' distributions shown in the Figure \ref{fig:sample_comp}.

For a more comprehensive comparison of the properties of preRQEs, RQEs, and postRQEs, we present Table \ref{tab:comptable}. This table provides detailed statistics for key physical and observational characteristics, including stellar mass, halo mass, redshift, central concentration ($C_r$), and light-weighted stellar age (Age$_{lw}$). Additionally, it includes the distribution of galaxy types (E, pE) and emission types (HII, Quiescent, Seyfert, LINER, Y06). The Y06 classification refers to galaxies with weak-LINER emission, as defined by M14 following \citet{2006ApJ...648..281Y}. The table presents median values, interquartile ranges ($Q1-Q3$), and full ranges ($min-max$) for continuous variables, as well as percentage distributions for categorical variables. These comprehensive statistics enable a thorough comparison of the different galaxy populations within our study.

{\setlength{\tabcolsep}{10pt} 
\begin{deluxetable*}{l l c c c}
\tablecaption{Key Properties of preRQE, RQE, and postRQE Galaxies \label{tab:comptable}}
\tablewidth{\textwidth}
\tablehead{
\colhead{\textbf{Quantity}} & \colhead{\textbf{Statistic}} & \colhead{\textbf{preRQE}} & \colhead{\textbf{RQE}} & \colhead{\textbf{postRQE}}
}
\startdata
\textbf{Sample Size} & count & 239 & 155 & 301 \\
\hline
\multirow{3}{*}{\textbf{log(M$_{\star}$) [M$_{\odot}$]}} & Median & 10.61 & 10.73 & 10.69 \\
 & IQR & 10.44 - 10.82 & 10.50 - 10.86 & 10.58 - 10.81 \\
 & min - max & 10.31 - 11.32 & 10.31 - 11.48 & 10.32 - 11.09 \\
\hline
\multirow{3}{*}{\textbf{log(M$_{halo}$) [M$_{\odot}$]}} & Median & 12.04 & 12.19 & 12.14 \\
 & IQR & 11.86 - 12.30 & 12.00 - 12.49 & 11.98 - 12.28 \\
 & min - max & 11.76 - 13.22 & 11.76 - 13.69 & 11.76 - 13.37 \\
\hline
\multirow{3}{*}{\textbf{Redshift ($z$)}} & Median & 0.0656 & 0.0675 & 0.0605 \\
 & IQR & 0.056 - 0.074 & 0.053 - 0.075 & 0.049 - 0.078 \\
 & min - max & 0.018 - 0.080 & 0.020 - 0.080 & 0.010 - 0.080 \\
\hline
\multirow{3}{*}{\textbf{\boldmath{$C_r$} ($R_{90}/R_{50}$)}} & Median & 2.92 & 3.03 & 2.96 \\
 & IQR & 2.76 - 3.07 & 2.85 - 3.25 & 2.81 - 3.11 \\
 & min - max & 2.60 - 3.63 & 2.61 - 3.67 & 2.60 - 3.68 \\
\hline
\multirow{3}{*}{\textbf{\boldmath{$t_{lw, age}$} [Gyr]}} & Median & 9.52 & 9.29 & 9.80 \\
 & IQR & 9.44 - 9.62 & 9.23 - 9.41 & 9.75 - 9.85 \\
 & min - max & 8.99 - 10.00 & 8.97 - 9.48 & 9.50 - 9.98 \\
\hline
\multirow{2}{*}{\textbf{Galaxy Type}} & E (\%) & 93.7 & 84.5 & 100.0 \\
 & pE (\%) & 6.3 & 15.5 & 0.0 \\
\hline
\multirow{5}{*}{\textbf{Emission Type}} & HII (\%) & 100.0 & 0.0 & 0.0 \\
 & Quiescent (\%) & 0.0 & 67.7 & 49.2 \\
 & LINER (\%) & 0.0 & 20.0 & 38.2 \\
 & Seyfert (\%) & 0.0 & 6.5 & 5.0 \\
 & Y06 (\%) & 0.0 & 5.8 & 7.6 \\
\enddata
\tablecomments{Comparative analysis of key properties for preRQE, RQE, and postRQE galaxies. The table provides statistics for stellar mass, halo mass, redshift, central concentration ($C_r$), and light-weighted stellar age (Age$_{lw}$). It also includes distributions for galaxy types (E, pE) and emission types (HII, Quiescent, Seyfert, LINER, Y06). The Y06 classification represents galaxies with weak-LINER emission, following \citet{2006ApJ...648..281Y}. Values are reported as medians, interquartile ranges ($Q1-Q3$), and full ranges ($min-max$) for continuous variables, with percentage distributions for categorical variables.}
\end{deluxetable*}}

\subsubsection{Plausible RQE descendants selection}
PostRQEs represent the natural transition of RQEs into classic \textit{red-and-dead} elliptical galaxies, assuming passive evolution over a few billion years after their initial quenching phase (e.g., Figure 1 of \citet{2008ApJS..175..356H}). For our comparative study, we selected postRQEs from a subset of galaxies identified by M14 as plausible RQE descendants. 

M14's motivation for identifying RQE descendants stemmed from the anticipated redward evolution of RQEs. They hypothesized that if RQEs evolved passively, there should exist red galaxies with similar properties but older light-weighted stellar ages at late cosmic times. 
To find these, M14 searched the red-sequence population of early-type galaxies (ETGs) for non-star-forming galaxies that matched RQEs in terms of small velocity dispersions and high stellar metallicities, but exhibited older stellar populations. They selected redETGs that were either spectroscopically quiescent or exhibited AGN (LINER or Seyfert) emission, with round morphologies and a long period of quiescence.

Our sample was further refined by selecting central ellipticals (Es or pEs) with no HII emission and no significant H$\alpha$ emission, matching the stellar mass, halo mass, and redshift range of RQEs. M14 utilized an empirical color cut described in section \ref{sec:rqesel} on \textit{sample1} to identify red-sequence galaxies. From these, M14 applied an $r$-band concentration cut ($C_r = R_{90}/R_{50} \ge 2.6$) to isolate spheroid-dominated redETGs. RedETGs with light-weighted stellar ages greater than 3 Gyr and fulfilling the non-SF criteria (indication of being long-quenched) established by \cite{Holden_2012}  with a round shape ($b/a > 0.6$) similar to typical RQEs were considered plausible RQE descendants.

The study by M14 resulted in the identification of 475 plausible postRQEs from \textit{sample1}. Utilizing the group catalog developed by \cite{2007ApJ...671..153Y}, these postRQEs were divided into central and satellite galaxies. By applying the same criteria for environment (selecting only centrals), redshift, stellar mass, and halo mass ranges as for the RQEs, the sample was refined to 306 plausible postRQE candidates.

\subsubsection{Morphology confirmation using DECaLS}
To visually confirm the elliptical nature of these postRQEs, the recently released (December 2022) DECaLS DR10 (The Dark Energy Camera Legacy Survey Data Release 10) images were used instead of the SDSS images. DECaLS which uses the Dark Energy Camera (DECam) on the Blanco 4m telescope, located at the Cerro Tololo Inter-American Observatory, boasts twice the resolution and can observe, on average, 1 magnitude fainter in the r-band compared to SDSS [\cite{2019AJ....157..168D}]. Additionally, the DECaLS images are known to capture tidal, spiral arm, or weak bar features that may be missed in SDSS (\cite{2022MNRAS.509.3966W}), making it a superior choice for confirming the elliptical nature of postRQEs, which are expected to be round and elliptical like RQEs.

\begin{figure*}
    \centering
    \includegraphics[width = 0.97\textwidth]{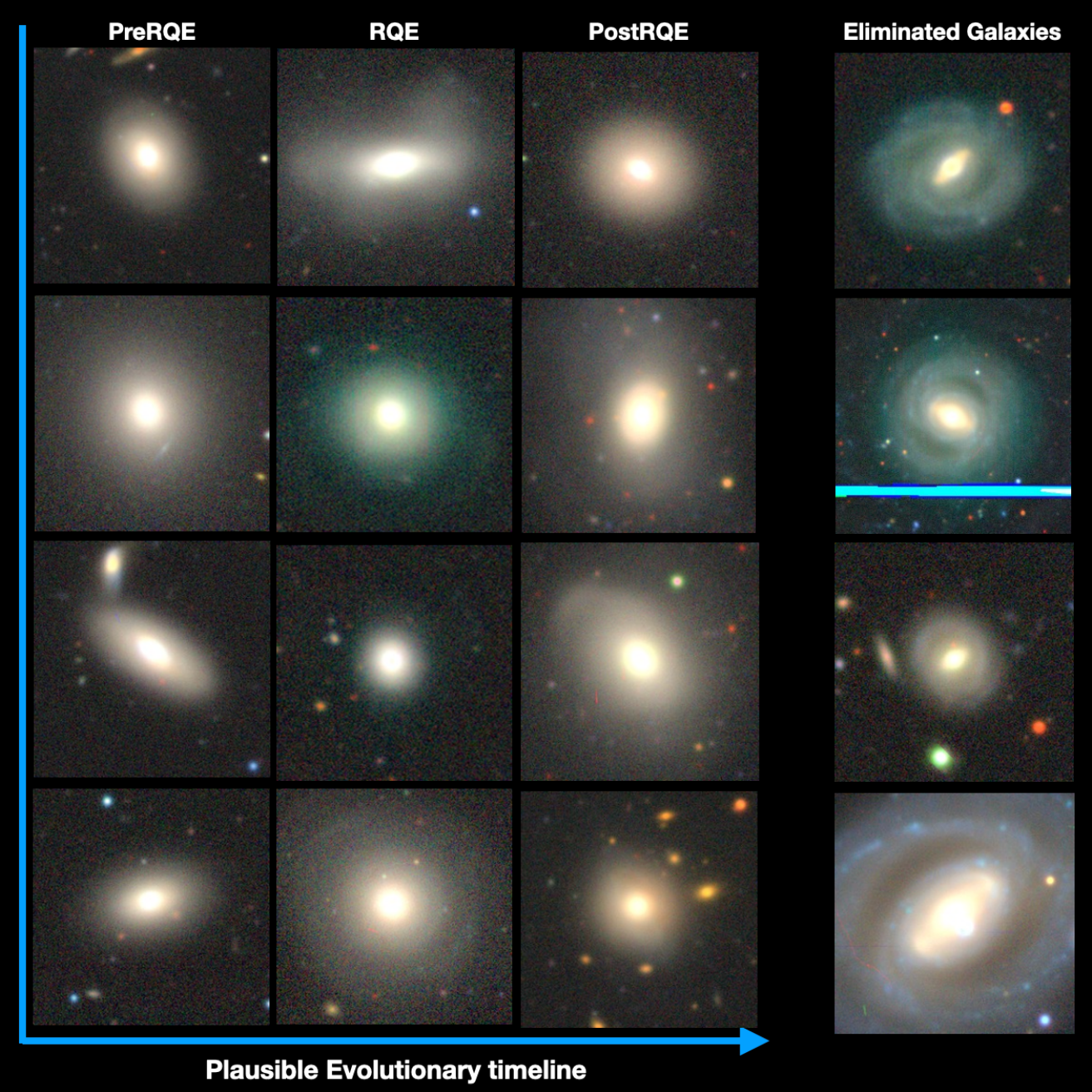}
    \caption{A sample representation of our sample (plausible RQE precursors, RQEs, plausible RQE descendants) along with those that got eliminated from the sample after reviewing their DECaLS DR10 images. The cutout images are taken from DECaLS DR10.}
    \label{fig:cutouts}
\end{figure*}

Out of the 306 postRQE candidates, 301 were confirmed to be ellipticals, and the remaining were eliminated as they had S (spiral) or iD (inclined Disk) features. 
The images of the RQEs and preRQEs (see section \ref{sec:preRQEsel} for preRQE selection criteria) were also revisited to check for contamination in their selection as ellipticals, and the selection was found to be 100\% consistent with the DECaLS images. A sample representation of our sample (preRQE, RQE, postRQE) along with those that got eliminated is shown in the Figure \ref{fig:cutouts}. The cutout images are taken from DECaLS DR10.

\subsubsection{Plausible RQE precursors selection} \label{sec:preRQEsel}
In our search for plausible RQE precursors (preRQEs), we aimed to identify central elliptical galaxies with stellar mass, halo mass, and redshift similar to RQEs, but exhibiting spectroscopic evidence of ongoing star formation. These preRQEs represent star-forming ellipticals expected to soon quench their star formation and evolve into RQEs. Below we describe their selection process in detail.

Our selection process drew from two sources in M14's analysis: star-forming blue early-type galaxies (blueETGs) and star-forming red early-type galaxies (redETGs). M14 identified these populations from a volume-limited sample of high-mass galaxies using color-mass cuts to separate blue-cloud and red-sequence galaxies (see Section \ref{sec:rqesel}), followed by an r-band concentration cut ($C_r \geq 2.6$) to isolate early-type galaxies. 

M14 employed visual inspection on blueETGs to confirm E and pE morphologies and used the BPT diagram along with methods from \cite{2010ApJ...719..415P} to identify spectroscopically star-forming E/pE populations. We further used Y07 to identify central galaxies within these populations that matched RQEs in halo mass, stellar mass, and redshift, resulting in our initial preRQE population.

\begin{figure}
    \includegraphics[width = \columnwidth]{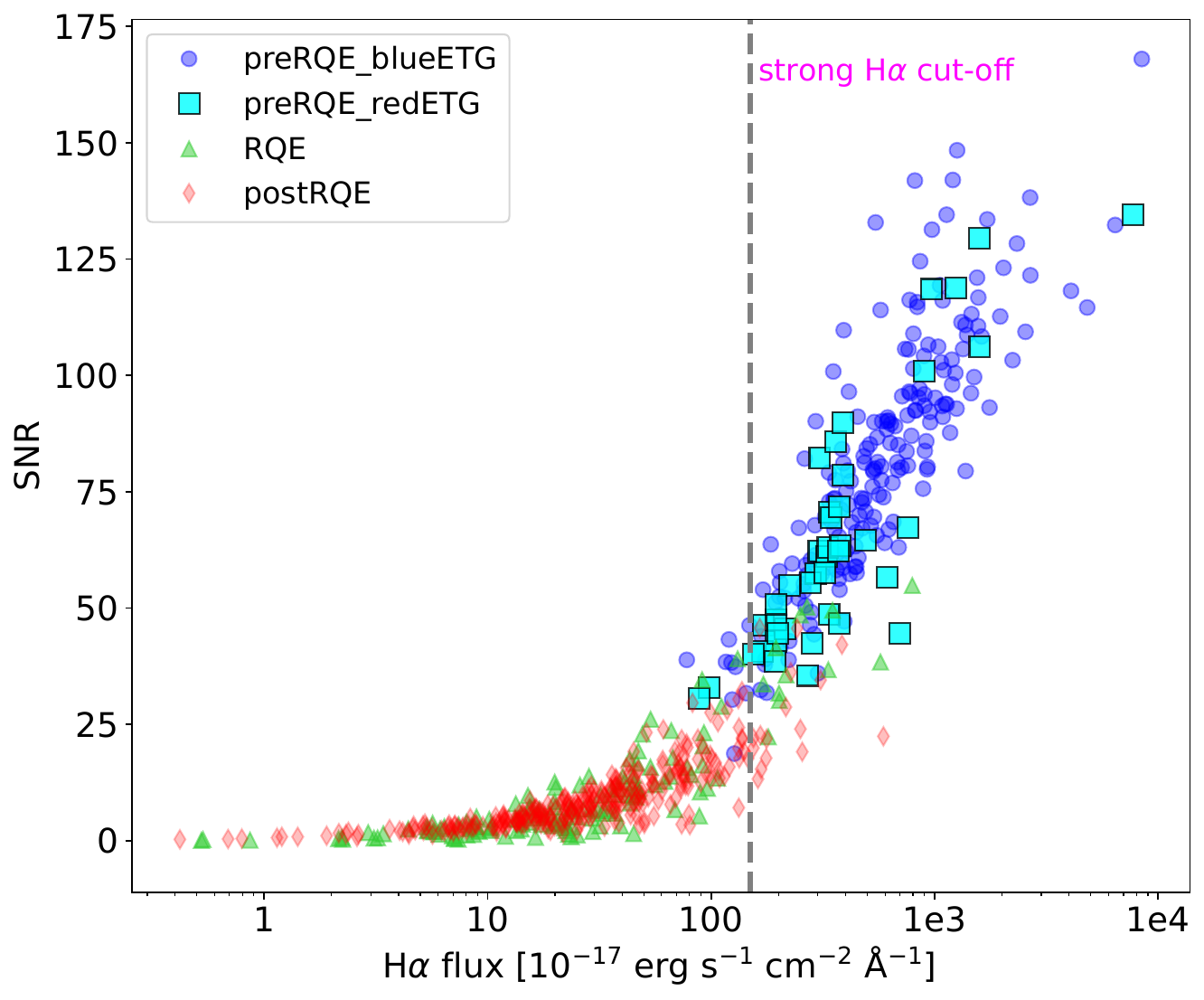}
    \caption{ Comparison of SDSS H$\alpha$ flux and its SNR (Signal-to-Noise Ratio) for RQEs and comparable subpopulations (preRQE\_blueETG, preRQE\_redETG, postRQE). A clear distinction between star-forming preRQEs and quiescent RQEs \& postRQEs is observed at H$\alpha$ flux $=150 \times 10^{-17} \mathrm{erg\,s^{-1}\,cm^{-2}\,\text{\AA}^{-1}}$. Potential plausible precursors from redETGs (preRQE\_redETG) with strong H$\alpha$ emission ($\geq$ 150 $\times 10^{-17} \mathrm{erg\,s^{-1}\,cm^{-2}\,\text{\AA}^{-1}}$), similar to preRQE\_blueETG and exhibiting E/pE morphology, are included in our preRQE control sample.}
    \label{fig:strongHa}
\end{figure}

We also investigated the star-forming redETGs for potential preRQEs. From this subset, we identified galaxies exhibiting strong H$\alpha$ emission with E/pE morphology to include in our preRQE population. 
To determine the strong H$\alpha$ emission cut, we plotted SDSS H$\alpha$ flux against its Signal-to-Noise Ratio (SNR), calculated as the ratio of H$\alpha$ flux to its error (values obtained from \textit{sample1}), for each galaxy in our subset of star-forming redETGs, alongside RQEs, postRQEs, and preRQEs from the blueETG population. 
We observed a clear divide between the actively SF population (preRQEs) and the quiescent population (postRQEs and RQEs) at an H$\alpha$ flux of $150 \times 10^{-17} \mathrm{erg\,s^{-1}\,cm^{-2}\,\text{\AA}^{-1}}$ (see Figure \ref{fig:strongHa}). 
In this figure, we distinguish between preRQEs from the blueETG population (labeled as preRQE$\_$blueETG) and potential preRQEs from the redETG population (labeled as preRQE$\_$redETG). 
Based on this analysis, we adopted $150 \times 10^{-17} \mathrm{erg\,s^{-1}\,cm^{-2}\,\text{\AA}^{-1}}$ as our threshold for strong H$\alpha$ emission. We then visually inspected the SF redETG population with H$\alpha$ flux $\geq 150 \times 10^{-17} \mathrm{erg\,s^{-1}\,cm^{-2}\,\text{\AA}^{-1}}$ using DECaLS DR10 images to select galaxies with E/pE morphology as potential preRQEs.

In total, we obtained 239 preRQEs: 198 from the SF blueETG population and 41 from the SF redETG population. Detailed statistics are provided in Table \ref{tab:comptable}.

\subsection{Sample with AGN and LINER}
We emphasize that a portion of our sample, specifically the recently quenched elliptical (RQE) galaxies and their plausible descendants (postRQE), host significant emission regions such as Active Galactic Nuclei (AGN) and Low-Ionization Nuclear Emission-line Regions (LINERs). These active regions are particularly important as they may play crucial roles in the quenching process \citep{2005Natur.433..604D, 2007MNRAS.382.1415S, 2013MNRAS.433.3297D, 2017MNRAS.466.2570B, 2022MNRAS.512.1052P}. Understanding the distribution and characteristics of these emission regions within our sample is critical for exploring their potential influence on the quenching process.

Among the 155 RQE centrals, 31 are classified as LINERs and 10 as AGNs (specifically Seyferts). Similarly, out of 301 postRQEs, there are 115 LINERs and 15 Seyfert galaxies. Additionally, there is a population of weak-LINERs, defined by M14 following \cite{2006ApJ...648..281Y}. For subsequent analyses in this paper, we categorize Seyfert samples as AGNs and combine LINER and weak-LINER (also referred as Y06) samples as LINERs.

\begin{figure}
    \centering
    \includegraphics[width=1\linewidth]{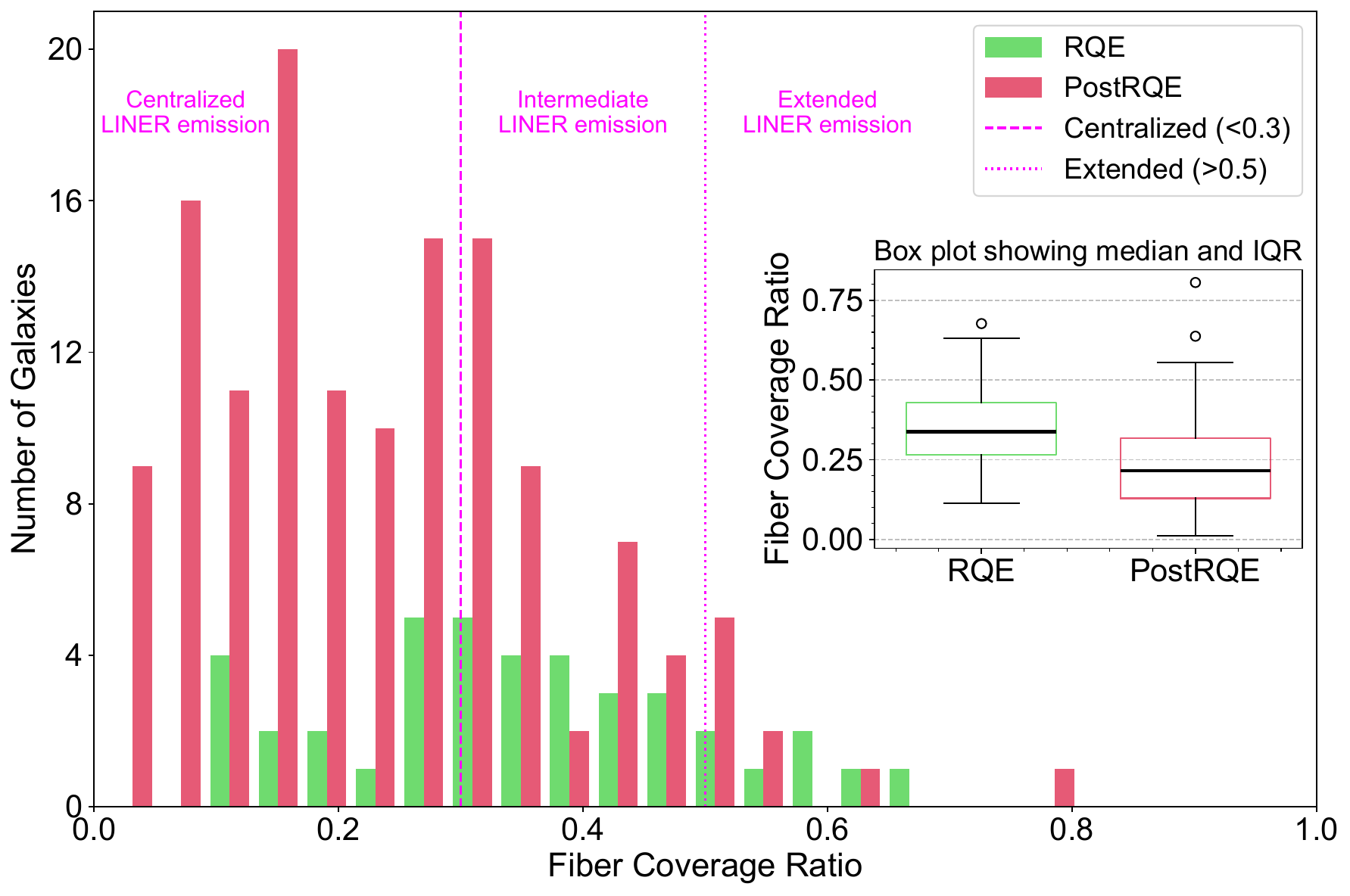}
    \caption{Spatial distribution of LINER emission in RQE and post-RQE galaxies, as indicated by the fiber coverage ratio, which is defined as the ratio of the physical size corresponding to the width of the SDSS fiber spectra to the physical size of the galaxy defined by twice the SDSS Petrosian radius in the r-band ($2 \times R90$). The histogram shows the distribution of fiber coverage ratios, with thresholds marked to distinguish between centralized ($< 0.3$), intermediate (0.3-0.5), and extended ($> 0.5$) emission regions. The box plot inset compares the median fiber coverage ratios and the interquartile ranges (IQR) between RQE and postRQE galaxies. LINER emission appears to be predominantly centralized in postRQE galaxies, with a higher percentage of RQE galaxies showing intermediate or extended emission.}
    \label{fig:sizeratio}
\end{figure}

Figure \ref{fig:sizeratio} illustrates a key characteristic of our LINER population (including both traditional LINERs and Y06): the spatial distribution of their emission relative to the galaxy size. This figure compares the fiber coverage ratio, which is the ratio of the physical size corresponding to the width of the SDSS fiber spectra at the galaxy’s redshift, to the physical size of the galaxy as defined by twice the SDSS Petrosian radius in the r-band ($2 \times R90$).
Our analysis reveals that approximately 62\% of the total LINER population exhibits centralized emission (fiber coverage ratio $< 0.3$), while 30\% show intermediate emission (ratio between 0.3 and 0.5), and only 8\% display extended emission (ratio $> 0.5$). Interestingly, distinct patterns emerge between the RQE and post-RQE populations. In RQE galaxies, 37.5\% show centralized emission, 47.5\% intermediate, and 15\% extended. In contrast, post-RQE galaxies exhibit a stronger tendency towards centralized emission, with 68.8\% centralized, 25.4\% intermediate, and only 5.8\% extended.
This distribution highlights the predominantly central nature of LINER emission in our sample, particularly in the more evolved post-RQE galaxies, while also revealing a notable fraction of RQEs with more spatially extended LINER activity.

Typically, studies of quiescent galaxies exclude those with Seyfert or LINER-like emission to maintain a sample that is clean and purely quiescent, thereby avoiding potential ambiguities introduced by active regions (e.g., \citealt{2007MNRAS.382.1415S}). However, in our study of HI content in galaxies, we intentionally include these samples (noAGN, LINER, AGN) to thoroughly investigate the quenching process in RQEs. By incorporating these emission regions, we can more accurately assess how the presence or absence of AGNs and LINERs correlates with HI content and, consequently, with the quenching pathways of these galaxies.

This inclusive approach allows us to explore several key questions about the quenching process. By comparing the HI content of RQEs with and without LINER or AGN activity, we aim to identify significant differences in their cold gas properties. This approach enables us to explore whether the evolution from star-forming to quenched status follows different routes and involves distinct quenching mechanisms in these three groups, or if they represent different phases along a continuous timeline of RQE evolution. Studying HI content in relation to these active regions can provide crucial insights into the gas depletion processes \citep{2011MNRAS.416.1739F}, thereby helping us discuss quenching pathways in our RQE population more robustly. We will discuss these relationships and their implications in greater detail in the discussion section of this paper.

\section{HI data}\label{sec:HIdata}
The neutral hydrogen (HI) content of galaxies plays a crucial role in understanding the quenching processes that govern their evolution. HI gas serves as the primary reservoir for star formation, and its presence (or absence) can provide significant insights into the mechanisms driving quenching in elliptical galaxies. In the following sections we discuss the kind of HI data we have obtained on our sample from literature for further analyses. 

\subsection{Direct HI mass measurements from observations}
To obtain HI data for our sample of elliptical galaxies, we utilized a variety of catalogs and data sources. We first searched the \textit{vizier} database (\url{https://vizier.cds.unistra.fr/viz-bin/VizieR}) for available HI catalogs that matched our sample, focusing on those that included elliptical galaxies with stellar masses greater than $10^{10}M_{\odot}$ and redshifts less than 0.08. This search resulted in 24 relevant HI catalogs, out of which five contained direct HI mass measurements for at least one galaxy in our sample.

Additionally, we cross-matched our sample with the extended GALEX Arecibo SDSS Survey (xGASS; \citealt{2018MNRAS.476..875C}) obtained from \url{https://xgass.icrar.org/data.html}. xGASS provides a gas fraction-limited, representative sample of nearby galaxies ($0.01 < z < 0.05$) with targeted HI observations from the Arecibo telescope. 
We also cross-matched our sample with the HI-MaNGA survey \citep{2019MNRAS.488.3396M, 2021MNRAS.503.1345S} and the recently released first catalog from The FAST All-Sky HI (FASHI) survey \citep{2024SCPMA..6719511Z}. The HI-MaNGA survey, a follow-up to the Mapping Nearby Galaxies at Apache Point Observatory (MaNGA) survey \citep{2015ApJ...798....7B}, provides HI mass measurements for a subset of MaNGA galaxies using the Arecibo Observatory's L-band Wide receiver and the Robert C. Byrd Green Bank Telescope (GBT). The FASHI survey, conducted with the Five-hundred-meter Aperture Spherical radio Telescope (FAST), offers unprecedented blind extragalactic HI measurements with broader sky and frequency coverage and deeper detection sensitivity compared to the Arecibo Legacy Fast ALFA (ALFALFA) survey.

In total, we found HI detections for 24 galaxies (including 2 RQEs, 10 PreRQEs, and 12 PostRQEs) in our sample (see Table \ref{tab:HIinfotable}). These detections are summarized in Table \ref{tab:HIdetections}, including details on their HI mass and other properties. Although these detections cover only a small fraction (3.5\%) of our sample, they offer valuable direct insights into the HI content and its variation across different evolutionary stages of elliptical galaxies.

\begin{deluxetable*}{lcccc}
\tablecaption{ HI data availability \label{tab:HIinfotable}}
\tablehead{
\colhead{Catalog name (author)} & \colhead{Total galaxies} & \colhead{RQE matched} & \colhead{preRQE matched} & \colhead{postRQE matched}
}
\startdata
HI-WISE catalog (Parkash+, 2018) & 3158 & 0 & 0 & 1 \\
HI spectral properties of galaxies (Springob+, 2005) & 9500 & 0 & 0 & 1 \\
xGASS catalog (Catinella+, 2018) & 874 & 0 & 0 & 1 \\
HI-MaNGA DR3.1 catalog (Masters+, 2019; Stark+, 2021) & 6820 & 2 & 4 & 5 \\
ALFALFA catalog (Haynes+, 2018) & 31502 & 0 & 4 & 3 \\
RESOLVE catalog (Kannappan+, 2008; Stark+, 2016) & 2289 & 0 & 2 & 0 \\
FASHI catalog (Zhang+, 2024) & 41741 & 0 & 1 & 2 \\
\hline
\textbf{Total Detection} & 24 & 2 & 10 & 12 \\
\hline
Catalog of ANN estimated gas fraction (Teimoorinia+, 2017) & 561585 & 144 & 226 & 283 \\
\enddata
\tablecomments{
    Summary of HI data availability and detections across various catalogs for the sample of 695 galaxies. The table presents the number of RQE, preRQE, and postRQE galaxies with HI detections from different surveys. Data from \citet{2017MNRAS.464.3796T} are also included, which provide indirect estimates of HI gas fractions for 94\% of the sample using Artificial Neural Network (ANN) techniques. The "Total Detection" row represents unique HI detections, avoiding duplication of galaxies detected in multiple surveys.
    }
\end{deluxetable*}

\begin{deluxetable*}{lccccccc}
\tablecaption{Properties of 24 HI-Detected Galaxies from the Sample \label{tab:HIdetections}}
\tablehead{
\colhead{Name (nyuID)} & \colhead{Redshift} & \colhead{Identity} & \colhead{Emission Type} & \colhead{logM$_{\rm halo}$ [M$_{\odot}$]} & \colhead{logM$_{\rm star}$ [M$_{\odot}$]} & \colhead{logM$_{\rm HI}$ [M$_{\odot}$]} & \colhead{Reference}
}
\startdata
nyu324678 & 0.022997 & preRQE & HII & 11.7624 & 10.322 & 9.150; 9.085 & a); f) \\
nyu815049 & 0.030163 & preRQE & HII & 11.8290 & 10.332 & 9.200 & a) \\
nyu648398 & 0.054899 & preRQE & HII & 11.8777 & 10.437 & 9.830 & b) \\
nyu392636 & 0.010133 & postRQE & LINER & 11.8924 & 10.424 & 9.300 & g) \\
nyu398366 & 0.029157 & preRQE & HII & 11.9237 & 10.472 & 9.531 & a) \\
nyu1085356 & 0.046117 & postRQE & Quiescent & 11.9328 & 10.471 & 9.874 & a) \\
nyu809 & 0.019442 & preRQE & HII & 11.9441 & 10.491 & 9.526 & f) \\
nyu847227 & 0.054453 & preRQE & HII & 11.9507 & 10.528 & 10.170 & b) \\
nyu802000 & 0.028385 & postRQE & LINER & 11.9807 & 10.523 & 9.725 & a) \\
nyu544101 & 0.038241 & RQE & LINER & 12.0252 & 10.549 & 10.179 & a) \\
nyu193320 & 0.030774 & postRQE & LINER & 12.1158 & 10.668 & 8.761 & a) \\
nyu818663 & 0.032207 & postRQE & LINER & 12.1295 & 10.681 & 9.385 & a) \\
nyu596119 & 0.041656 & preRQE & HII & 12.1300 & 10.660 & 9.730 & b) \\
nyu729044 & 0.025839 & postRQE & LINER & 12.1315 & 10.665 & 9.830 & b) \\
nyu507490 & 0.028964 & postRQE & LINER & 12.1540 & 10.693 & 9.804 & a) \\
nyu100885 & 0.031036 & postRQE & Quiescent & 12.1600 & 10.694 & 8.869 & e) \\
nyu992633 & 0.044315 & postRQE & Quiescent & 12.2515 & 10.786 & 9.830 & b) \\
nyu76779 & 0.034580 & preRQE & HII & 12.2994 & 10.813 & 9.760 & a) \\
nyu199174 & 0.050810 & postRQE & LINER & 12.3136 & 10.834 & 10.050 & b) \\
nyu15039 & 0.010687 & postRQE & LINER & 12.3311 & 10.818 & 9.440; 9.869 & c); d) \\
nyu192837 & 0.079229 & preRQE & HII & 12.3439 & 10.811 & 10.390 & g) \\
nyu100426 & 0.041761 & preRQE & HII & 12.3800 & 10.830 & 10.080 & b) \\
nyu816340 & 0.044436 & postRQE & Seyfert & 12.4242 & 10.880 & 9.610 & g) \\
nyu397622 & 0.046250 & RQE & weak-LINER & 12.6460 & 11.046 & 9.800 & a) \\
\enddata
\tablecomments{Properties of 24 galaxies from the sample of 695 detected in HI surveys. The table is sorted by descending halo mass and includes HI mass, halo mass, stellar mass, redshift, classification (RQE, preRQE, or postRQE), and emission type for each galaxy. The last column provides references to the HI surveys where detections occurred. Multiple HI mass entries indicate detections in more than one survey. The references correspond to: (a) HI-MaNGA \citet{2021MNRAS.503.1345S}, (b) ALFALFA \citet{2018ApJ...861...49H}, (c) HI-WISE \citet{2018ApJ...864...40P}, (d) \citet{2005ApJS..160..149S}, (e) xGASS \citet{2018MNRAS.476..875C}, (f) RESOLVE \citet{2016ApJ...832..126S}, and (g) FASHI \citet{2024SCPMA..6719511Z}.}
\end{deluxetable*}

\subsection{HI non-detections from ALFALFA and FASHI}
\begin{figure*}
    \includegraphics[width = \textwidth]{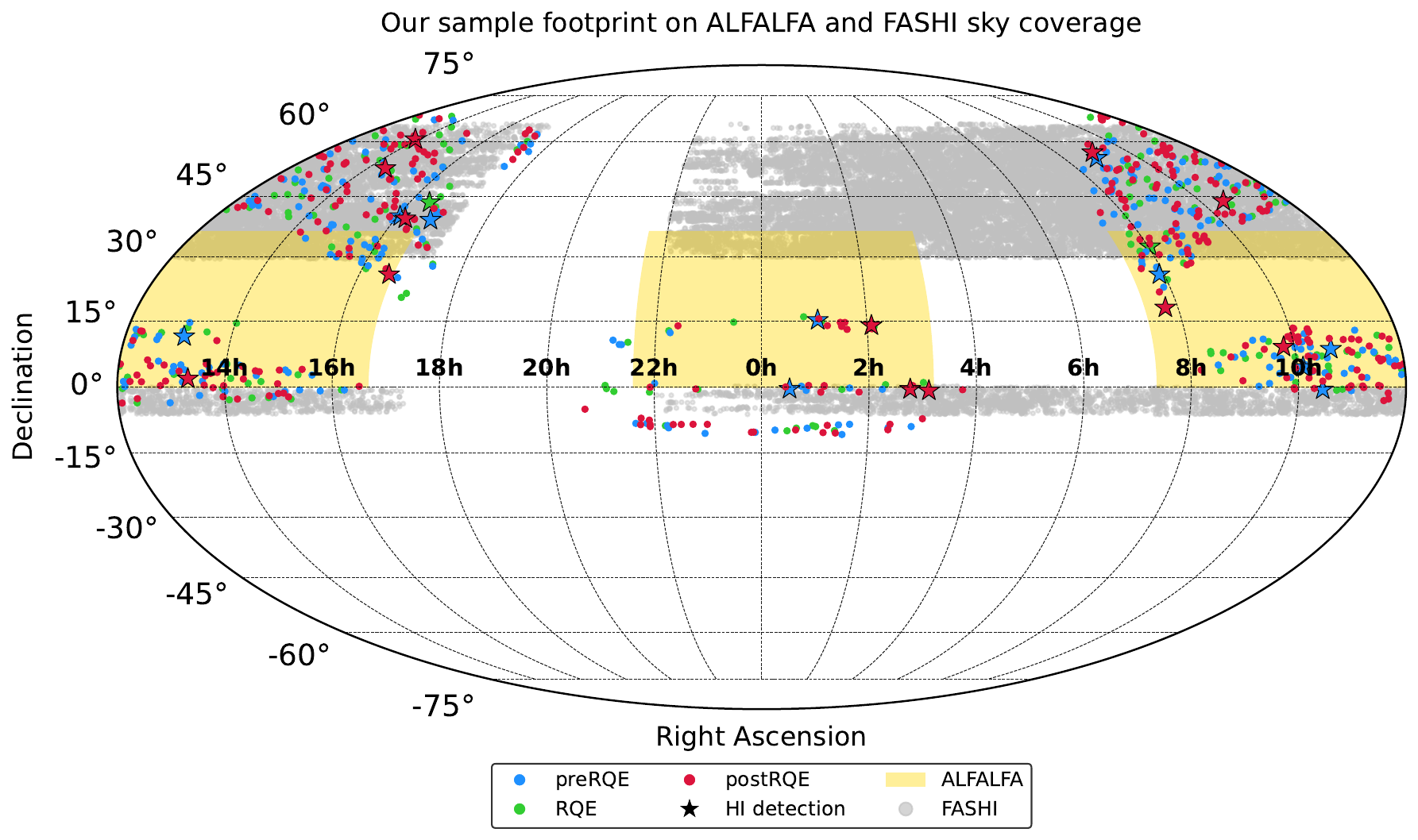}
    \caption{Sky coverage of our sample within the footprints of the ALFALFA and FASHI HI surveys. Approximately 40\% of our sample, consisting of 279 galaxies, lies within the ALFALFA survey footprint, while 52\% of the sample, comprising 364 galaxies, falls within the FASHI survey footprint. Among these, a small fraction of 2.5\% (7 galaxies) show HI detections in the ALFALFA data, and less than 1\% (3 galaxies) show detections in the FASHI data. The majority of the galaxies within these footprints remain undetected in HI.}
    \label{fig:coverage}
\end{figure*}

A significant portion of our sample falls within the sky coverage of two of the most sensitive untargeted HI surveys available—ALFALFA and FASHI, as shown in Figure \ref{fig:coverage}. ALFALFA, a blind extragalactic HI survey, covers 6,300 square degrees of high Galactic latitude sky (0$^{\circ}$ to 30$^{\circ}$ declination), targeting the local universe up to a redshift of $z \sim 0.06$. The survey was conducted using the 305-meter Arecibo radio telescope in Puerto Rico, equipped with the seven-beam Arecibo L-Band Feed Array (ALFA) \citep{2018ApJ...861...49H}.

FASHI, an unprecedented blind HI survey conducted by the FAST in China, surpasses ALFALFA in both breadth and depth. FASHI covers 22,000 square degrees within declinations from $-14^{\circ}$ to $66^{\circ}$ and frequencies between 1050 and 1450 MHz. From August 2020 to June 2023, FASHI successfully surveyed over 7,600 square degrees (35\% of the total sky observable by FAST, with coverage shown in Figure \ref{fig:coverage}), achieving a median sensitivity of approximately 0.76 mJy beam$^{-1}$ and a spectral line velocity resolution of about 6.4 km s$^{-1}$ at 1.4 GHz, representing an approximately 2.5-fold improvement in detection sensitivity over ALFALFA. Results from this initial FASHI survey, as published in \citet{2024SCPMA..6719511Z}, reported the detection of 41,741 extragalactic HI sources within the frequency range of 1305.5-1419.5 MHz, corresponding to a redshift limit of $z \leq 0.09$.

A total of 272 galaxies, representing 39.1\% of our sample, were observed but not detected in ALFALFA, while 361 galaxies (51.9\%) remained undetected in FASHI. Of these, 56 galaxies were non-detections in both surveys, collectively covering 83\% of our sample. Despite these non-detections, the observations from these surveys are still highly valuable for our research. They allow us to calculate upper limit (UL) HI mass values, which provide critical constraints on the maximum possible HI content in these galaxies. 

\subsubsection{Upper limit HI mass estimation}\label{sec:UL}
To calculate the UL HI masses, we employ methods based on the sensitivity limits and specific characteristics of each survey. The general principle involves determining the minimum detectable integrated flux density for a given noise level, assumed spectral line width, and distance to the galaxy. This integrated flux is then converted to HI mass using established relations that consider the luminosity distance of the galaxy and the redshift \citep{1962AJ.....67..437R, 2018ApJ...864...40P}. 
We use slightly different approaches for FASHI and ALFALFA data, reflecting the distinct properties and data products of these surveys.

For FASHI data, we use a formula typically applied to single-dish telescopes (see equation 3 in \citealt{2021MNRAS.503.1345S}):
\begin{equation}
M_{\mathrm{HI}} = \mathrm{S/N} \times \frac{2.36 \times 10^{5}}{(1+z)^2} \times D^{2} \times \mathrm{rms} \times \sqrt{W \times \mathrm{dV}}
\end{equation}
For ALFALFA data, we use a modified approach based on \citet{2018ApJ...861...49H}, as employed by \citet{2021MNRAS.503.1345S}:
\begin{equation}
M_{\mathrm{HI}} = \mathrm{S/N} \times \frac{2.36 \times 10^{5}}{(1+z)^2} \times D^{2} \times W \times \mathrm{rms} \times w_{\mathrm{smo}}^{-1/2}
\end{equation}
In both equations, $M_{\mathrm{HI}}$ represents the HI mass upper limit in $M_{\odot}$, $\mathrm{S/N}$ is the signal-to-noise ratio, $z$ is the redshift, $D$ is the luminosity distance to the galaxy in Mpc, $\mathrm{rms}$ is the root mean square noise of the observation in Jy, and $W$ is the assumed HI linewidth in km s$^{-1}$. The FASHI equation incorporates $\mathrm{dV}$, the spectral resolution in km s$^{-1}$, while the ALFALFA equation uses $w_{\mathrm{smo}} = W/20$, representing the smoothing width expressed as the number of spectral resolution bins.

For our calculations, we adopt specific parameters for each survey based on their respective characteristics and detection thresholds. We assume a fiducial HI linewidth $W$ of 200 km s$^{-1}$ for all undetected galaxies in our sample. For FASHI, we use an $\mathrm{rms}$ of 1.5 mJy and a spectral resolution $\mathrm{dV}$ of 6.4 km s$^{-1}$, as reported in Table 1 of \citet{2024SCPMA..6719511Z}. For ALFALFA, we adopt an $\mathrm{rms}$ of 2 mJy, based on Table 1 of \citet{2018ApJ...861...49H}.

Untargeted HI surveys typically require relatively high signal-to-noise ratios to ensure reliable detections. Accordingly, we adopt the minimum S/N thresholds as required by the original surveys for confirmed detections: 5.0 for FASHI and 4.5 for ALFALFA. These thresholds represent the actual detection criteria used in these surveys, ensuring our upper limit calculations are consistent with the surveys' methodologies.

\begin{figure}
    \centering
    \includegraphics[width=1\linewidth]{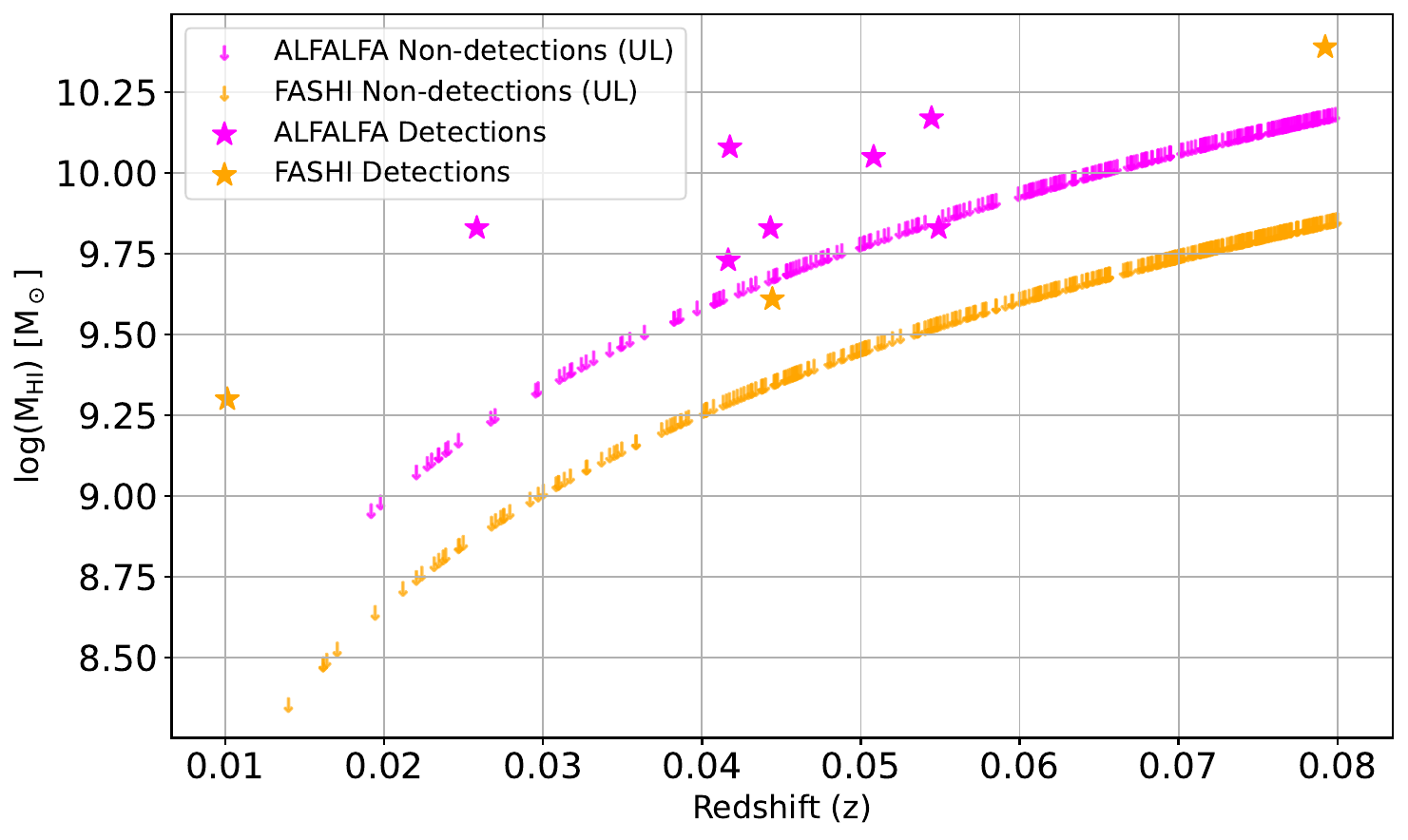}
    \caption{Distribution of upper limit (UL) HI mass values as a function of redshift for galaxies undetected in the ALFALFA and FASHI surveys, shown as downward arrows. The plot also displays HI detections from FASHI (orange stars) and ALFALFA (magenta stars) within our sample. The lower UL values from FASHI, compared to ALFALFA, indicate the higher sensitivity of the FASHI survey, with UL HI mass values generally 0.3 dex lower.}
    \label{fig:compUL}
\end{figure}

Figure \ref{fig:compUL} illustrates the distribution of upper limit (UL) HI mass values as a function of redshift for galaxies undetected in both the ALFALFA and FASHI surveys. The plot also includes HI detections from these surveys (3 from FASHI and 7 from ALFALFA) within our sample. This comparison highlights the difference in UL values between the two surveys, with FASHI’s UL HI mass values generally lower than those from ALFALFA by approximately 0.3 dex, reflecting FASHI’s higher sensitivity.

\subsection{Indirect HI measurements}\label{sec:T17}
The scarcity of direct HI detections and the limitations posed by upper limit HI values in our sample prompted us to explore available indirect HI measurements. Indirect methods for estimating HI mass or gas fractions generally fall into three categories: scaling relations, machine learning models, and HI stacking techniques.

Scaling relations leverage observable galaxy properties—such as stellar mass, color, and surface brightness—to predict HI content. Early studies, such as that by \citet{2004ApJ...611L..89K}, established correlations between optical colors (e.g., $u - K$) and HI richness, achieving a root mean square (rms) scatter of 0.37 dex in log(M${\mathrm{HI}}$/M${\star}$), where M${\mathrm{HI}}$ is the HI mass and M${\star}$ is the stellar mass. \cite{2009MNRAS.397.1243Z} refined these relations by incorporating i-band surface brightness alongside $g - r$ color, reducing the scatter to 0.31 dex. Subsequent studies 
\citep{2010MNRAS.403..683C, 2012ApJ...756..113H, 2012MNRAS.424.1471L, 2013MNRAS.436...34C, 2015ApJ...810..166E} further refined these relations by incorporating parameters such as stellar mass surface density, NUV-$r$ color, and axial ratios, achieving a scatter as low as 0.3 dex. 

Machine learning models have emerged as a powerful tool in recent years, particularly for their ability to manage complex, multidimensional datasets. For example, \citet{2017MNRAS.464.3796T} (hereafter T17) employed an artificial neural network (ANN) trained on 15 galaxy parameters derived from SDSS spectroscopy and photometry to predict HI gas fractions, achieving a remarkably low scatter of 0.22 dex, substantially outperforming previous photometric gas fraction techniques.
Other machine learning approaches, such as those by \citet{2018MNRAS.479.4509R}, have also demonstrated significant promise, although with slightly higher scatter values of 0.25 to 0.3 dex depending on the algorithm used. They estimate gas mass fraction for observed and simulated galaxy samples by using a variety of machine learning algorithms, such as random forests, gradient-boosted trees, and deep neural networks. 
More recent studies, such as \citet{2020ApJ...900..142W}, explored computer vision algorithms like deep convolutional neural networks (CNNs) to predict HI mass fractions directly from optical imaging (using SDSS $gri$ image cutouts), achieving a scatter of 0.20 dex compared to HI detections from ALFALFA. 

HI stacking is another technique used to enhance the detection of HI signals by averaging the spectra from multiple galaxies \citep{2001A&A...372..768C, 2007ApJ...668L...9V, 2011MNRAS.411..993F, 2023MNRAS.518.4646R}. This method, as described by \citet{2011MNRAS.411..993F}, is particularly useful for studying galaxy populations with weak individual signals, allowing for the detection of HI in cases where individual observations might fall below the sensitivity threshold. HI stacking can achieve sensitivities up to $\sqrt{N}$ times better than individual observations, where $N$ is the number of stacked spectra. However, it provides only average HI content for galaxy populations rather than individual estimates, making it less versatile than machine learning models.

Each of these methods presents unique advantages and limitations. Scaling relations, while straightforward to apply, may not fully capture the complexity of gas content across diverse galaxy types, particularly for transitional objects such as RQEs. Machine learning models, despite their impressive accuracy, may face challenges when applied to galaxies significantly different from their training sets. HI stacking, while effective for population studies, cannot provide the individual galaxy measurements crucial for understanding the diversity within our sample.

Given these considerations, the study by T17 stands out as particularly promising, not only because it covers $\sim$95\% of our sample, offering a near-complete coverage where direct measurements are unavailable, but also because it offers gas fraction values with one of the lowest scatters (0.22 dex) which is significantly lower than that of traditional scaling relations and comparable to or better than more recent machine learning approaches.

To assess the reliability of the T17 estimates for our sample, we conducted a comprehensive comparison between these predictions and direct HI measurements from ALFALFA, as well as upper limit (UL) HI masses derived from ALFALFA and FASHI non-detections (see Section \ref{sec:UL}). This analysis aims to evaluate the suitability of T17 values as proxies for HI content in galaxies lacking direct detections within our sample.

\begin{figure}
    \centering
    \includegraphics[width=1\linewidth]{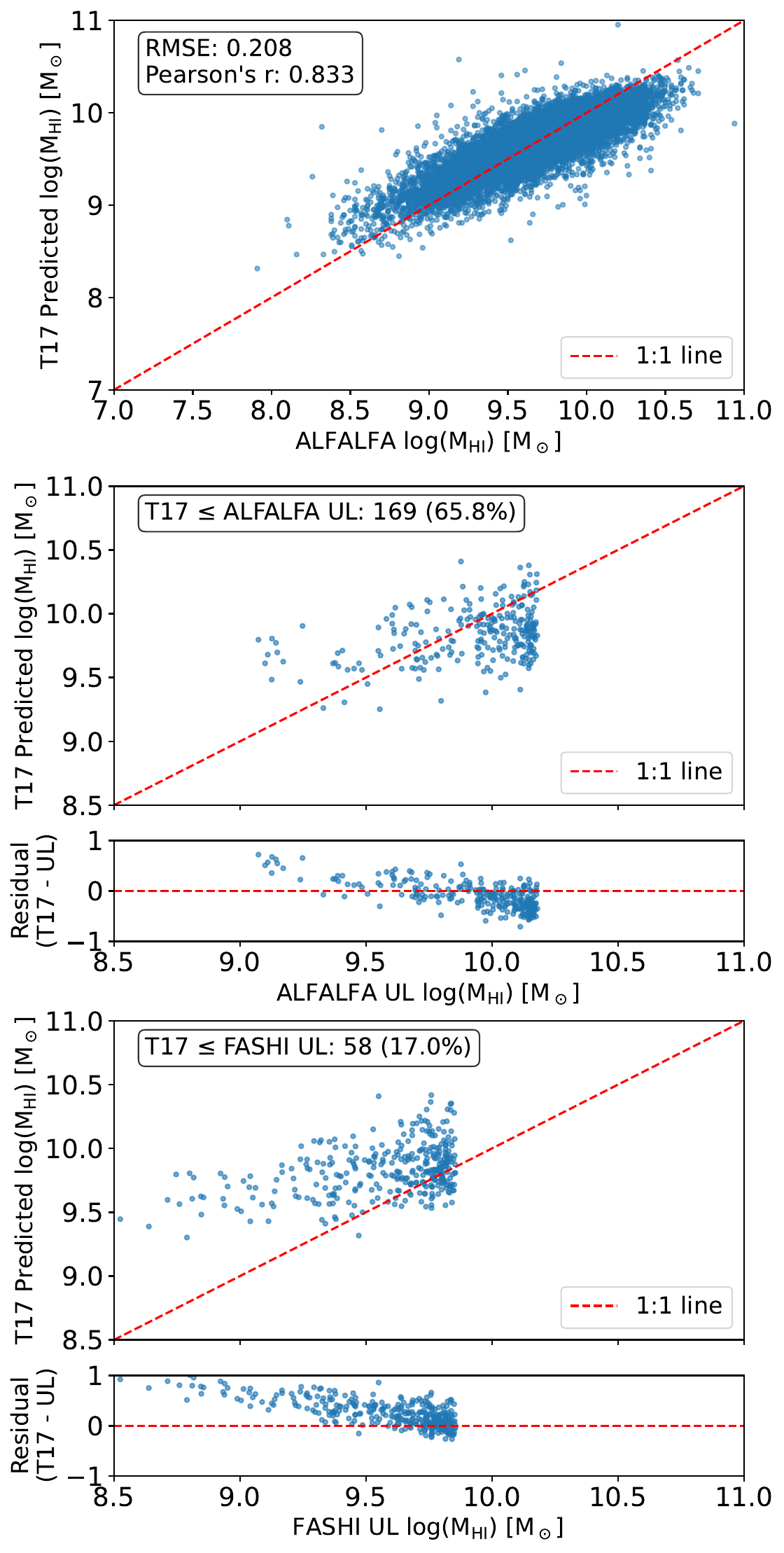}
    \caption{Comparison of predicted HI mass values from Teimoorinia et al. 2017 (T17) with observed and upper limit HI mass values from ALFALFA and FAST surveys.\textit{Top panel:} T17 predicted $\log(M_\mathrm{HI})$ vs. ALFALFA observed $\log(M_\mathrm{HI})$ values for detected galaxies. The red dashed line represents the 1:1 relationship. RMSE and Pearson's r correlation coefficient are shown.
    \textit{Middle and bottom panels:} T17 predicted $\log(M_\mathrm{HI})$ vs. ALFALFA (middle) and FAST (bottom) upper limit $\log(M_\mathrm{HI})$ values. The red dashed lines represent the 1:1 line, and the residuals (T17 - UL) are plotted below each comparison. A significant portion of the T17 predictions are below the upper limits, with 65.8\% for ALFALFA and 17.0\% for FAST.}
    \label{fig:T17plot}
\end{figure}

Figure \ref{fig:T17plot} presents a three-panel comparison of T17 predictions with observed and upper limit HI mass values. To quantify the agreement between T17 predictions and observed HI masses, we employ two key statistical metrics: the Root Mean Square Error (RMSE) and Pearson's correlation coefficient (r). These metrics are calculated as follows:

The RMSE is given by:
\begin{equation}
    \text{RMSE} = \sqrt{\frac{1}{n}\sum_{i=1}^n (y_i - \hat{y}_i)^2}
\end{equation}
where $y_i$ are the observed values, $\hat{y}_i$ are the predicted values, and $n$ is the number of observations. RMSE provides a measure of the typical magnitude of prediction errors in the same units as the original data (in our case, dex of $\log(M_\mathrm{HI})$).

Pearson's correlation coefficient is calculated using:
\begin{equation}
    r = \frac{\sum_{i=1}^n (x_i - \bar{x})(y_i - \bar{y})}{\sqrt{\sum_{i=1}^n (x_i - \bar{x})^2} \sqrt{\sum_{i=1}^n (y_i - \bar{y})^2}}
\end{equation}
where $x_i$ and $y_i$ are the individual sample points indexed with i, $\bar{x}$ is the mean of $x_i$, and $\bar{y}$ is the mean of $y_i$. Pearson's r ranges from -1 to 1, with values closer to $\pm$ 1 indicating stronger linear correlations.

These metrics provide complementary information about the performance of the T17 predictions. While RMSE quantifies the absolute deviation of predictions from observations, Pearson's r measures the strength and direction of the linear relationship between predicted and observed values.

\begin{figure}
    \centering
    \includegraphics[width=1\linewidth]{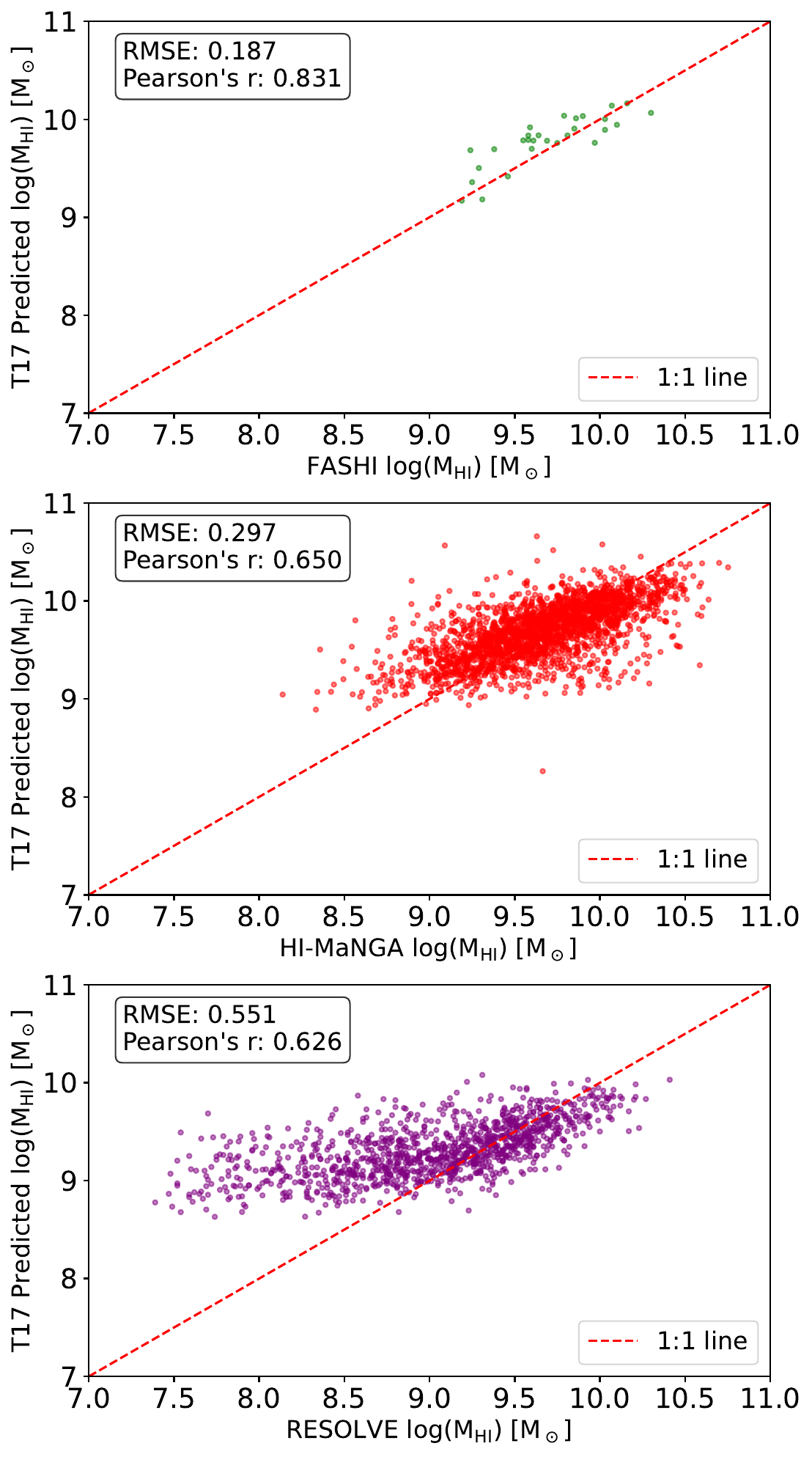}
    \caption{Comparison of predicted $\log(M_\mathrm{HI})$ values for SDSS galaxies from T17 to that of observed values from large HI surveys - FASHI, HI-MaNGA and RESOLVE. Despite being trained on ALFALFA measurements T17 predicts mass estimates within 0.3 dex RMSE scatter for FASHI and HI-MaNGA SDSS detections and performs poorly in predicting measurements from RESOLVE.}
    \label{fig:T17scatter}
\end{figure}

The top panel of Figure \ref{fig:T17plot} demonstrates the relationship between T17 predicted $\log(M_\mathrm{HI})$ values and ALFALFA detections. Consistent with T17's claims, we observe a strong correlation between predicted and observed values, characterized by a low RMSE of 0.208 dex and a high Pearson correlation coefficient (r) of 0.833. These results confirm the considerable reliability of T17 predictions for SDSS galaxies with ALFALFA detections.

While the T17 model demonstrates high accuracy for ALFALFA detections, it's important to note that this performance may be partially attributable to the model's training on ALFALFA data. Figure \ref{fig:T17scatter} illustrates comparisons with other HI surveys (FASHI, HI-MaNGA, and RESOLVE), revealing varying degrees of agreement. The RMSE and Pearson's r for these comparisons range from 0.187 to 0.551 and 0.626 to 0.831 respectively showing least scatter with FASHI and maximum scatter with RESOLVE HI detection measurements. However, these variations should be interpreted cautiously, given that the T17 model was primarily optimized for ALFALFA measurements.

The middle and bottom panels of Figure \ref{fig:T17plot} compare T17 predictions with upper limit (UL) values from ALFALFA and FASHI, respectively, with residual plots (T17 - UL) included beneath each scatter plot. These residual plots offer a clearer view of discrepancies, highlighting cases where T17 predictions exceed the UL values. For ALFALFA, 65.8\% of the T17 predictions are below the ULs, suggesting reasonable consistency. In contrast, only 17.0\% of T17 predictions fall below the FASHI ULs, indicating a tendency for T17 to overestimate HI masses when compared to FASHI's more sensitive observations. The residual plots thus reveal systematic differences that might not be immediately apparent from the scatter plots alone, providing critical insight into the reliability of T17 predictions relative to these UL datasets.

Based on these findings, we propose the following strategy for incorporating indirect HI measurements in our analysis:
a) For galaxies where T17 predictions are below ALFALFA or FASHI ULs, we will adopt the T17 values as proxies.
b) For galaxies where T17 overestimates relative to ULs, we will use the more conservative ALFALFA UL or FASHI UL values.
c) In cases where both ALFALFA and FASHI ULs are available, we will preferentially use the FASHI UL due to its higher sensitivity.
d) For galaxies with no detections but valid T17 values, we will include these estimates in our analysis while applying a conservative uncertainty of $\pm 0.2$ dex in $\log(M_\mathrm{HI})$ to account for potential systematic errors in the predictions.

The credibility of the T17 estimates is further reinforced by a comparative study conducted by \citet{2020ApJ...900..142W}, which demonstrated that their convolutional neural network (CNN) method provides a modest improvement of 0.07 dex in gas fraction estimates over the fully connected neural network (FCNN) approach used by T17. Both of these machine learning techniques significantly outperform traditional scaling relations based on optical properties, such as those relying on (g-r) color and stellar surface mass density \citep{2010MNRAS.403..683C, 2012MNRAS.424.1471L, 2015ApJ...810..166E}. 

Despite this improvement, the lack of available data from \citet{2020ApJ...900..142W} necessitates our reliance on the T17 estimates for indirect HI mass measurements throughout our analysis. Importantly, while the CNN method's improvement is noteworthy, it remains relatively modest and is unlikely to significantly alter the overall conclusions of our study, particularly given the broader uncertainties inherent in HI mass estimation for undetected sources \citep{2017MNRAS.464.3796T}. Future work could potentially refine our findings by applying the CNN derived values to our sample, should the data become available.

This approach allows us to maximize the use of available data while acknowledging the limitations and uncertainties associated with indirect HI mass estimates. By carefully combining direct measurements, upper limits, and validated T17 predictions, we assemble a comprehensive dataset for a detailed investigation of the HI content in RQEs and their potential evolutionary trajectories.

\begin{deluxetable*}{lr}
\tablecaption{Categorization of galaxies by HI data sources \label{tab:HIcounts}}
\tablehead{
\colhead{Condition} & \colhead{Number of Galaxies}
}
\startdata
HI detections & 24 \\
T17 values $\leq$ UL, excluding detections & 190 \\
T17 values only, no detections or ULs & 91 \\
UL-only galaxies and UL $<$ T17, excluding detections & 383 \\
\hline
Total with HI data (direct or indirect) & 688 \\
\hline
No HI data & 7 \\
\hline
\textbf{Total sample} & 695 \\
\enddata
\tablecomments{This table summarizes the galaxies included in the HI mass analysis (Figure~\ref{fig:HImassnGF} and subsequent figures). Galaxies are categorized based on their HI data sources, which include direct detections, HI mass values from T17, and upper limits (ULs) derived from non-detections in ALFALFA and FASHI.}
\end{deluxetable*}

\section{HI analyses}\label{sec:HIanalyses}
The comprehensive data compilation described in Section \ref{sec:HIdata} has yielded HI data for 688 out of 695 galaxies, effectively covering 99\% of our sample. 
This data encompasses actual detections, upper limits, and estimates derived from indirect methods, as summarized in Table \ref{tab:HIcounts}.

Ideally, a thorough investigation of HI gas properties—including mass, morphology, kinematics, asymmetry, and temperature—would require spatially and spectrally resolved HI data. Such detailed data enables the construction of various moment maps that can uncover critical information about the distribution, dynamics, and state of the HI gas reservoirs within galaxies.
However, for our sample, we are constrained to HI mass or HI gas fraction (defined as the ratio of HI mass to stellar mass) data. Despite this limitation, these global HI properties still offer valuable information about the gas content of our sample galaxies and its potential role in the quenching process.

In this section, we present a comprehensive analysis of the HI content in our sample, comparing the preRQE, RQE, and postRQE populations. Our primary objectives are to determine whether these galaxies retain any significant HI gas, to quantify the extent of their HI content, and to assess how this content varies across different evolutionary stages. We will also explore whether the observed HI content suggests that these galaxies are gas deficient and examine the relationship between HI gas fraction and current star formation rates. 
Through these analyses, we aim to understand how HI content evolves through the preRQE, RQE, and postRQE phases and provide crucial insights into the mechanisms driving the transition from star-forming to quiescent states in RQEs.

\subsection{HI mass and gas fraction}
Understanding the distribution of HI mass and gas fraction within our sample of RQE, preRQE, and postRQE galaxies is crucial for exploring the role of gas in the quenching processes. HI mass provides insight into the amount of cold gas available for star formation, while the HI gas fraction ($f_{gas}$), defined as the ratio of HI mass to stellar mass, offers a measure of the relative abundance of gas in relation to the galaxy's overall stellar content. By examining these properties across the three subpopulations, we aim to uncover patterns that might suggest different evolutionary paths and quenching mechanisms.

\begin{figure*}
    \includegraphics[width=\textwidth]{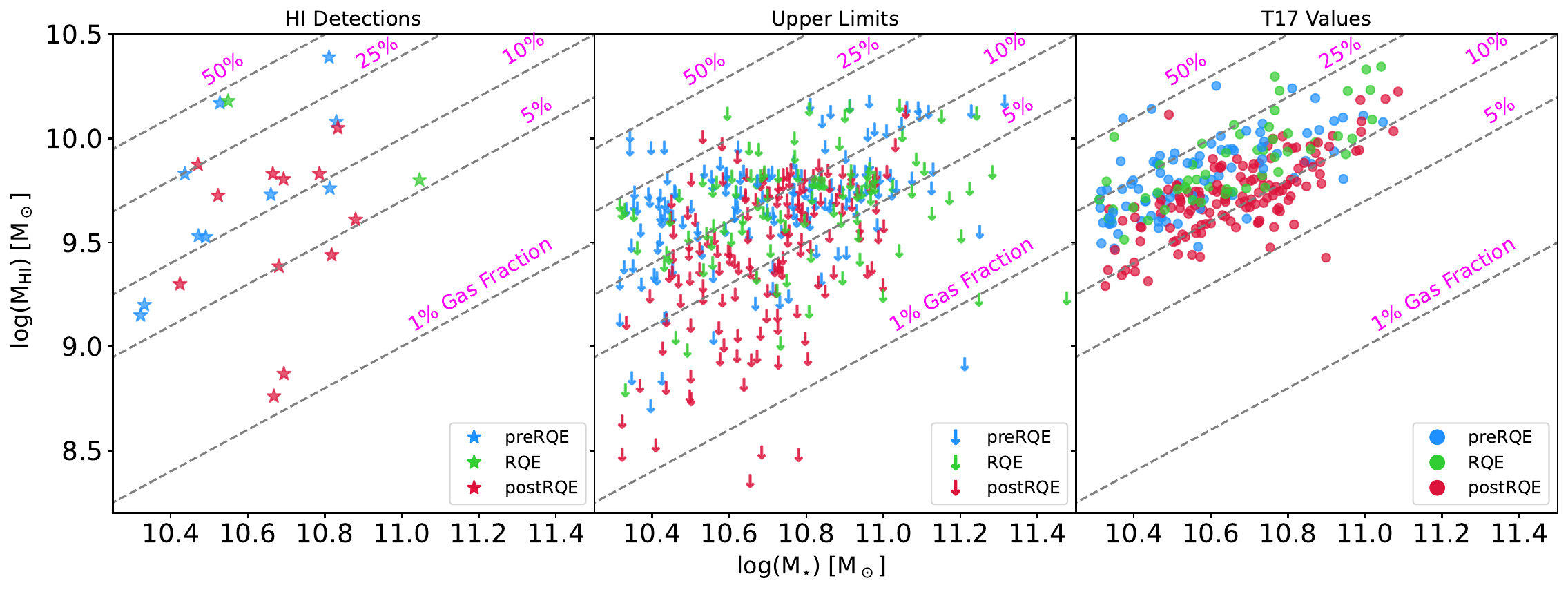}
    \caption{Log HI mass vs. log stellar mass for preRQEs, RQEs and postRQEs. The left panel shows galaxies with HI detections, the middle panel displays galaxies with upper limits on HI mass obtained from ALFALFA and FASHI surveys, and the right panel shows galaxies with T17 estimated HI masses. Gas fraction lines (1\%, 5\%, 10\%, 25\%, 50\%) are indicated by dashed lines, illustrating the typical HI gas content relative to stellar mass for these galaxy types.}
    \label{fig:HImassnGF}
\end{figure*}

Figure \ref{fig:HImassnGF} presents the distribution of log(M$_{HI}$) versus log(M$_\star$) for our sample, separated into three panels based on data source: HI detections, upper limits (ULs), and estimates from T17. Overlaid gas fraction lines allow for direct comparison of gas content across our subsamples.

The detection panel of Figure \ref{fig:HImassnGF} presents HI observations for our sample, utilizing the most recent data for galaxies with multiple measurements. The two RQEs with HI detections exhibit markedly different gas fractions of 6\% and 43\%, spanning a wide range. This disparity suggests that some RQEs may retain substantial gas reservoirs despite quenched star formation, although the limited sample size precludes definitive conclusions about the typical gas content of RQEs as a population.
Plausible precursors (preRQEs) exhibit a significant range of HI gas fractions, from 7\% to 50\%, with majority possessing above 10\%, which is consistent with the expectation. Plausible descendants (postRQEs), however, display lower gas fractions, ranging from 1\% to 25\%. reflecting the depletion of gas that likely accompanied their transition out of the star-forming phase. The presence of some gas-rich postRQEs ($f_{gas} >$ 10\%) likely reflects an observational bias towards higher HI masses in detection-limited surveys.
The upper limits derived from FASHI and ALFALFA non-detections (middle panel) provide constraints on the maximum possible HI content. The median UL gas fractions are 10.2\%, 8.5\%, and 6.5\% for preRQEs, RQEs, and postRQEs, respectively, with RQEs showing gas fraction ULs as high as 33\% and as low as 0.6\%. This trend is consistent with a scenario of progressive gas depletion from preRQE through RQE to postRQE stages.
T17 estimates (right panel) suggest that RQEs possess significant HI gas, with a median gas fraction of 17\%. This is comparable to preRQEs (18\%) and notably higher than postRQEs (11\%). The similarity between RQE and preRQE gas fractions is intriguing, given the cessation of star formation in RQEs, and may indicate that factors beyond mere gas availability play a role in quenching.

Overall, Figure \ref{fig:HImassnGF} indicates that RQEs do contain significant amounts of HI gas and, in terms of gas content, are more similar to star-forming preRQEs than to postRQEs. This observation raises intriguing questions about the mechanisms quenching star formation in RQEs despite the presence of substantial gas reservoirs. For our subsequent analysis, we will focus on the direct HI detections and T17 estimates, as shown in the left and right panels of Figure \ref{fig:HImassnGF}, respectively. These data collectively provide HI information for approximately 44\% of our sample (N = 305). We will avoid including UL values, as they primarily serve to provide an upper bound on possible HI mass and are less useful for drawing definitive conclusions.

\begin{figure*}
\plottwo{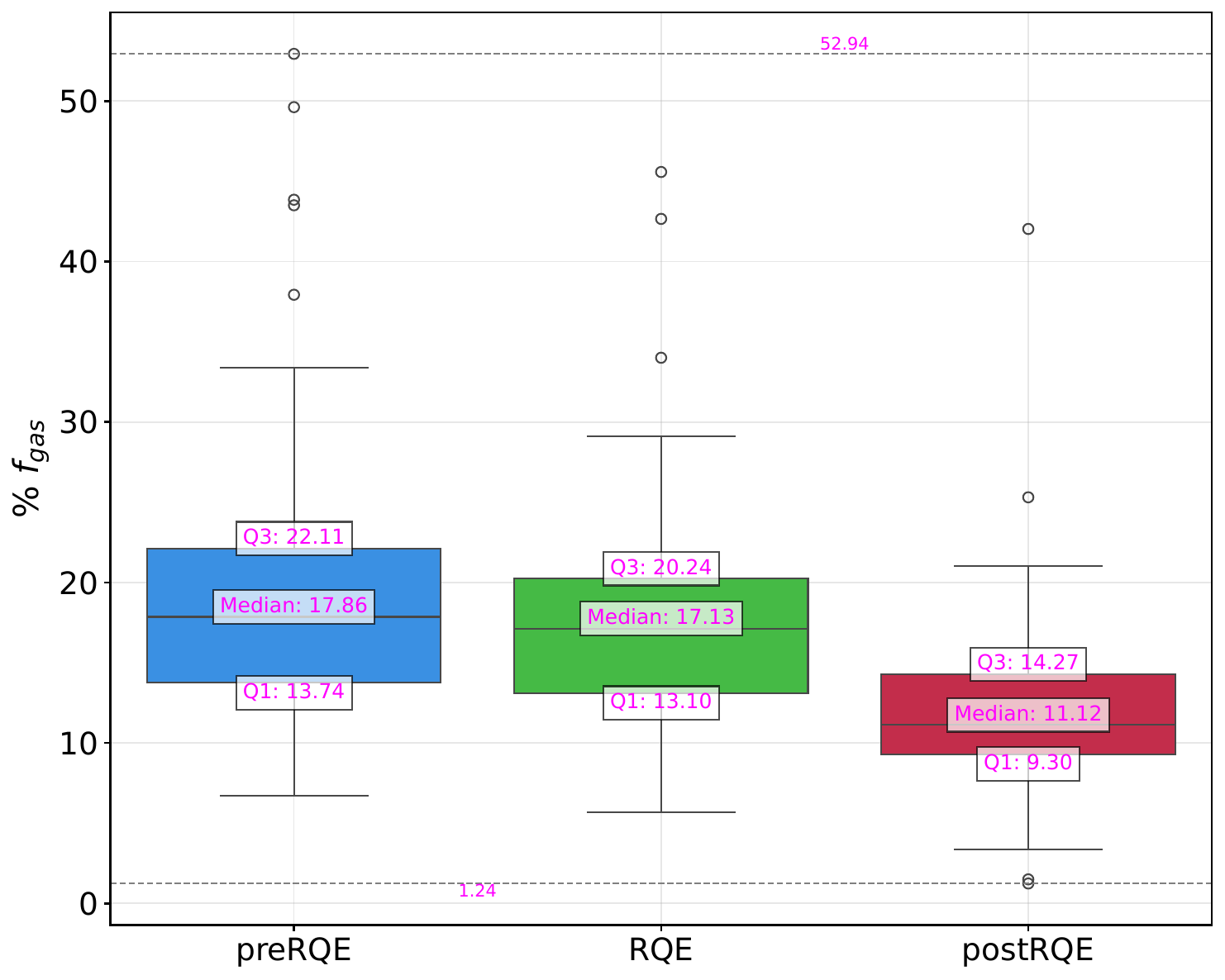}{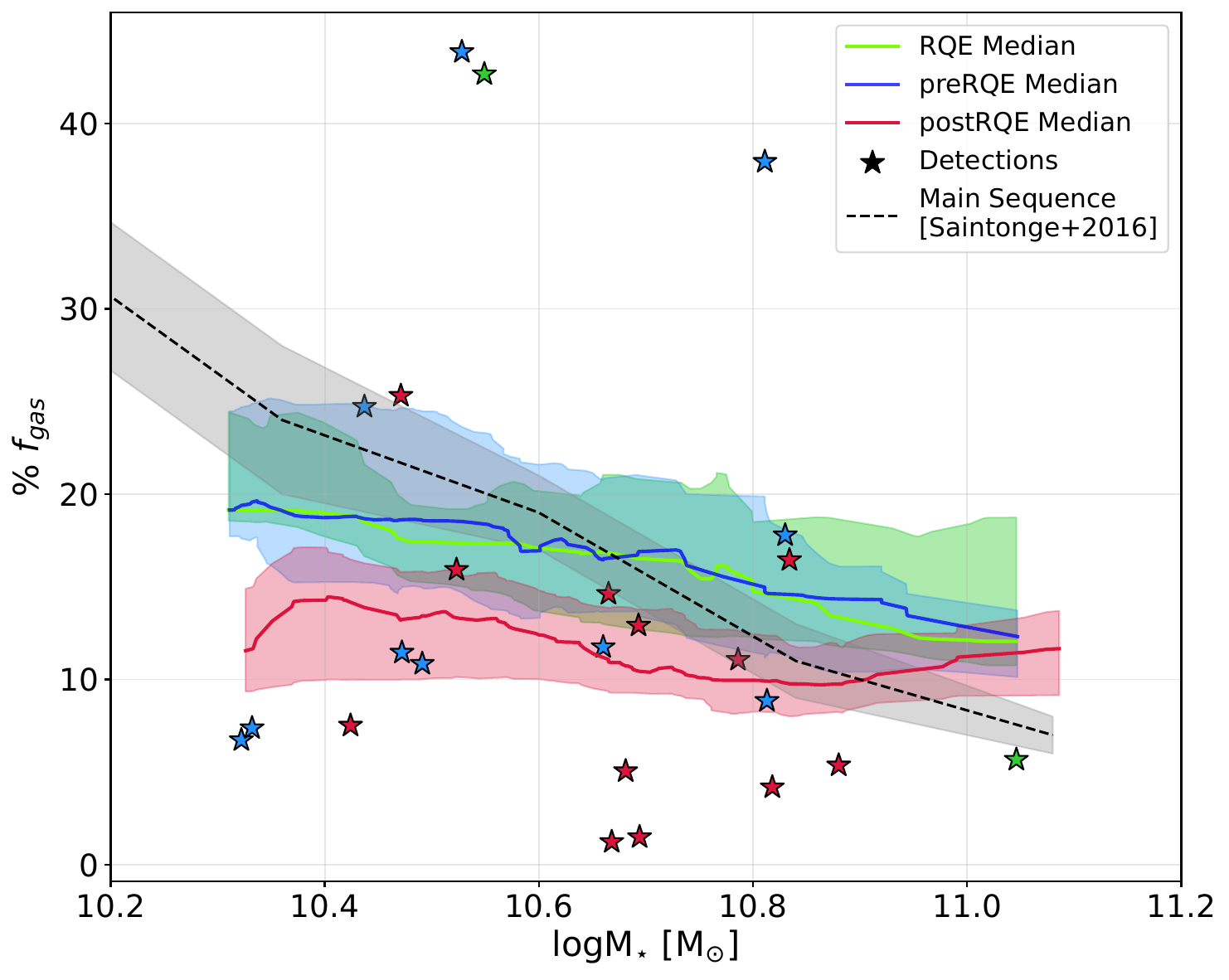}
\caption{(\textit{Left}): Box plot comparing the central tendency of percentage gas fraction for preRQEs, RQEs, and postRQEs. The boxes represent the interquartile range (IQR), with the median ($Q2$) shown as a horizontal line. Whiskers extend to $Q1 - 1.5\times IQR$ and $Q3 + 1.5\times IQR$. Outliers are plotted as individual points. Median, $Q1$, and $Q3$ values are annotated for each galaxy type. (\textit{Right}): A plot of percentage gas fraction versus stellar mass ($\log M_{\star}$), highlighting the running median and IQR for each galaxy type. Individual HI detections are plotted as stars. The dashed black line represents the mean gas fraction for main-sequence star-forming galaxies, as derived from Saintonge et al. (2016), with the associated error shades showing the uncertainty in those estimates.}
\label{fig:GFdist}
\end{figure*}

To further investigate the gas content trends observed in Figure \ref{fig:HImassnGF}, we conducted a more detailed analysis of the gas fractions across our galaxy types, as presented in Figure \ref{fig:GFdist}. This analysis focuses on the 305 galaxies (44\% of our sample) with either direct HI detections or reliable T17 estimates, excluding upper limits to ensure robust comparisons.

The left panel of Figure \ref{fig:GFdist} provides a statistical summary of the percentage gas fractions for preRQEs, RQEs, and postRQEs. The box plot reveals that RQEs exhibit a median gas fraction (17.1\%) remarkably similar to that of preRQEs (17.9\%), while postRQEs show a notably lower median (11.1\%). This observation is consistent with our earlier findings but provides a more nuanced view of the distribution. The interquartile ranges (IQRs) reflecting the central tendencies further illustrate this trend, with RQEs (IQR: 13.1\% - 20.2\%) closely mirroring preRQEs (IQR: 13.7\% - 22.1\%), while postRQEs show a lower and tighter distribution (IQR: 9.3\% - 14.3\%).

The right panel of Figure \ref{fig:GFdist} presents the relationship between gas fraction and stellar mass for our galaxy types (RQEs, preRQEs, and postRQEs), compared against the main sequence relation from \cite{2016MNRAS.462.1749S} (hereafter S16). To facilitate a consistent comparison, we calculated running medians for each galaxy type using a fixed stellar mass window of 0.24 dex, the same bin size employed by S16. This method highlights the central trends in gas fraction while the shaded regions indicate the IQR, reflecting the spread of gas fractions within each mass bin. Gaussian smoothing ($\sigma = 1$) was applied to the running medians and IQRs to reduce noise while preserving underlying trends. 
The main sequence relation (represented by the dashed black line) and the accompanying grey shaded area are based on the mean gas fractions and their associated uncertainties reported in Table 2 of S16.

We observe that the running median lines for preRQEs, RQEs, and postRQEs suggest a decline in gas fraction with increasing stellar mass, consistent with findings for main-sequence star-forming galaxies as indicated by the comparison line from \citet{2016ApJ...830...51S}. However, RQEs and postRQEs generally have lower gas fractions at a given stellar mass compared to preRQEs.
The IQR intervals indicate substantial overlap between RQEs and preRQEs, particularly at higher stellar masses ($\log M_{\star}\geq 10.6$), while postRQEs remain distinctly separated.

Our analysis has revealed that RQEs, despite their quenched state, retain substantial gas reservoirs, more similar to their star-forming precursors than to their quiescent descendants. The persistence of significant gas content across the stellar mass range suggests that the quenching mechanism in RQEs is not simply a result of gas depletion. However, the presence of gas alone does not fully account for the cessation of star formation in these galaxies. To gain a more comprehensive understanding, it is crucial to assess whether the observed gas content in RQEs is consistent with their stellar mass and recent star formation history.

In the following section, we will investigate the HI deficiency of our sample, focusing on how RQEs compare to both preRQEs and postRQEs in this regard. HI deficiency, defined as the deviation of a galaxy's HI mass from that expected for its stellar mass, provides a critical diagnostic for understanding the quenching process. By correlating gas fraction with $NUV-r$ color—a reliable tracer of recent star formation—we aim to determine whether RQEs are indeed gas-deficient relative to expectations. This examination will allow us to ascertain whether RQEs are genuinely gas-poor compared to typical star-forming galaxies or if they maintain normal gas reservoirs but experience less efficient star formation.

\subsection{HI deficiency}
HI gas deficiency refers to the lack of neutral atomic hydrogen gas (HI) in galaxies compared to the expected HI content based on their morphological type, luminosity, and other properties. The concept of HI deficiency was first introduced by \cite{1984AJ.....89..758H} in their study of spiral galaxies in the Virgo cluster. They observed that a large fraction of spiral galaxies in the cluster had HI gas content significantly lower than field galaxies with similar properties, suggesting an environmental effect on the gas content of these galaxies. A galaxy is considered HI deficient if its HI gas content is lower than the expected value. Since HI gas is the raw material for star formation in galaxies, its deficiency can lead to a decline in the star formation rate, eventually resulting in the quenching of star formation altogether.

The HI deficiency parameter ($\textit{Def}_{\text{HI}}$) is typically quantified as the logarithmic difference between the expected and observed HI mass \citep{2014MNRAS.444..667D}: 
\begin{equation}
    \textit{Def}_{\text{HI}} = \log(M_{\text{HI, expected}}) - \log(M_{\text{HI, observed}})
\end{equation}
where $M_{\text{HI, expected}}$ is the HI mass predicted by scaling relations, and $M_{\text{HI, observed}}$ is the actual measured or inferred HI mass. A positive \text{Def}{\text{HI}} value indicates a deficiency, implying that the galaxy has less HI than expected, which may inhibit its ability to sustain star formation.

Investigating HI deficiency in RQEs and their plausible precursors and descendants is crucial for understanding the processes that drive the cessation of star formation. While our previous analyses have shown that RQEs retain significant gas reservoirs, assessing whether they are HI deficient relative to their stellar mass provides insight into whether quenching in these galaxies is due to gas depletion or other factors, such as a reduced efficiency in converting gas into stars.

In this study, we focus on comparing HI deficiency across RQEs, preRQEs, and postRQEs. By correlating HI deficiency with $NUV-r$ color—a well-established tracer of recent star formation activity \citep{2007ApJS..173..267S}—we aim to determine whether RQEs are indeed gas-deficient relative to expectations, or if they retain typical gas reservoirs with suppressed star formation efficiency.

To perform this analysis, we utilized SDSS $r$ band data, which were extinction-corrected and $k-$corrected to $z=0.0$ using the methodology of \cite{2010MNRAS.405.1409C} and \cite{2012MNRAS.419.1727C}, respectively. We also made use of GALEX Near-UV (NUV) data from the GR6 data release, which were corrected for Galactic extinction using the method of \cite{2005ApJ...619L..15W} and $k-$corrected to $z=0.0$ using the same resources as the SDSS data. We then classified our sample into star-forming, green valley, and quenched galaxies based on their extinction-corrected, $k-$corrected, and AB system $(NUV-r)$ values, following \cite{2014SerAJ.189....1S}.

\begin{figure}
    \includegraphics[width = \columnwidth]{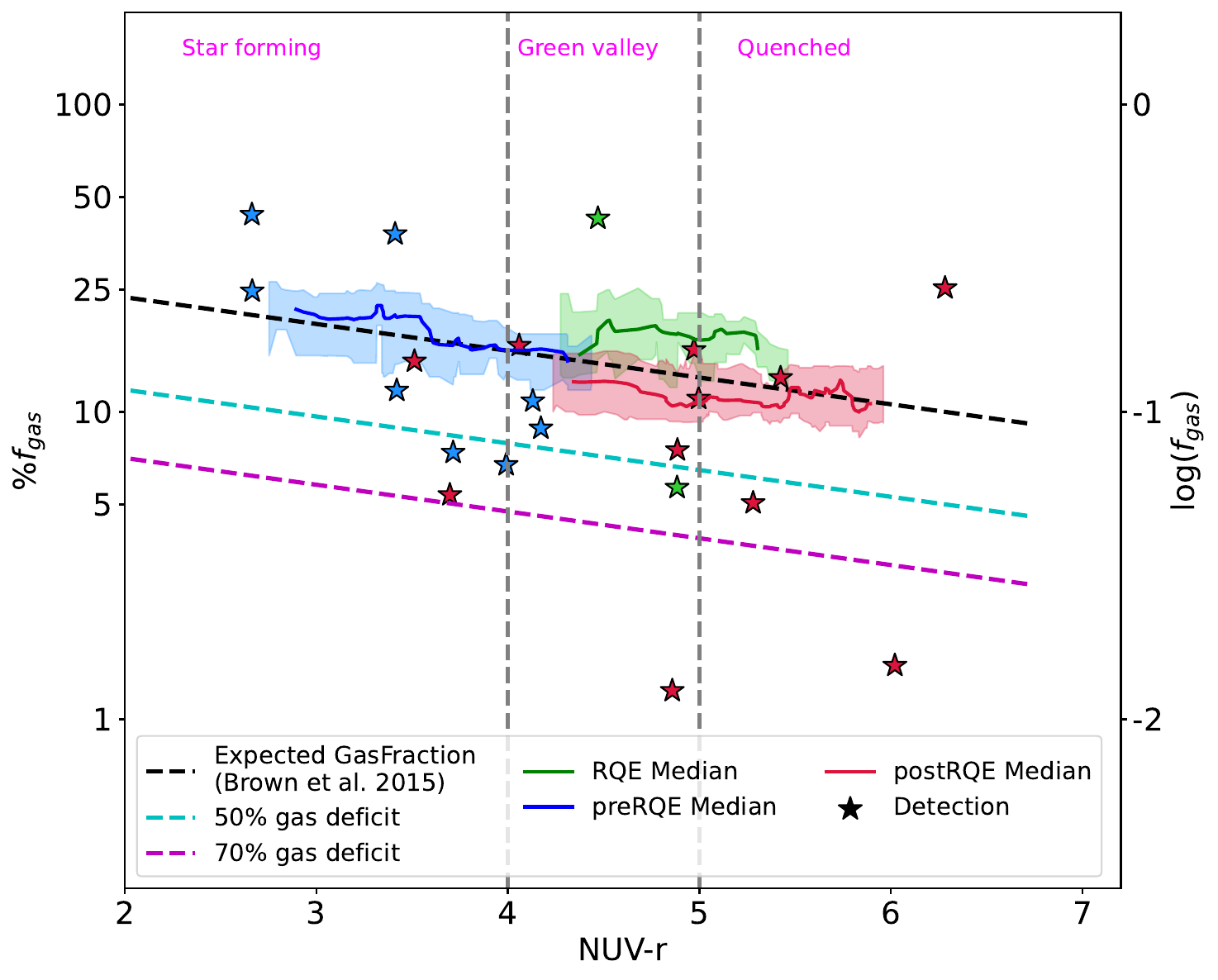}
    \caption{The relationship between HI gas fraction and $NUV-r$ color index for preRQEs, RQEs, and postRQEs. Data points represent individual galaxies, with running medians and interquartile ranges (25-75\%) shown for each galaxy type. The solid lines represent the running medians, while the shaded regions indicate the interquartile ranges. Overlaid dashed lines represent the expected gas fractions for star-forming, green valley, and quenched galaxies, derived from \citet{2015MNRAS.452.2479B}, with additional dashed lines indicating 50\% and 70\% HI deficiency levels. The vertical dashed lines mark the boundaries between the star-forming, green valley, and quenched regions, as classified by the $NUV-r$ color index.}
    \label{fig:HIdef}
\end{figure}

Using the HI gas fraction information and the $(NUV-r)$ values of our galaxies, and expected gas fraction values for star-forming, green valley, and quenched galaxies from \cite{2015MNRAS.452.2479B}, we were able to determine the state of HI deficiency in our galaxies. 
\citet{2015MNRAS.452.2479B} employed the HI spectral stacking technique to estimate gas fractions for a large sample of approximately 25,000 galaxies, selected based on stellar mass and redshift from the Sloan Digital Sky Survey (SDSS). The HI spectral stacking method allowed them to determine average gas fractions for galaxies grouped by stellar mass and $(NUV-r)$ color, providing a robust baseline for our analysis. To quantify the expected gas fraction for our galaxies, we derived an equation from the data points in Figure 3(c) of \citet{2015MNRAS.452.2479B}, which correlates gas fraction with $(NUV-r)$ color. The comparison of expected versus observed gas fractions is illustrated in Figure  \ref{fig:HIdef}.

Figure \ref{fig:HIdef} presents gas fraction as a function of the $(NUV-r)$ color index, with data points representing detections for preRQEs, RQEs, and postRQEs. Galaxies lacking $NUV-r$ information are excluded from the plot. The running median gas fraction, along with IQR, is shown for each galaxy type, providing a comprehensive view of the gas content distribution.

Overlaid on the data are the expected gas fractions for galaxies in different evolutionary stages—star-forming, green valley, and quenched—as derived from the equation we formulated based on \citet{2015MNRAS.452.2479B}. The dashed line in the figure represents the expected gas fraction for a galaxy at a given $(NUV-r)$ color, while the other lines depict 50\% and 70\% gas deficiency levels. This allows for a direct comparison between the expected and observed gas fractions across our sample.

Our analysis reveals several intriguing trends. Contrary to expectations for quenched systems, RQEs do not appear to be HI deficient. In fact, RQEs situated in the green valley (GV) region tend to have slightly more gas than predicted by the Brown et al. relation. This finding is particularly noteworthy given the presumed quenched state of these galaxies. Among the two RQE detections in our sample, we observe a broad range in gas content, with one galaxy showing a significant surplus and another displaying a deficiency exceeding 50\%.

PreRQEs similarly do not appear to be gas-deficient overall. However, an interesting pattern emerges as these galaxies approach the GV region: over 60\% (5 out of 8) of HI-detected preRQEs appear to have less gas than expected. This trend is followed by an apparent increase in gas content as they enter the GV region, suggesting a complex relationship between gas fraction and the transition into the green valley.

PostRQEs, in contrast, tend to be gas deficient, particularly those in the green valley region. Approximately 60\% (7 out of 11) of postRQE detections in this region have less gas than expected based on the Brown et al. relation. This pattern aligns more closely with conventional understanding of galaxy quenching, where gas depletion accompanies the cessation of star formation.

The progression of gas content across our three galaxy types in the GV region is particularly intriguing. PreRQEs align with expected gas fraction values, RQEs exhibit an excess, while postRQEs display a deficit. This sequence may hint at an evolutionary link among these galaxy types, which is most evident in the GV region - often considered a transition zone where blue star-forming galaxies are either evolving into red and quiescent systems, or quiescent galaxies are undergoing rejuvenation towards bluer colors. These possible evolutionary connections, especially in relation to gas content, will be further explored in the discussion section.

Given that RQEs are not experiencing gas deficiency and, in some cases, even exhibit a surplus of gas despite their quenched state, this raises important questions about their current star formation rates and efficiency. This observation prompts a closer investigation into how these factors relate to their gas content. In the following section, we will delve into the star formation properties of RQEs, comparing them with the subpopulations, particularly in the context of their HI content.

\subsection{Star Formation and HI Gas Content Relationship} \label{sec:sf_higas}
The relationship between star formation and HI gas content is fundamental to our understanding of galaxy evolution. While cold HI gas typically fuels star formation, our analysis of RQEs reveals a more complex scenario where substantial gas reservoirs coexist with quenched star formation. This paradox prompts a deeper investigation into the star formation properties of RQEs in relation to their HI content.

To explore this relationship, we analyzed the star formation rates (SFRs) of our sample using data from the GALEX-SDSS-WISE Legacy Catalog 2 (GSWLC-2) developed by \cite{2018ApJ...859...11S, 2016ApJS..227....2S}. We primarily used the GSWLC-M2 catalog, which covers 49\% of SDSS galaxies and is recommended for passive or off-main sequence galaxies. For galaxies not present in M2, we used the GSWLC-X2 catalog, which covers 90\% of SDSS galaxies. These catalogs provide SFRs computed by modeling the UV, optical, and mid-IR broadband Spectral Energy Distribution (SED). For the remaining 77 galaxies, we could find SFR information from \cite{2015ApJS..219....8C}, but we chose not to use it for our analysis due to its differences in the method of estimating SFRs compared to the GSWLC-2 catalog.

\begin{figure}
    \centering
    \includegraphics[width=\columnwidth]{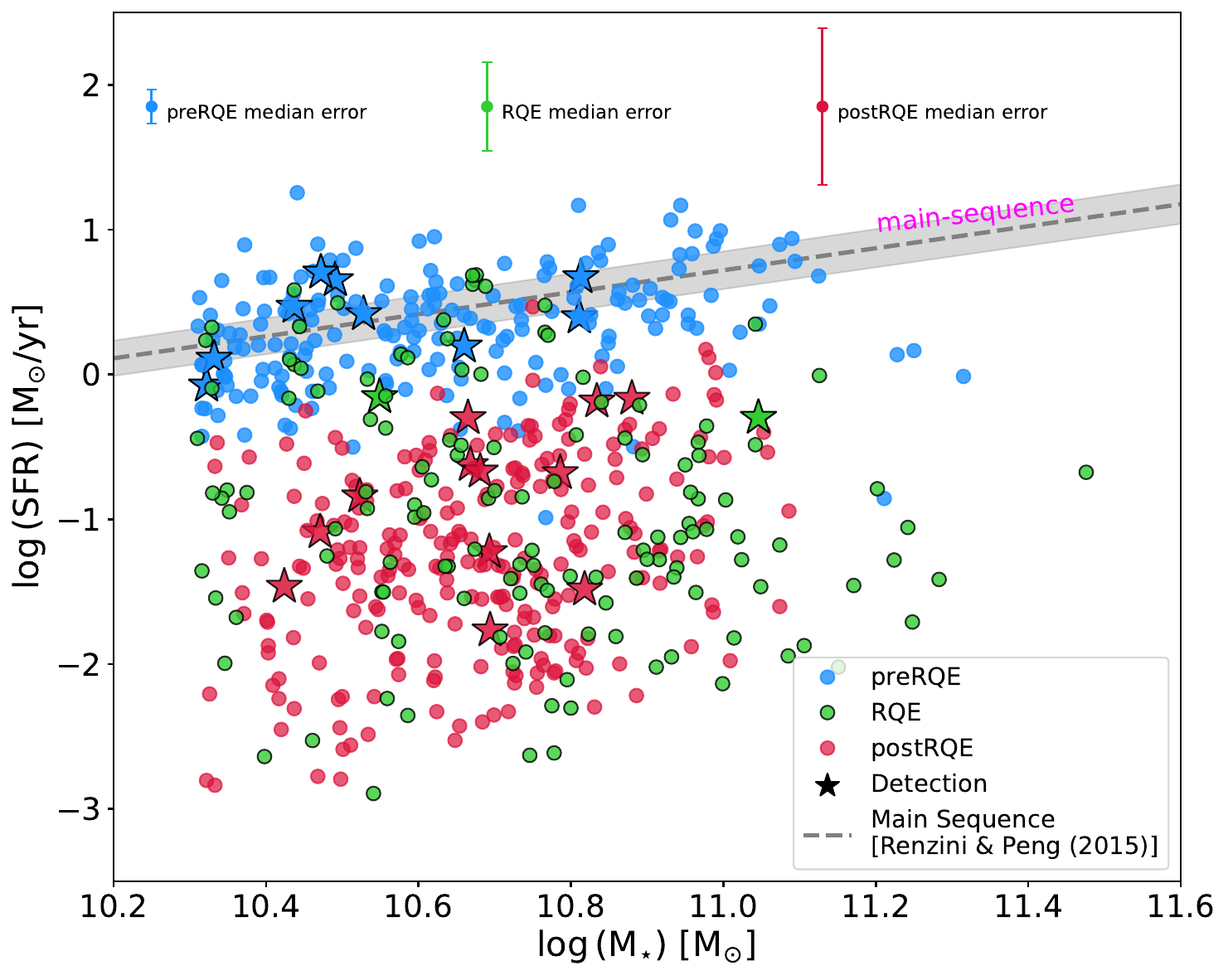}
    \caption{logSFR-logM$_{\star}$ comparison of RQE with comparable subpopulations. The data points represent total SFR values obtained from the GSWLC-M2 catalog, with supplementary values from the GSWLC-X2 catalog where necessary. Median error bars for each galaxy type are indicated, highlighting the statistical uncertainties in the sample. The star-forming main sequence (SFMS) relation, derived from \citet{2015ApJ...801L..29R}, is overlaid as a dashed line, providing a reference for typical star-forming galaxies in the local universe. Galaxies with HI detections are marked with star symbols, emphasizing their positions relative to the SFMS and their respective galaxy types.}
    \label{fig:SF}
\end{figure}

Figure \ref{fig:SF} presents the relationship between log(SFR) and log(M$_{\star}$) for our sample of preRQEs, RQEs, and postRQEs. The plot includes the main sequence relation for star-forming galaxies derived by \cite{2015ApJ...801L..29R}, given by the equation:
\begin{equation}
\log(\text{SFR}) = (0.76 \pm 0.01)\log(\text{M}_{\star}) - (7.64 \pm 0.02)
\end{equation}
This relation serves as a benchmark for the star formation activity of normal star-forming galaxies in the local universe, providing a reference point against which we can assess the star formation rates of our galaxy subpopulations.

The distribution of galaxies in Figure \ref{fig:SF} reveals several key features. PreRQEs predominantly occupy the region near and slightly below the main sequence, consistent with their classification as star-forming galaxies. RQEs, on the other hand, are found well below the main sequence, confirming their quenched status. However, their distribution shows considerable scatter, with some RQEs exhibiting SFRs closer to those of preRQEs and even approaching the main sequence line. PostRQEs clearly occupy the lowest region in the SFR-M$_{\star}$ plane, indicating a more advanced stage in the quenching process. The majority of HI detections (marked as stars) for all three galaxy types tend to have higher SFRs within their respective populations, particularly notable for RQEs.

To quantify these observations and the alignment of the highly scattered log(SFR) values for RQEs towards comparable subpopulations, we performed two types of statistical tests: the independent samples $t-$test and the Kolmogorov-Smirnov ($K-S$) test. The $t-$test is designed to compare the means of two independent groups, allowing us to determine if RQEs have significantly different average SFRs compared to preRQEs and postRQEs. The results revealed significant differences between RQEs and preRQEs ($t-$statistic = -19.025, $p-$value < 0.001), and between RQEs and postRQEs ($t-$statistic = 3.903, $p-$value < 0.001). These findings confirm that RQEs have significantly lower SFRs than preRQEs but higher SFRs than postRQEs.

The $K-S$ test, which compares the distributions of two samples, further supports these results. The $K-S$ test yielded a $KS-$statistic of 0.711 ($p-$value $<$ 0.001) when comparing RQEs and preRQEs, indicating a substantial difference in their SFR distributions. A $KS-$statistic of 0.190 ($p-$value = 0.002) was found when comparing RQEs and postRQEs, suggesting that while the SFR distribution of RQEs is more similar to postRQEs, it is still significantly different from them. The significant differences in SFR and its distribution between these galaxy populations underscore the transitional nature of RQEs, likely representing a stage in which star formation is being quenched, but not yet fully extinguished.

\begin{figure*}
    \includegraphics[width=\textwidth]{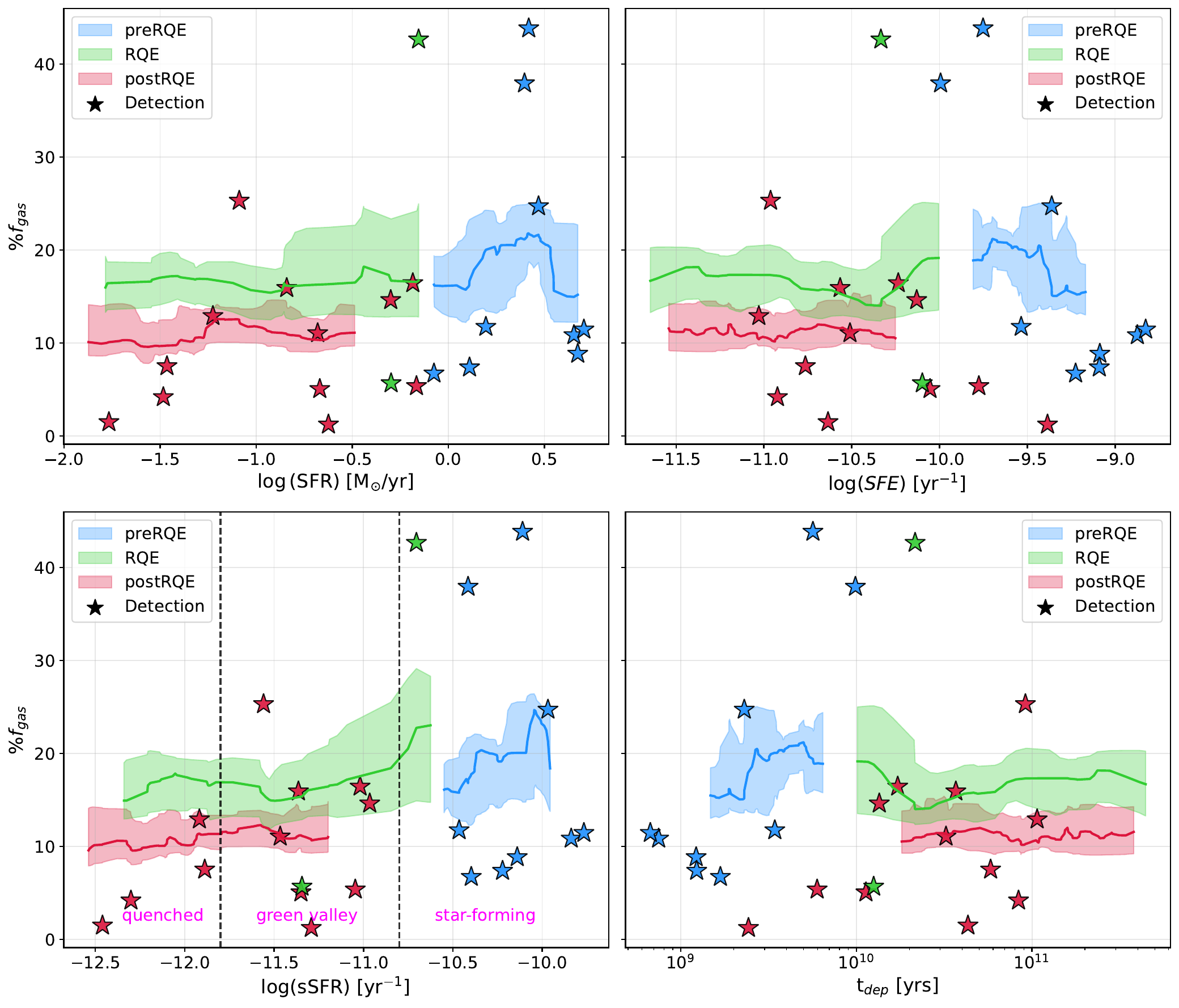}
    \caption{Relationships between gas fraction ($f_{\text{gas}}$) and various star formation parameters for preRQEs (blue), RQEs (green), and postRQEs (red). Top left: $f_{\text{gas}}$ vs. log(SFR). Top right: $f_{\text{gas}}$ vs. log(SFE), where SFE = SFR/M$_{\text{HI}}$. Bottom left: $f_{\text{gas}}$ vs. log(sSFR), with vertical dashed lines delineating quenched, green valley, and star-forming regimes. Bottom right: $f_{\text{gas}}$ vs. gas depletion timescale (t$_{\text{dep}}$ = 1/SFE). Solid lines represent running medians, while shaded regions show the interquartile ranges (IQRs). Star symbols indicate HI detections. The running medians and IQRs were calculated using a window size of 20\% of the total dataset for each galaxy type, with Gaussian smoothing applied. This figure illustrates the complex interplay between gas content and star formation properties across different evolutionary stages, highlighting the unique characteristics of RQEs as transitional objects between star-forming and quenched states.}
    \label{fig:GFvsAll}
\end{figure*}

To further elucidate the relationship between star formation and gas content, we analyzed the correlation between gas fraction (f${gas}$), specific star formation rate (sSFR), star formation efficiency (SFE), and gas depletion timescale (t$_{dep}$) for our sample, as presented in Figure \ref{fig:GFvsAll}. 

Figure \ref{fig:GFvsAll} presents four critical relationships, offering insights into the interplay between star formation and gas content for preRQEs, RQEs, and postRQEs. The panels illustrate \(f_{\text{gas}}\) as a function of SFR (top left), SFE (top right), sSFR (bottom left), and \(t_{\text{dep}}\) (bottom right). The running medians and interquartile ranges (IQRs) were calculated using a window size defined as 20\% of the total dataset for each category, with Gaussian smoothing applied to ensure the trends are representative and smooth. This method ensures that the calculated medians and IQRs accurately reflect the central tendencies and variability within the dataset.

In the upper left panel \(f_{\text{gas}}\) is compared with log(SFR). This panel reveals distinct patterns for each galaxy type. PreRQEs show a positive correlation between gas fraction and SFR, with gas fractions ranging from $\sim$15\% to 25\% across the observed SFR range. RQEs, however, display a flatter trend, with gas fractions consistently around 15-20\% despite having significantly lower SFRs compared to preRQEs. This suggests that while RQEs retain substantial gas reservoirs, their ability to convert this gas into stars has diminished, consistent with their quenched state. PostRQEs exhibit the lowest gas fractions, typically around 10\%, with minimal variation across their SFR range, reflecting a more advanced stage of gas depletion and quenching.

The detection data points (represented by stars) often deviate from these trends, particularly for RQEs and postRQEs. This deviation may be due to the small sample size or the possibility of these detections being outliers, indicating that galaxies with detectable HI could have unique properties within their respective populations.

In the bottom left panel, we examine the relationship between specific star formation rate (sSFR = \text{SFR}/M$_{\star}$) and gas fraction ($f_{\text{gas}}$). This panel is particularly informative, as it includes demarcations for quenched, green valley, and star-forming regimes based on sSFR thresholds established in the literature \citep{2014SerAJ.189....1S, rathore2022star, 2017MNRAS.471.2687B}. A clear trend emerges: higher log(sSFR) correlates with higher \(f_{\text{gas}}\) across all galaxy types. 

PreRQEs predominantly occupy the star-forming region, exhibiting a positive correlation between \(f_{\text{gas}}\) and sSFR. This trend aligns with expectations for actively star-forming galaxies, where higher gas fractions fuel increased star formation. The median $f_{\text{gas}}$ for preRQEs ranges from approximately 15\% to 25\% across the observed sSFR range.

RQEs, in contrast, span both the green valley (GV) and quenched regions, revealing a more complex relationship between gas content and star formation. Notably, RQEs maintain relatively high gas fractions (median $f_{\text{gas}}$ $\sim$15-20\%) despite their reduced sSFRs. RQEs in the GV region exhibit $f_{\text{gas}}$ values and trends similar to preRQEs in the star-forming region, suggesting that the initial stages of quenching in these galaxies are not primarily driven by gas depletion.

PostRQEs cluster predominantly in the quenched regime, characterized by consistently lower gas fractions (median \(f_{\text{gas}}\) $\sim$10\%) and sSFRs. This pattern is consistent with expectations for galaxies in advanced stages of quenching, where both star formation activity and gas content have significantly diminished.

Interestingly, the two RQE detections in our sample follow the general trend of higher \(f_{\text{gas}}\) with higher log(sSFR). One of these detections lies in the GV region, exhibiting gas content and star-forming properties akin to postRQEs, while the other is positioned in the star-forming region near the boundary with the GV. Despite this latter RQE's gas content resembling that of preRQEs, it is spectroscopically quiescent, highlighting the complexity of the quenching process.

In the top right panel, which plots \(f_{\text{gas}}\)
against SFE (defined as ration of SFR and HI gas mass, SFR/M$_{\text{HI}}$), we observe that RQEs have a lower star forming efficiency compared to preRQEs, despite having comparable gas fractions. This difference in SFE is consistent with the idea that RQEs are in a transitional state, where the mechanisms driving star formation have been suppressed. The postRQEs, with even lower SFEs, indicate a more advanced stage of quenching, where both gas content and star formation activity are significantly reduced.

The \(f_{\text{gas}}\) versus log(SFE) plot (top right panel) offers insights into the efficiency of star formation across our sample. SFE is defined as the ratio of SFR and HI gas mass, SFR/M$_{\text{HI}}$. We observe that RQEs have a lower star forming efficiency compared to preRQEs, despite having comparable gas fractions. This difference in SFE is consistent with the idea that RQEs are in a transitional state, where the mechanisms driving star formation have been suppressed. The postRQEs, with even lower SFEs, indicate a more advanced stage of quenching, where both gas content and star formation activity are significantly reduced.
Interestingly, preRQEs display a slight negative correlation, indicating that higher SFEs are associated with lower gas fractions, possibly due to the consumption of gas driven by efficient star formation.

Finally, the bottom right panel of Figure \ref{fig:GFvsAll} presents the gas depletion timescale ($t_{dep} = 1/SFE$) as a function of \(f_{\text{gas}}\). This panel illustrates that RQEs have significantly longer depletion timescales ($\geq10$ Gyr) compared to preRQEs, implying that the gas in RQEs would take much longer to be consumed if star formation were to continue at its current, reduced rate. The extended $t_{dep}$ in RQEs suggests that their quenching is not driven by a lack of available gas but rather by a reduced efficiency in utilizing this gas for star formation. In contrast, preRQEs, with their higher SFE, have shorter depletion timescales, consistent with active star formation. PostRQEs, which have low gas fractions and low SFE, exhibit long depletion timescales as well, indicating that they are in the final stages of gas consumption.

These results suggest that RQEs occupy a transitional phase between the actively star-forming preRQEs and the fully quenched postRQEs. The presence of substantial gas reservoirs in RQEs, coupled with their lower SFE and prolonged gas depletion timescales, indicates that the quenching of star formation in these galaxies is not merely a consequence of gas exhaustion. Instead, it points to more complex mechanisms at play—potentially involving feedback processes, heating of gas, dynamical stabilization, changes in gas properties, or environmental effects—that suppress star formation while leaving the gas reservoir largely intact. This nuanced understanding challenges simple quenching models and underscores the need to consider additional factors beyond just gas availability in the quenching process of RQEs.

In the following discussion section, we will explore the implications of these results for our understanding of galaxy quenching processes and the evolutionary pathways that lead to the formation of RQEs.

\section{Discussion}\label{sec:dis}
Our analysis of neutral hydrogen (HI) content in Recently Quenched Elliptical galaxies (RQEs) reveals a complex relationship between gas content and star formation, challenging conventional paradigms of galaxy quenching and evolution. A key finding is the retention of substantial HI reservoirs in RQEs, despite their spectroscopically quiescent state. Contrary to expectations for quenched systems, RQEs are not HI deficient; rather, many exhibit an HI surplus compared to predictions from HI stacking studies \citep{2015MNRAS.452.2479B}. This phenomenon is observed across both green valley and fully quenched regimes, suggesting that the quenching process in RQEs suppresses star formation without significantly depleting their gas reservoirs.

Further examination of star formation properties—including star formation rates (SFR), star formation efficiency (SFE), and gas depletion timescales—reveals that RQEs demonstrate lower SFE and significantly extended gas depletion timescales compared to their star-forming counterparts. These findings indicate that the quenching of star formation in RQEs is not primarily due to a lack of gas but likely involves more intricate processes that inhibit the efficient conversion of gas into stars.

These results raise fundamental questions about the quenching mechanisms at play in RQEs. Specifically, what processes are responsible for suppressing star formation in these gas-rich galaxies? How do RQEs fit into the broader context of galaxy evolution, particularly in terms of their transition from star-forming to quiescent states? Investigating the roles of internal processes, such as feedback mechanisms, and external environmental influences is critical for advancing our understanding of these galaxies.

In the following discussion, we will contextualize our findings within the broader empirical and theoretical frameworks. We begin by reviewing empirical data on HI content in galaxies similar to RQEs, particularly focusing on isolated or central blue early-type galaxies (ETGs) in low-density environments. This review will be followed by an examination of theoretical perspectives that could explain the unique combination of quenched star formation and significant gas content observed in RQEs. 
Finally, we will attempt to constrain the possible quenching scenarios in RQEs by comparing and analyzing our results with plausible quenching mechanisms in similar environments, as well as discussing relevant scenarios proposed in the literature, including those from \citet{2014MNRAS.442..533M}.
By synthesizing our findings with existing literature, we aim to constrain the possible evolutionary pathways leading to the formation of RQEs. 

\subsection{Empirical Information on HI in RQE-like Galaxies}
Empirical studies have consistently shown that galaxies, including both early-type (ETGs) and late-type galaxies (LTGs), in lower-density environments (LDEs) tend to have higher gas fractions compared to their counterparts in high-density environments \citep{1985ApJ...292..404G, 1986A&A...165...15C, 1991AJ....102..572E, 2011MNRAS.415.1797C, 2014MNRAS.444..667D}. This trend is particularly relevant for understanding the gas content of central RQEs, which are predominantly found in LDEs.

Early observations of HI in ETGs indicated that those with bluer optical colors typically retained significant HI reservoirs \citep{1976ApJ...209..710B}. In a seminal study using the National Radio Astronomy Observatory's 43m telescope, \cite{1976ApJ...209..710B} detected HI in the seven bluest ETGs, as classified by their $B-V$ photometry, while redder ETGs generally lacked detectable HI. This correlation between blue optical colors and substantial HI content suggested that blue ETGs, potentially akin to RQEs, could harbor considerable gas reservoirs.

Subsequent studies confirmed this trend, indicating that blue ETGs in LDEs or galaxy groups often retain considerable HI reservoirs and may exhibit ongoing star formation \citep{1991AJ....102..572E, 1996AJ....112..937D, 2009A&A...498..407G}. For example, \cite{1991AJ....102..572E}, using the Arecibo radio telescope, found that most ETGs with detected HI displayed $H_\alpha$ emission, a signature of active star formation. This suggests that the presence of HI in blue ETGs is closely tied to their star formation potential.

\cite{1996AJ....112..937D} proposed that the high HI content in LDE ETGs might result from interactions with HI-rich dwarf companions. They suggested that more massive ETGs could tidally disrupt smaller neighbors and accrete their HI gas. While this hypothesis offers a plausible explanation for the gas-rich nature of RQEs, it does not fully account for the suppression of star formation observed in these galaxies.

A comprehensive study by \cite{2009A&A...498..407G} reinforced the idea that ETGs in LDEs are likely to maintain significant HI gas reservoirs, with over 60\% of such ETGs showing evidence of ongoing star formation. However, it is important to recognize that not all blue ETGs in LDEs exhibit significant HI content; for instance, \cite{1976ApJ...204..365F} reported several cases of LDE ETGs with negligible or absent HI gas.

Given these empirical results, the presence of significant HI reservoirs in our RQE sample is consistent with the behavior of other blue ETGs in LDEs. However, the suppression of star formation in RQEs remains a puzzling aspect that distinguishes them from typical blue, gas-rich ETGs. This discrepancy underscores the need to consider additional factors beyond gas content in understanding the quenching process in RQEs.
In the next section, we will compare our results with theoretical studies in the literature to gain further insights into the mechanisms that might be responsible for the unique properties of RQEs.

\subsection{Theoretical Perspectives on HI in RQE- like Galaxies}
The empirical evidence outlined earlier aligns well with our observations regarding the HI content in RQEs. In this section, we explore theoretical models and predictions to further contextualize and interpret our findings, with a focus on those relevant to galaxies in low-density environments.

Theoretical studies, such as those by \citet{2014Natur.509..177V}, predict that galaxies with characteristics similar to RQEs in terms of stellar mass should harbor significant gas reservoirs. Their models estimate the present-day HI content of galaxies in the local universe, predicting gas fractions ranging from 2\% to 50\% for stellar masses comparable to RQEs. Notably, these predictions follow the trend of decreasing gas fraction with increasing stellar mass, which aligns well with our observations of $f_{gas}$ values between 6\% and 40\% in RQEs. This consistency supports the notion that RQEs retain substantial gas reservoirs despite being quiescent.

The retention of gas in RQEs can be further understood through the lens of gas accretion models. \cite{2005MNRAS.363....2K} and \cite{2012MNRAS.425.2027K} emphasize the importance of the halo mass in determining the mode of gas accretion. Central galaxies in halo masses comparable to those of RQEs are expected to accrete gas through different modes depending on the mass of their halos. For halo masses below $10^{11.4} M_{\odot}$, cold-mode accretion dominates, while for halos between $10^{11.4} M_{\odot}$ and $10^{13} M_{\odot}$, hot-mode accretion becomes prevalent. Since all our RQEs reside in halos with masses greater than $10^{11.7} M_{\odot}$, it is plausible that they have been accreting gas through the hot mode, which could have significant implications for their star formation activity.

\citet{2013MNRAS.429.3353N} further elaborates on the impact of gas accretion modes, particularly in the absence of feedback mechanisms such as AGN activity. This study highlights that central galaxies in halos less massive than $10^{13} M_{\odot}$, where hot-mode accretion is prevalent, can sustain efficient star formation unless additional feedback mechanisms intervene to quench the inflow of star-forming fuel to the central regions. 
For our RQEs, which reside in $<10^{13} M_{\odot}$ halos and lack active AGN, this would suggest that they should be forming stars actively. However, the observed quenching in RQEs indicates that other factors are likely suppressing star formation despite the availability of gas.

Recent studies have highlighted the differences between cold and hot accretion modes. Cold-mode accretion delivers high-density gas streams directly onto galaxies without shock heating, fueling bursts of star formation. In contrast, hot-mode accretion involves gas being shock-heated to the virial temperature before cooling, resulting in a slower, more gradual supply of gas that fuels steady, lower-level star formation \citep{1978MNRAS.183..341W, Kacprzak2017, 2011MNRAS.410.2653B, 2020MNRAS.492.6042S}. This difference in accretion modes may contribute to the quenching observed in RQEs.

AGN feedback is recognized as a critical factor that can suppress hot-mode gas accretion, particularly in massive halos. \citet{2011MNRAS.415.2782V} demonstrated that AGN feedback becomes particularly effective in halos with masses above $10^{12} M_{\odot}$, where hot-mode accretion dominates. The more diffuse nature of hot-mode gas makes it more susceptible to heating and removal by AGN-driven outflows. Furthermore, \cite{2011MNRAS.412.1965M} showed that early AGN feedback can raise gas entropy at later times, reducing the density of hot halo gas at low redshift and further inhibiting its ability to cool and accrete onto the central galaxy.

Given their halo masses, it is evident that RQEs can harbor substantial gas reservoirs accreted through the hot mode. Unlike cold mode accretion, where cold streams can reach the central region of the galaxy where star formation takes place, gas accreted through hot mode is shock-heated to high temperatures ($> 10^{5.5}$ K) near the virial radius. \cite{2009MNRAS.393...99W} notes that much of this gas cannot cool within a Hubble time and therefore cannot accrete onto the central galaxy. This suggests that much of the gas accreted onto RQEs via hot-mode accretion may remain hot for extended periods, potentially explaining their quenched state despite significant gas content.

This theoretical perspective, combined with our empirical findings, suggests that the quenching of star formation in RQEs is likely driven by complex interactions between gas accretion modes, halo mass, and feedback mechanisms. While the presence of substantial gas reservoirs in RQEs challenges the notion of quenching through gas exhaustion, it also highlights the importance of considering how the state of the gas—whether hot or cold, dense or diffuse—affects its ability to contribute to star formation.

\subsection{Constraining Quenching Scenarios in RQEs}
M14 conducted a comprehensive study of RQEs in the optical regime, exploring various quenching scenarios based on their halo mass, stellar mass, color, environment, and emission line properties. While M14 considered RQEs to be first-generation ellipticals and suggested that major mergers could be a likely explanation for the observed quenching, they did not arrive at a definitive conclusion regarding the exact mechanisms responsible. In this section, we build upon the insights from M14's optical analysis by integrating them with our findings from HI observations. By doing so, we aim to further constrain the plausible quenching scenarios and explore the evolutionary pathways that these enigmatic galaxies may have followed, offering a deeper understanding of the processes that led to their quenching.

When examining environmental quenching, M14 proposed that halo quenching \citep{2012MNRAS.427.1816G} could plausibly explain the quenching observed in central galaxies within small groups, particularly if additional feedback mechanisms—such as those from AGNs or LINERs—prevent the gas from cooling and subsequently forming stars. They suggested that the emissions from LINERs and the cyclical activity of AGNs might be effective in maintaining the high temperatures of the galactic halo, thereby facilitating hot halo quenching as a primary mechanism for the quenching observed in central RQEs within small groups. However, since AGNs or LINERs are present in only a subset of the RQE population, this theory falls short of explaining the quenching in RQEs that lack AGN activity. Thus, while this mechanism may be effective, it is not universally applicable to all RQEs.

M14 also argued that other heating processes not involving AGNs, such as gravitational heating and supernova (SN) feedback, are insufficient to explain the quenching in RQEs without AGN activity. Gravitational heating is effective only in halos larger than $10^{12.85}M_{\odot}$, while SN feedback is relevant primarily in smaller halos below $10^{11}M_{\odot}$. Given that no RQEs are observed with halo masses less than $10^{11}M_{\odot}$ and only a few (precisely five) RQEs have halo masses greater than $10^{12.85}M_{\odot}$, these quenching processes do not seem to be universally applicable across the RQE population.

In their exploration of energetic feedback as a quenching mechanism, M14 analyzed the impacts of energy released from the accretion of supermassive black holes, termed AGN feedback, and from supernovae and massive-star winds during periods of heightened star formation, referred to as SN or stellar feedback. They highlighted that radio mode AGN feedback could sustain a heated state in the atmospheres of galaxy groups, while mechanical AGN feedback might expel or significantly disrupt the gas reserves in isolated elliptical galaxies \citep{2012MNRAS.424..190G}. However, this framework does not explain the quenching in the majority of RQEs, which do not display active AGN features—approximately 94\% of the RQE population.

Furthermore, within the subset of RQEs with significant emission regions, some are Seyfert AGNs, while the majority are classified as LINERs. Seyferts have a relatively short active period, typically less than or equal to 100 million years \citep{2004cbhg.symp..169M, 2006ApJS..166....1H}), which is brief compared to the RQE phase—assumed to last about as long as an $A-$type star’s lifetime. M14 suggested that Seyfert activity is likely not connected to the quenching process.
Additionally, LINERs might represent the very tail end of AGN activity that is in the process of quenching, as schematically illustrated in Figure 1 of \cite{2008ApJS..175..356H}. Alternatively, their emissions may not be linked to nuclear ionization, as pointed out in \cite{2010A&A...519A..40A}, casting doubt on their role in sustaining widespread quenching in RQEs. However, as shown in section 2.3, about 37.5\% of the RQE population does show centralized LINER emission from the nuclear region, potentially playing a role in quenching the RQEs harboring them.

Star formation processes, including supernovae or stellar feedback, have also been dismissed as likely causes for quenching since the rate at which star formation consumes gas is insufficient to account for the rapid transition from blue to red observed in local early-type galaxies (ETGs) \cite{2007ApJS..173..512S}.

In their analysis of quenching mechanisms linked to major mergers, M14 discussed the major merger model as a compelling explanation for the transition of blue, star-forming galaxies into red, quiescent ellipticals, emphasizing that such mergers can rapidly form stars and then quench this formation, supporting a natural evolution from blue to red galaxies \citep{2006ApJS..166....1H, 2008ApJS..175..356H}. They propose that RQEs are key to understanding this process due to their characteristics being potentially indicative of recent major merger activity. M14 conducted four key tests to assess the major merger-quenching hypothesis in RQEs.

The first test involved looking for morphological disturbances, such as signs of tidal activity indicative of recent mergers, in RQEs. However, only 15\% of RQEs display such features, a finding that suggests either a swift post-merger evolution or variable timescales in detecting these disturbances, as supported by studies from \citep{2008MNRAS.391.1137L, 2010MNRAS.404..575L, 2010MNRAS.404..590L, 2010ApJ...714L.108S}. The second test examined AGN activity within RQEs, where they found significant AGN presence, including 9\% showing Seyfert emissions and 5\% with post-starburst signatures. This aligns with theories by \citep{1991ApJ...370L..65B, 1996ApJ...471..115B, 2006MNRAS.373.1013C, 2008MNRAS.384..386C}, which suggest that mergers can trigger inward gas flows, fueling AGN activity.

The third aspect of their study assessed the environmental positioning of RQEs, primarily found at the centers of smaller galactic groups, a preferred location for major mergers as suggested by \cite{2008ApJS..175..356H}. Finally, they compared the frequency of RQEs with the rate of major spiral-spiral interactions, discovering a congruence with the lower estimates of such interactions. This comparison further bolsters the likelihood that a considerable number of RQEs result from recent major mergers. Overall, while not all RQEs are products of major mergers, M14 concluded that these events play a crucial role in the quenching of star formation in a significant fraction of elliptical galaxies, presenting a complex yet coherent picture of galaxy evolution influenced by major merging events.

In light of our study’s findings on the HI content in RQEs and the insights from M14, we seek to constrain the quenching scenarios in these galaxies, delving into the complexities of their evolutionary processes. Our results, exhibiting substantial HI reservoirs in RQEs, showing HI surplus, may provide a unique perspective on the quenching mechanisms in play.

\begin{figure*}
    \includegraphics[width=\textwidth]{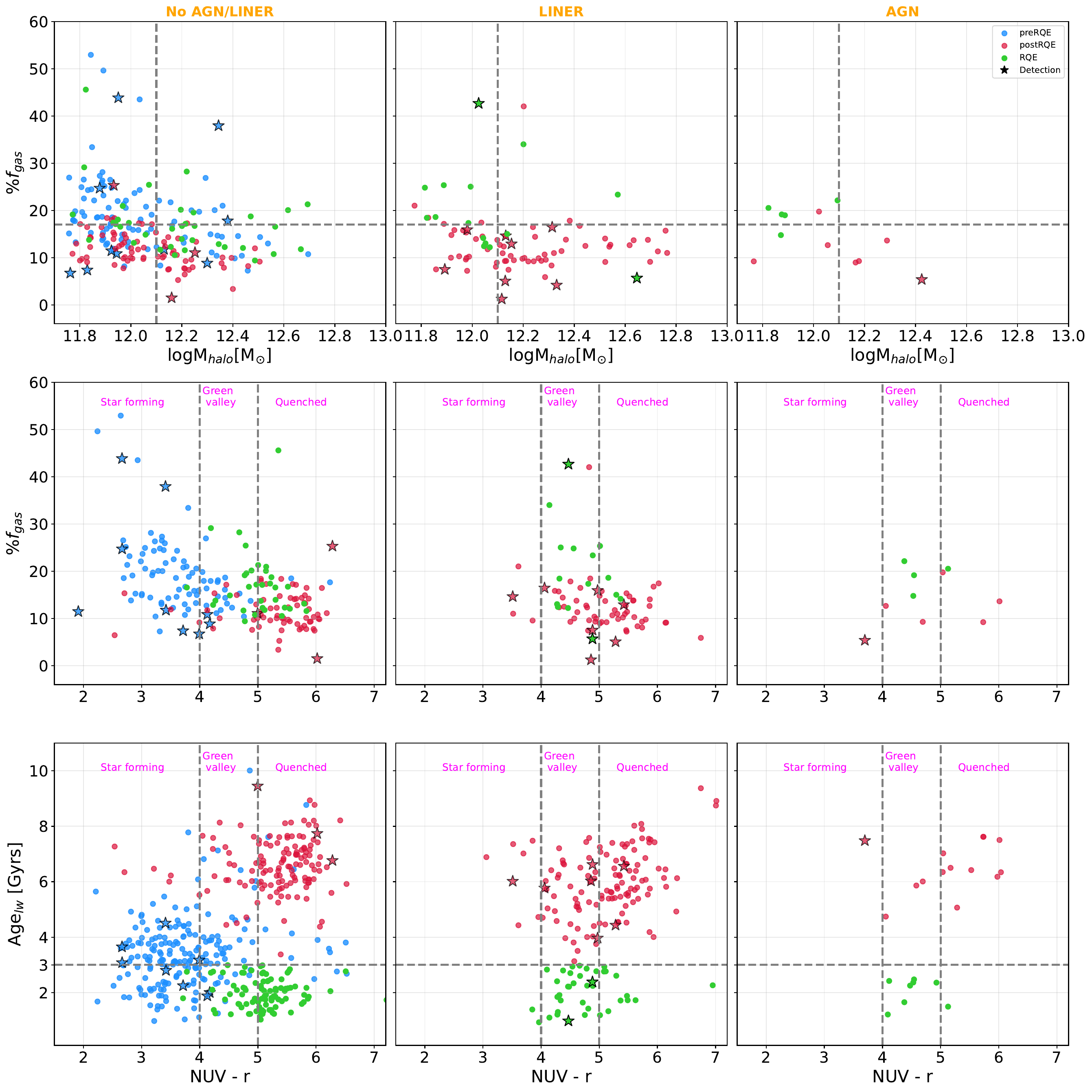}
    \caption{Multi-panel comparison of gas fraction ($f_{gas}$) and light-weighted stellar age (Age$_{lw}$) as a function of halo mass (log$M_{\text{halo}}$) and $NUV-r$ color for preRQE, RQE, and postRQE. The three panels correspond to galaxies without AGN/LINER (left), those with LINER (middle), and those with AGN (right). The top row shows $f_{gas}$ versus log$M_{\text{halo}}$, illustrating the relationship between gas content and halo mass across different emission categories. The middle row plots $f_{gas}$ versus $NUV-r$ color and the bottom row presents stellar age (Age$_{lw}$) as a function of $NUV-r$ color. The number of data points is higher in the bottom row plots because they include all galaxies with valid age and color measurements, regardless of the availability of HI data used in the top and middle row plots. Vertical dashed lines in the top row is at log$M_{\text{halo}}$ = 12.1 M$_{\odot}$, while the horizontal dashed line is at 17\% $f_{gas}$. In the middle and bottom rows, vertical dashed lines at $NUV-r$ = 4 and 5 delineate the star-forming, green valley, and quenched regions. The horizontal dashed line in the bottom row at 3 Gyr shows the light-weighted stellar age cutoff used for RQE identification.}
    \label{fig:disc_plot}
\end{figure*}

M14’s optical analysis suggests that major mergers are a probable scenario for quenching. However, this does not directly address the state of neutral HI gas, which, if heated, expelled, or depleted, could lead to quenching. We delve into the condition of the HI gas in RQEs, comparing it with plausible precursors and descendants (preRQEs \& postRQEs) as illustrated in Figure \ref{fig:disc_plot}. 
This comparison illuminates the relationship between the state of HI gas and the cessation of star formation in RQEs, offering insights into the potential quenching scenarios responsible for the observed HI gas properties.

In Figure \ref{fig:disc_plot}, we present a multi-panel $3 \times 3$ plot that examines the distribution of gas fraction ($f_{gas}$) as a function of halo mass (log$M_{\text{halo}}$) in the top row, $NUV-r$ color in the middle row, and light-weighted stellar age (Age$_{lw}$) over $NUV-r$ color in the bottom row for preRQE, RQE, and postRQE galaxies. The three columns correspond to galaxies without AGN/LINER (left), those with LINER (middle), and those with AGN (right). The top row displays $f_{\text{gas}}$ versus log$M_{\text{halo}}$, illustrating the relationship between gas content and halo mass across different emission categories.

By plotting LINER and AGN emissions separately, we aim to assess whether these emission classes exhibit distinct trends in their gas fractions and distributions over halo mass compared to RQEs that lack such high emission regions. This comparison could provide valuable insights into the role of AGN or LINER activity in influencing gas retention and the quenching processes in RQEs. Specifically, by analyzing these trends, we seek to determine if the presence of AGN or LINER emission correlates with more significant gas depletion or if these emissions signify different evolutionary pathways that contribute to quenching. The middle and bottom rows, which plot $f_{\text{gas}}$ and Age$_{\text{lw}}$ against $NUV-r$ color, respectively, will further our understanding of the relationship between gas content, recent star formation activity, and the evolutionary trajectory from star-forming preRQEs to quenched RQEs. Through this comprehensive analysis, we aim to uncover patterns that could elucidate the mechanisms driving quenching in these gas-surplus RQEs, thus contributing to a more detailed understanding of their evolution.

In the top row plot of $f_{\text{gas}}$ vs. log$M_{\text{halo}}$, a notable distinction emerges at log$M_{\text{halo}}$ = 12.1 M$_{\odot}$ and $f{\text{gas}}$ = 17\%. The majority of RQEs without AGN/LINER emission appear to show $f_{\text{gas}} \geq$ 17\% when residing in halos with log$M_{\text{halo}} \leq$ 12.1 M$_{\odot}$. A similar pattern is observed in RQEs with LINER or AGN emission, where most of these galaxies also reside in lower halo masses. Notably, log$M_{\text{halo}}$ = 12.1 M$_{\odot}$ closely matches the median halo mass of our RQE and postRQE sample, while $f_{\text{gas}}$ = 17\% aligns with the median gas fraction observed in the preRQEs and RQEs.

\begin{figure*}
\gridline{\fig{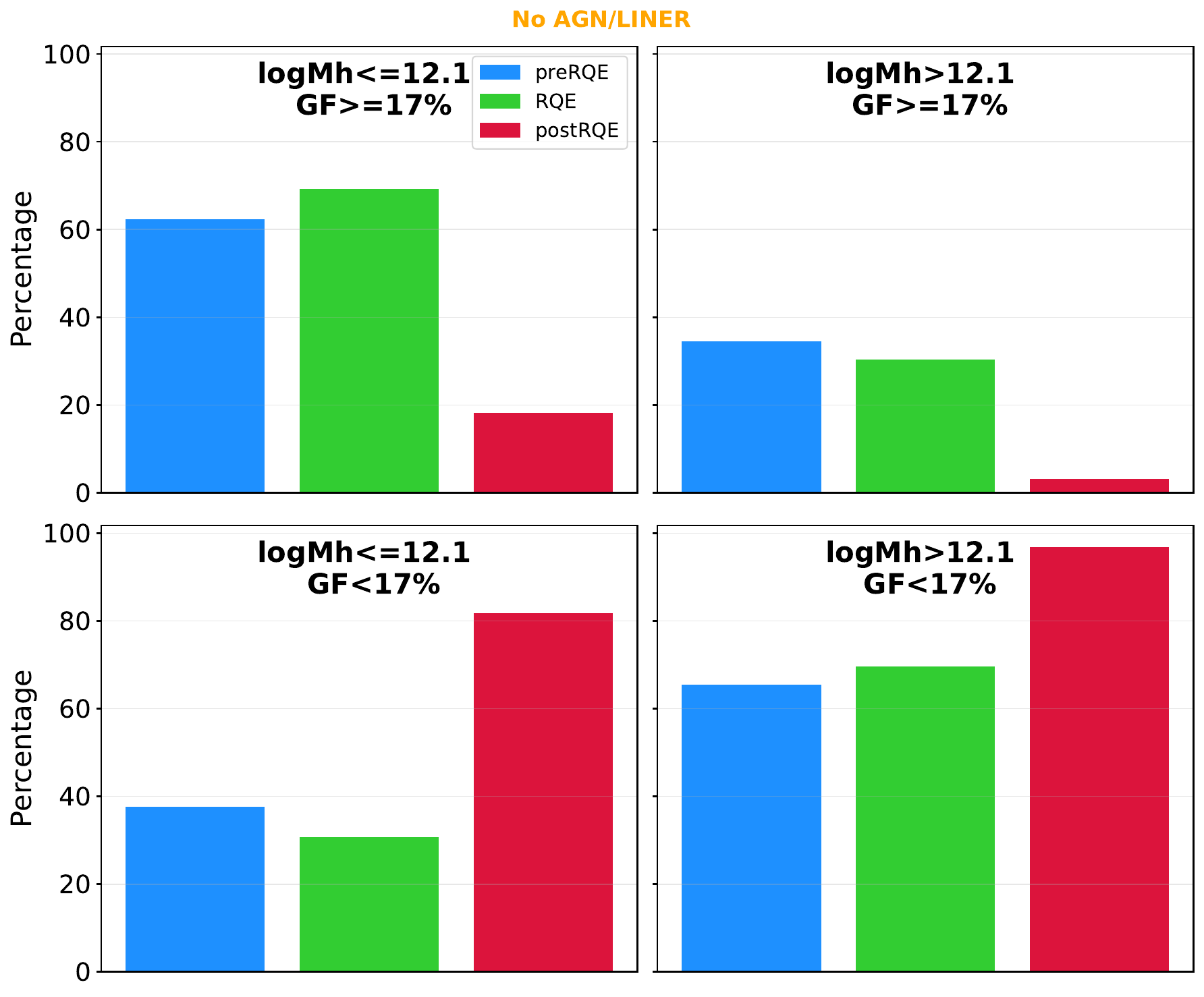}{0.32\textwidth}{}\fig{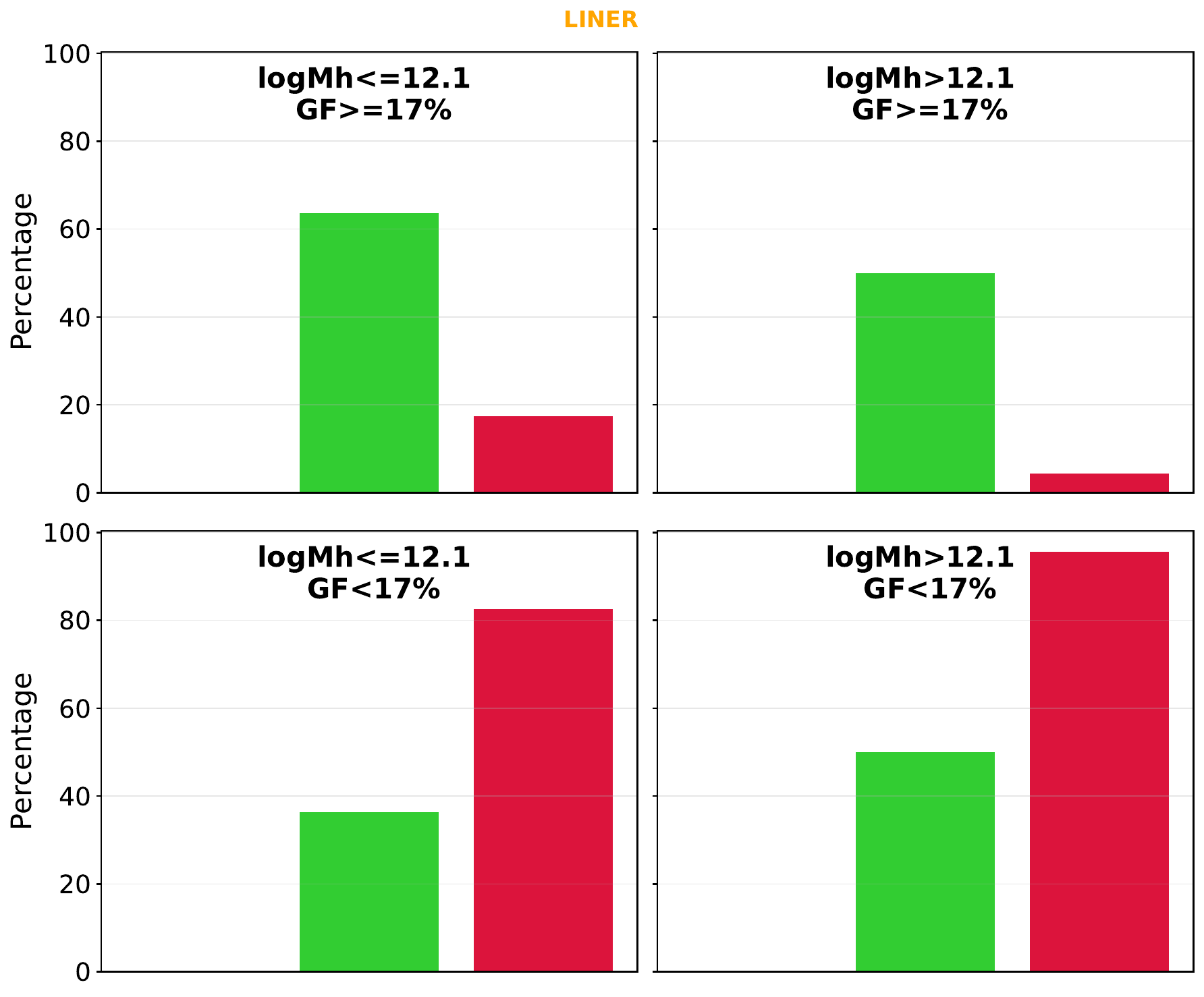}{0.32\textwidth}{}\fig{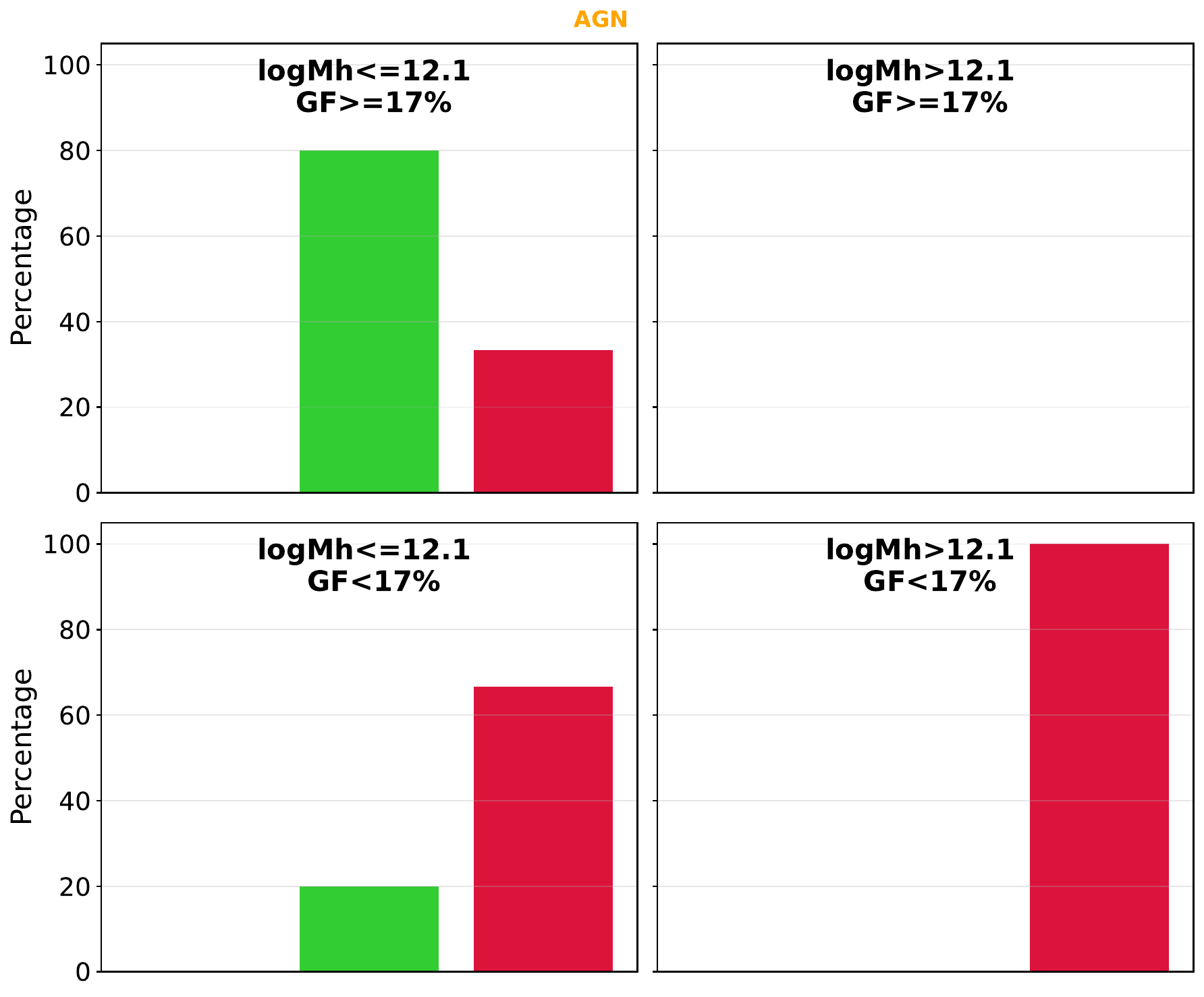}{0.32\textwidth}{}}
\caption{Statistics on the distribution of RQEs across gas fraction and halo mass categories for different emission line classifications. The plot is divided into three columns representing No AGN/LINER, LINER, and AGN galaxies. Each column is subdivided into four quadrants based on halo mass (log$M_{\text{halo}}$) and gas fraction ($f_{\text{gas}}$) thresholds. The y-axis shows the percentage of galaxies in each category. Blue bars represent preRQEs, green bars indicate RQEs, and red bars show postRQEs. Each quadrant displays the percentage of RQEs in that quadrant relative to the total RQEs in the same halo mass range, along with a comparison of similar statistics for preRQE and postRQE populations. Statistics are based on a log$M_{\text{halo}}$ cut of 12.1 and an $f_{\text{gas}}$ cut of 17\%, chosen based on a visual distinction observed in either the amount of gas in RQEs or their location in the gas fraction vs. halo mass space. Notably, this halo mass and gas fraction cut also corresponds to the median halo mass and median gas fraction for RQEs.}
\label{fig:stat_plot}
\end{figure*}

Moreover, in the no AGN/LINER population, a decrease in gas fraction with increasing halo mass is evident for both preRQEs and postRQEs. This trend, however, is not observed in postRQEs with LINER or AGN activity, where most candidates are found in log$M_{\text{halo}} >$ 12.1 M$_{\odot}$. Interestingly, much like the RQEs without AGN/LINER, a significant portion of preRQEs shows $f_{\text{gas}} \geq$ 17\% when residing in halos with log$M_{\text{halo}} \leq$ 12.1 M$_{\odot}$. These quantitative details are also illustrated in the multi-panel quadrant plot in Figure \ref{fig:stat_plot}, where the top left quadrant in the no AGN/LINER panel compares the percentage of RQEs with $f_{\text{gas}} \geq$ 17\% residing in log$M_{\text{halo}} \leq$ 12.1 M$_{\odot}$ to that of preRQEs and postRQEs, while similar comparisons are made in the other quadrants.

These observations suggest that RQEs residing in lower-mass halos (log$M_{\text{halo}} \leq 12.1 M_{\odot}$) likely underwent different quenching pathways compared to those in more massive halos. Notably, log$M_{\text{halo}} = 12.1 M_{\odot}$ corresponds closely with the transition mass ($\sim 10^{12}$ M$_{\odot}$), where galaxies are theorized to shift from cold-mode to hot-mode accretion \citep{2006MNRAS.368....2D, 2011MNRAS.415.2782V, 2012MNRAS.427.1816G}. This transition could be crucial in driving the different quenching mechanisms observed in RQEs on either side of log$M_{\text{halo}} = 12.1 M_{\odot}$. For RQEs in less massive halos (log$M_{\text{halo}} \leq 12.1 M_{\odot}$), the observation of gas fractions comparable to their plausible precursors (preRQEs) with $f_{\text{gas}} \geq$ 17\% hints at a potential rapid quenching pathway, possibly through major mergers. This mechanism is prevalent in small group-scale halos \citep{2008ApJS..175..356H}, and the fact that all RQEs residing below log$M_{\text{halo}} = 12.1 M_{\odot}$ fall within this scale further suggests major mergers as a plausible quenching pathway. Additionally, RQEs with LINER or AGN emission in this halo mass range exhibit similarly high gas fractions, implying they may represent different phases of the same evolutionary sequence. The presence of AGN/LINER activity in some of these gas-rich RQEs might indicate a phase where the active nucleus maintains the quenched state by heating the gas or inducing turbulence, thereby preventing the gas from cooling and condensing into the dense clouds necessary for star formation.

\begin{figure}
    \centering
    \includegraphics[width=\columnwidth]{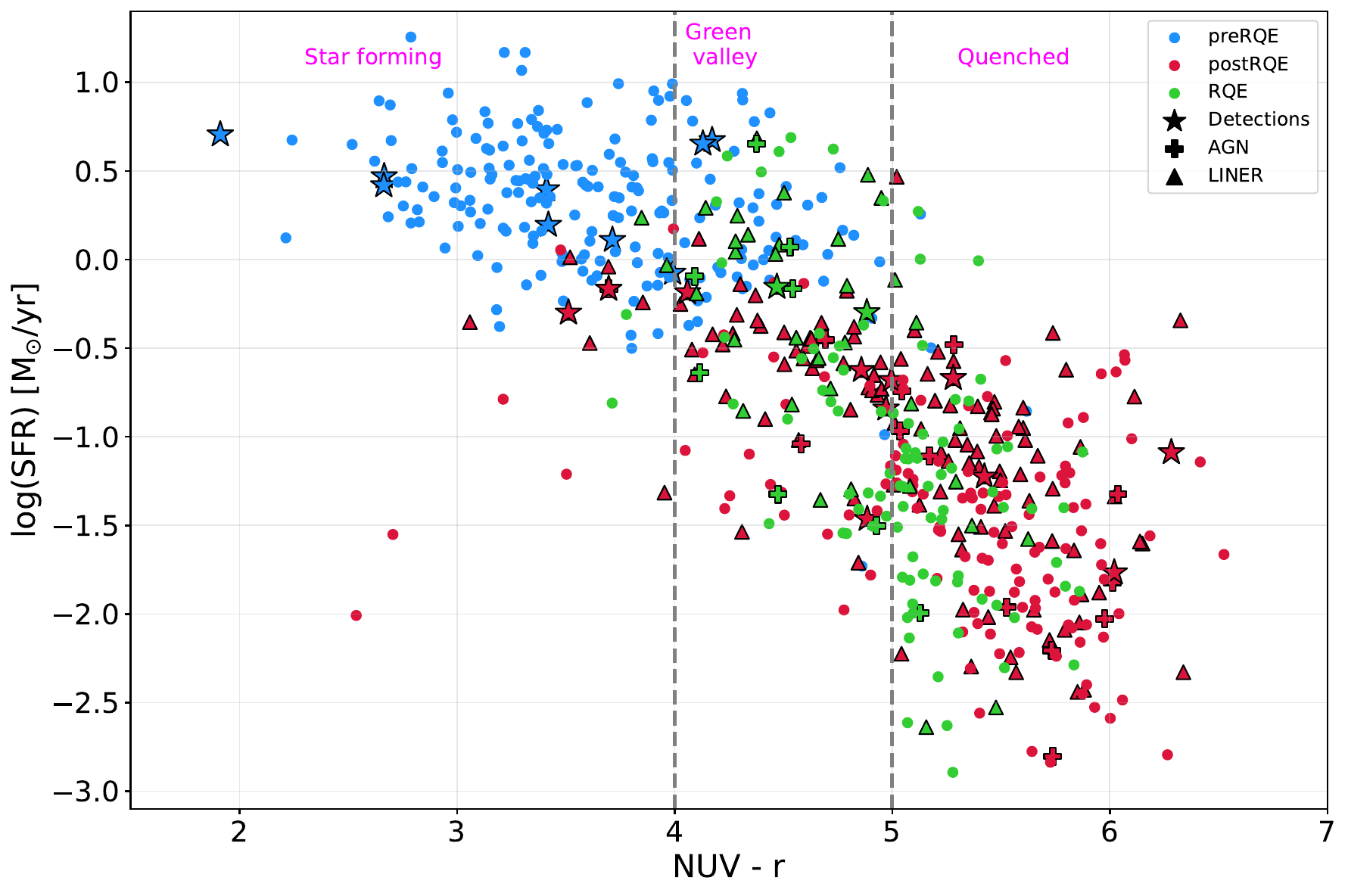}
    \caption{logSFR vs (NUV-r) comparison of RQE with preRQEs and postRQEs. Total SFR values are taken from GSWLC-M2 and X2 catalogs. X2 values were used where M2 was not available. The plot reveals that RQEs located within the Green Valley (GV) exhibit higher SFRs compared to those situated near the boundary of the GV and quenched regions, with a difference of at least 0.5 dex. This gradient in SFR underscores the transitional nature of RQEs as they evolve from star-forming to quiescent states, highlighting their intermediate position in the quenching process.}
    \label{fig:SFR-color}
\end{figure}

In the middle row of Figure \ref{fig:disc_plot}, we explore the correlation between recent star formation activity ($NUV-r$) and gas fraction ($f_{\text{gas}}$). The left panel indicates that the majority of RQEs are situated at the edge of the Green Valley (GV) and the quenched region, seemingly transitioning from the GV toward a more quenched state (see also Figure \ref{fig:SFR-color}). For RQEs with LINER and AGN activity, most ($>80\%$) are located within the GV, including two HI detections. RQEs in the GV region across all emission types generally exhibit higher gas fractions, comparable to their star-forming counterparts, than those in the quenched zone. This could suggest an evolutionary sequence from star-forming preRQEs to quiescent RQEs that have undergone AGN and LINER phases.

In Figure \ref{fig:SFR-color}, we compare the current star formation rate (logSFR) with signatures of recent past star formation activity ($NUV-r$) and observe that RQEs in the GV region, regardless of their emission type, display SFRs similar to preRQEs. This observation suggests a trend of declining SFR as RQEs transition from the GV to the quenched regime. This trend implies a possible evolutionary trajectory for RQEs: starting from the preRQE phase with SFRs akin to star-forming galaxies, progressing through the GV as RQEs, moving toward the boundary of the GV and quenched regions with lower SFR, and ultimately evolving into fully quenched postRQEs. However, the transition from preRQEs to RQEs and then to postRQEs is not straightforward, as indicated by the bottom row plot in Figure \ref{fig:disc_plot}.

The bottom row of Figure \ref{fig:disc_plot} examines the relationship between light-weighted stellar age ($Age_{\text{lw}}$) and $NUV-r$ color. By definition, RQEs have a light-weighted age ($Age_{\text{lw}}$) of less than 3 Gyr, while postRQEs are characterized by ages greater than 3 Gyr. No specific age criteria were applied to preRQEs. In the no AGN/LINER category, the majority of RQEs are clustered near the boundary of the Green Valley (GV) with $Age_{\text{lw}} \leq 3$ Gyr. The ages of preRQEs are more dispersed, ranging from 1 Gyr to 5 Gyr, and are predominantly found in the star-forming region.

Most postRQEs appear to lie within the quenched region and show typical light-weighted ages greater than 5 Gyr. It is noteworthy that there are few, if any, postRQEs in the quenched region or near the boundary of the GV and quenched regions, suggesting that RQEs with no AGN or LINER emissions did not undergo a straightforward passive evolution to transition into postRQEs with similar emission properties. The age gap between Seyfert RQEs and postRQEs is also evident, with the majority of Seyfert-hosting postRQEs appearing to be long quenched.

Interestingly, there is no significant age gap between LINER postRQEs and preRQEs, particularly in the Green Valley (GV) region, suggesting a potential evolutionary pathway where GV LINER RQEs passively evolve into GV LINER postRQEs. During this evolution, several processes could contribute to their transformation: the consumption of gas through small-scale, localized star formation, influenced by the effectiveness and spatial extent of LINER emission; partial gas expulsion via supernova feedback; and potential rejuvenation through minor merger events. These processes, either individually or in combination, may facilitate the passive evolution of LINER RQEs into LINER postRQEs with halo masses greater than $10^{12.1}$ M$_{\odot}$. TThe persistence of LINER emission throughout this evolution indicates its role in regulating star formation, preventing a complete depletion of the gas reservoir, and maintaining the quiescent state of these galaxies.

The multi-panel plots presented in Figure \ref{fig:disc_plot}, contrasting RQEs with their plausible precursors and descendants across various gas and star formation properties, offer valuable insights into potential evolutionary pathways. These plots suggest two distinct quenching routes for RQEs: one for those residing in lower halo masses (log$M_{\text{halo}} \leq 12.1$ M$_{\odot}$) and another for those in higher halo masses (log$M_{\text{halo}} > 12.1$ M$_{\odot}$).

RQEs typically reside in halo masses where efficient gas accretion is expected to occur (log$M_{\text{halo}} < 13$ M$_{\odot}$). Below log$M_{\text{halo}} = 12$ M$_{\odot}$, accretion predominantly takes place via cold-mode accretion, allowing continuous star formation. In contrast, gas accretion onto central galaxies within halo masses of approximately log$M_{\text{halo}} \sim 12-13$ M$_{\odot}$ occurs through hot-mode accretion, where the gas is heated to temperatures around $10^5$ K. To maintain such high temperatures and prevent the gas from cooling and triggering star formation in the central region, additional heating mechanisms are necessary \citep{2015MNRAS.447..374G, 2003MNRAS.345..349B, 2005MNRAS.363....2K, 2010MNRAS.407..749G, 2013MNRAS.429.3353N}. Therefore, RQEs are likely to form stars unless halted by some heating mechanism that sustains their quiescent state.

Research by \cite{2011MNRAS.415.2782V} demonstrates that AGN feedback can significantly inhibit gas accretion in galaxies, especially in the hot mode due to the diffuse nature of the hot-mode gas. Additionally, \cite{2011MNRAS.412.1965M} show that AGN feedback, by removing low-entropy gas from high-redshift halos ($z > 2$), increases the entropy and reduces the cooling rates of hot gas in lower-redshift halos. This process effectively prolongs the quenched state in the central galaxy.

For RQEs with halo masses greater than 12.1 M$_{\odot}$, AGN feedback—particularly from Seyfert or LINER activity—can likely maintain high temperatures, quenching star formation. This quenching is particularly effective in Seyfert-hosting RQEs, given Seyferts' relatively short active period of about 100 million years. LINERs, however, may not be as effective in sustaining widespread quenching over extended periods. In a small population of RQEs with high halo masses (logM$_{\text{halo}} \geq 12.84$ M$_{\odot}$), quenching could be maintained through unresolved gravitational heating, which keeps gas temperatures high \citep{2008MNRAS.383..119D}. In the absence of any significant heating mechanism, RQEs without AGN/LINER activity might have accreted gas that is too diffuse and stable to collapse and form stars. Even as the gas cools over time, it might not reach the critical HI column density ($N_{\text{HI}} > 10^{21}$ atoms cm$^{-2}$) required for star formation.

\begin{figure*}
    \centering
    \includegraphics[width= \textwidth]{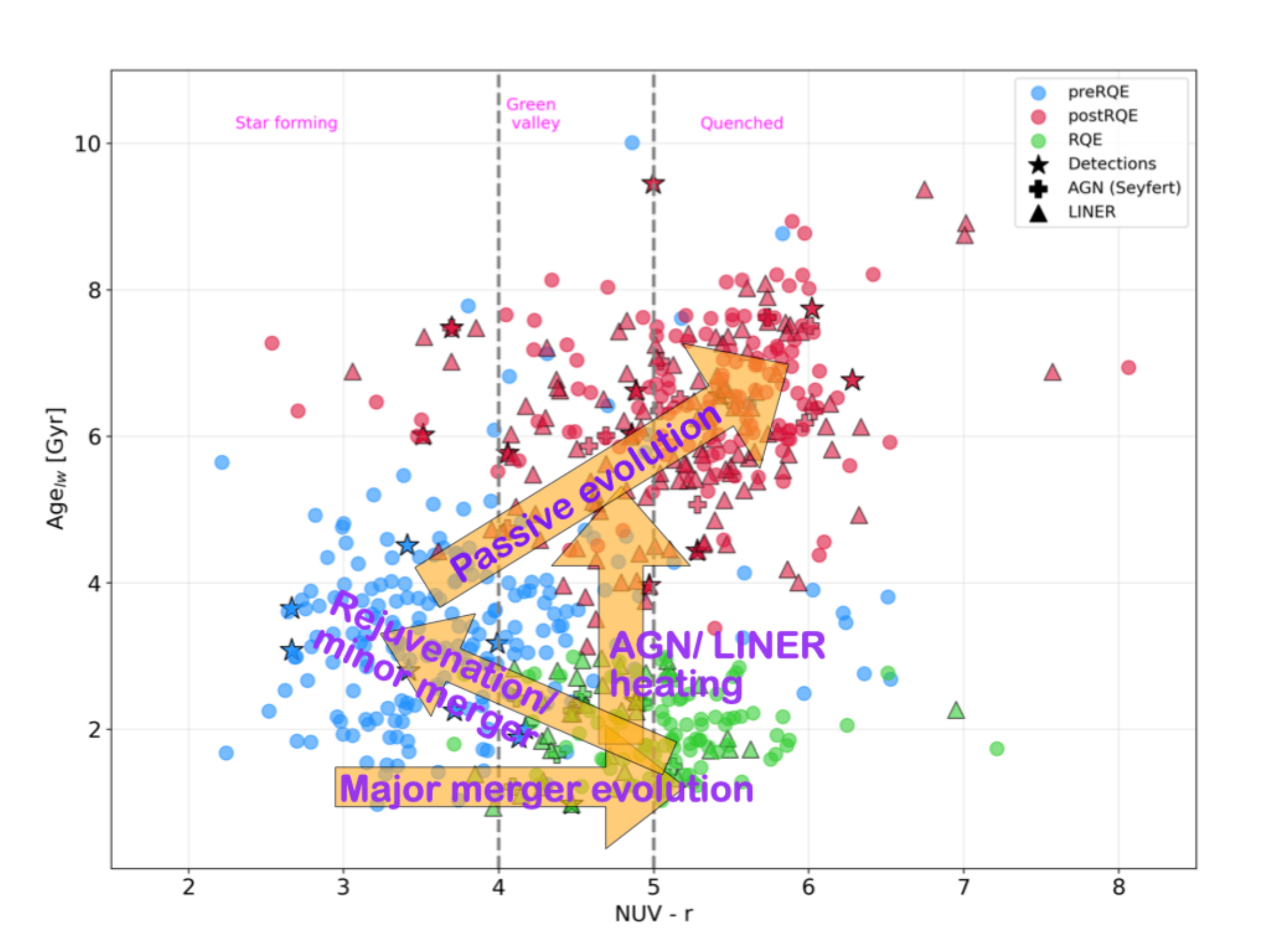}
    \caption{Illustration of potential evolutionary pathways RQEs based on their gas content, halo mass, star formation activity, and emission properties. The plot displays the light-weighted stellar age ($Age_{\text{lw}}$) versus $NUV-r$ color for preRQEs, RQEs, and postRQEs, with AGN (Seyfert), LINERS and HI detections indicated by separate markers. The vertical dashed lines at $NUV-r$ = 4 and 5 delineate the boundaries between the star-forming, Green Valley (GV), and quenched regions. The proposed evolutionary pathways for RQEs include: (a) rapid quenching via major merger, followed by a brief AGN phase with sustained LINER activity, keeping the galaxy in the green valley before passive evolution into long-term quenched ellipticals; and (b) rapid quenching via major merger, followed by brief rejuvenation in star formation due to minor mergers, before passive evolution into more massive, long-term quenched ellipticals.}
    \label{fig:evol_cartoon}
\end{figure*}

Based on the results and discussions so far, we propose a plausible evolutionary pathway that includes the quenching steps an RQE might have undergone. Our analysis reveals a clear distinction at a halo mass of 12.1 M$_{\odot}$, evident in our comparisons of gas content, star formation, and other properties. It is plausible that RQEs with halo masses below 12.1 M$_{\odot}$ underwent a rapid quenching process, possibly through major mergers—a scenario that is more applicable to RQEs in lower-mass halos, where major mergers are more common \citep{2008ApJS..175..356H}. The observed HI gas surplus in RQEs could have resulted from such a major merger event. After merging, these RQEs likely experienced a brief AGN phase (particularly Seyfert activity, which lasts less than 100 Myr) and a sustained period of centralized LINER emission. This sequence would have kept the intergalactic environment hot, preventing gas from cooling and falling into the center to form stars. Some LINER RQEs may have undergone small-scale, localized star formation, consuming some of their gas, and eventually passively evolved into LINER postRQEs with halo masses similar to their precursors. Others might have experienced rejuvenation through minor mergers with gas-rich companions \citep{2017A&A...598A..45G}, increasing their halo masses and eventually evolving into LINER postRQEs with high halo masses by depleting much of their star-forming gas. 
A conceptual illustration of these potential evolutionary pathways and the associated quenching scenarios is depicted in Figure \ref{fig:evol_cartoon}, which suggests two primary routes: (a) rapid quenching via major merger, followed by a brief AGN phase and sustained LINER activity, leading to passive evolution into long-term quenched ellipticals; and (b) rapid quenching followed by a brief rejuvenation in star formation due to minor mergers, eventually resulting in more massive, long-term quenched ellipticals.

While our neutral HI analysis supports the proposed evolutionary pathways of RQEs, definitive conclusions require further investigation. A comprehensive environmental study of these RQEs (Deo et al., \textit{in preparation}), which will assess their local and large-scale environments, is essential for contextualizing their evolution. This analysis will help determine whether external processes are contributing to star formation quenching in RQEs, despite their substantial HI reservoirs.

Moreover, high-resolution analysis of HI gas properties—including mass, morphology, kinematics, asymmetry, and temperature—is essential to fully understand the quenching mechanisms at play. Such data would enable the creation of various moment maps, revealing vital details about the HI gas reservoirs in RQEs. The zeroth moment map (integrated intensity, $\int I_{v}dv$) could quantify the total gas amount and its spatial distribution, highlighting regions of enhanced or depleted gas density. The first moment map (intensity-weighted velocity, $\int vI_{v}dv$) would illuminate the galaxy's kinematic structure, revealing rotation patterns, streaming motions, or signs of disturbances indicative of interactions or mergers. The second moment map ($\int v^2I_{v}dv$) could provide insights into gas temperature and turbulence, offering clues about energetic processes affecting the interstellar medium. 

These analyses would be instrumental in characterizing the properties of the cold, star-forming HI gas, thereby helping to identify the most likely quenching mechanisms. For instance, if the environmental study reveals that RQEs are not as isolated and truly central as previously thought, environmental quenching processes such as ram pressure stripping or galaxy harassment could be significant factors. Alternatively, environmental interactions like minor mergers could explain the presence of surplus HI by contributing additional gas to the galaxies.

Future work combining these detailed HI observations with our current findings will be pivotal in constructing a comprehensive picture of RQE evolution. This holistic approach promises to shed light not only on the quenching processes in RQEs but also on broader questions of galaxy evolution and the diverse pathways through which galaxies transition from star-forming to quiescent states, particularly those residing in low-density environments. By integrating environmental context with detailed gas properties, we can better understand the balance between internal and external factors in galaxy quenching, ultimately contributing to a more nuanced model of galaxy evolution.

\section{Conclusions}\label{sec:con}
This study presents a multi-wavelength investigation of Recently Quenched Elliptical galaxies, with a particular focus on their HI gas content and star formation properties. By comparing RQEs with their plausible evolutionary precursors (preRQEs) and descendants (postRQEs), we aim to uncover clues that may explain the unusual quenching of star formation in these galaxies. The subpopulations were carefully selected to explore potential evolutionary connections with RQEs, based on similarities in their HI content and related properties. Our analysis has revealed several key findings that challenge conventional understanding of galaxy quenching processes:

\begin{itemize}
    \item RQEs retain substantial HI gas reservoirs despite their quenched state, with gas fractions often comparable to their star-forming precursors. This unexpected gas richness suggests that quenching in these systems is not primarily driven by gas depletion or exhaustion.
    \item We identify a critical halo mass threshold at $\log M_{\text{halo}} = 12.1 M_{\odot}$, which appears to delineate different evolutionary pathways for RQEs. This mass closely corresponds to the transition between cold-mode and hot-mode gas accretion in theoretical models.
    \item RQEs in lower-mass halos ($\log M_{\text{halo}} < 12.1 M_{\odot}$) likely undergo rapid quenching, possibly triggered by major mergers. These events may be responsible for the observed HI gas surplus and the initiation of a brief AGN phase followed by sustained LINER activity.
    \item We propose two primary evolutionary pathways for RQEs: (a) rapid quenching via major merger, followed by AGN/LINER activity and passive evolution, and (b) rapid quenching followed by brief rejuvenation through minor mergers before final passive evolution.
    \item The presence of LINER emission in many RQEs suggests its potential role in regulating star formation without completely depleting the gas reservoir, possibly explaining the extended residence of these galaxies in the Green Valley.
    \item The presence of hot gas in RQEs, possibly due to hot mode accretion, might be a crucial factor in sustaining their quenched state in RQEs residing in high-mass halos ($\log M_{\text{halo}} > 12.1 M_{\odot}$). Hot gas can be diffuse in nature and have low HI column density which could be the reason to not form stars even if gas had the time to cool down.
   \item RQEs have a complex evolution and probably follow a path in which it is evolved from young light weighted age preRQEs via rapid quenching and then eolved in about $\sim 1$ Gyr to high light weighted age ($>3$ Gyr) preRQEs via rejuvenation due to minor merger and then passively evolved to long quenched postRQEs in few billion years. 
\end{itemize}

These findings collectively underscore the complex nature of RQE evolution, indicating a path from young light-weighted age preRQEs through rapid quenching, possible rejuvenation due to minor mergers, and eventual passive evolution to long-quenched postRQEs over several billion years. This evolutionary sequence highlights the intricate interplay between gas dynamics, star formation, and AGN activity in shaping galaxy evolution.
Looking ahead, future high-resolution observations with facilities like the Square Kilometre Array (SKA) will be crucial for mapping the spatial distribution, kinematics, and thermodynamic properties of HI gas in RQEs. Such detailed studies, combined with environmental analyses, will further constrain quenching mechanisms and contribute to a more comprehensive understanding of galaxy evolution in low-density environments.

\section*{Acknowledgments}
The authors would like to thank DM, SV, KM and RK for their comments and suggestions in the manuscript. 
DKD also thanks PR, OB, TD and EM for their support in various ways that helped in completion of this project.
This work made use of data from the Sloan Digital Sky Survey (SDSS). The SDSS Archive was created and distributed with the financial support of the Alfred P. Sloan Foundation, the Participating Institutions, the National Science Foundation, the U.S. Department of Energy, the National Aeronautics and Space Administration, the Japanese Monbukagakusho, the Max Planck Society, and the Higher Education Funding Council for England.
We also utilized data from the Galaxy Evolution Explorer (GALEX), a NASA Small Explorer launched in April 2003. GALEX is operated by the California Institute of Technology under NASA contract NAS5-98034 in cooperation with the Centre National d’Etudes Spatiales of France and the Korean Ministry of Science and Technology. GALEX data were obtained from the MAST data archive at the Space Telescope Science Institute, which is operated by the Association of Universities for Research in Astronomy, Inc., under NASA contract NAS5-26555.
This publication also made use of the Green Bank Telescope (GBT) Legacy Archive. The Green Bank Observatory is a facility of the National Science Foundation operated under cooperative agreement by Associated Universities, Inc.
We employed several software tools and services, including NASA’s Astrophysics Data System Bibliographic Services, TOPCAT (Tools for OPerations on Catalogues And Tables, \cite{2005ASPC..347...29T}), 
and ``K-corrections calculator" service available at \url{http://kcor.sai.msu.ru/} \citep{2010MNRAS.405.1409C, 2012MNRAS.419.1727C}. 
Additionally, this research benefited from the use of Astropy\footnote{http://www.astropy.org},
a community developed core Python package for Astronomy \citep{2013A&A...558A..33A, 2018AJ....156..123A}, Matplotlib \citep{2007CSE.....9...90H, matplotlib} and 
LLMs (Large Language Models) such as GPT-4 \citep{openai2023} and Claude 3.5 Sonnet \citep{anthropic2023claude}. 
          
\section*{Data Availability}
Value added catalog used in this paper can be made available through contacting main author whose email id is mentioned in the front page of the manuscript.

\software{astropy \citep{2013A&A...558A..33A,2018AJ....156..123A}, Numpy \citep{numpy}, Scipy \citep{scipy}, matplotlib \citep{2007CSE.....9...90H, matplotlib}, seaborn \citep{Waskom2021}, Pandas \citep{pandas}, TOPCAT \citep{2005ASPC..347...29T}, K-corrections calculator (\citep{2010MNRAS.405.1409C, 2012MNRAS.419.1727C}; \url{http://kcor.sai.msu.ru/})}

\bibliography{paper1}{}
\bibliographystyle{aasjournal}



\end{document}